\documentclass[12pt]{iopart}
\usepackage{graphicx}

\begin{document}

\title{Vibrating cavities - A numerical approach} 

\author{Marcus Ruser \footnote[2]{\tt Marcus.Ruser@physics.unige.ch}}
\address{D{\'e}partement de Physique Th{\'e}orique, Universit{\'e}
de Gen{\`e}ve, 24 quai Ernest Ansermet, CH-1211 Gen{\`e}ve 4, Switzerland}

\date{\today}

\begin{abstract}

We present a general formalism allowing for efficient numerical calculation of the production of massless scalar particles from vacuum in a one-dimensional dynamical cavity, i.e. the dynamical Casimir effect.  
By introducing a particular parametrization for the time evolution of the field modes inside the cavity we derive a coupled system of first-order linear differential equations. 
The solutions to this system determine the number of created particles and can be found by means of numerical methods for arbitrary motions of the walls of the cavity. To demonstrate the method which accounts for the intermode coupling we investigate the creation of massless scalar particles in a 
one-dimensional vibrating cavity by means of three particular cavity motions. We compare the numerical results with analytical predictions as well as a different numerical approach.

\end{abstract}

\pacs{03.65.-w, 03.70.+k, 12.20.Ds, 42.50.Lc}

\maketitle

\section{Introduction}

The decay of the ground state of quantum field theory, the vacuum, via the creation of real particles 
due to disturbances of the vacuum caused by time dependent external conditions demonstrates the highly non-trivial nature of the quantum vacuum. 

The dynamical or non-stationary Casimir effect (see \cite{Bordag:2001, Dodonov:2001} and references therein) represents one particular example out of the variety of fascinating phenomena occurring in 
the sector of quantum field theory under the influence of external conditions \cite{Bordag:1996,Bordag:2002,Grib:1994}. In this scenario the quantum vacuum responds to time-varying boundaries like moving mirrors by its decay via the creation of particles out of virtual quantum fluctuations. For calculations involving a single mirror see, e.g., \cite{Fulling:1976, Davis:1977, Ford:1982,  Neto:1996}.

In a dynamical (one-dimensional) cavity with one mirror residing at a fixed position, say $x_1=0$, and a second mirror changing its position $x_2$ according to a given dynamics $x_2(t)\equiv l(t)$ the source of particle creation is twofold. The so-called squeezing of the vacuum \cite{Schuetzhold:1998} due to the dynamical change of the quantization volume (the size of the cavity) yields time dependent  eigenfrequencies of the field modes. Furthermore, boundary conditions imposed on the field inside the cavity at the positions of the mirrors cause a time dependent coupling between all field modes. This is denoted as the acceleration effect \cite{Schuetzhold:1998}. Being proportional to the velocity $\dot{l}(t)$ of the boundary motion this coupling of the time evolution of all field modes inside a dynamical cavity distinguishes the vacuum decay in the dynamical Casimir effect from other quantum vacuum effects like particle creation in an expanding universe (see, e.g., \cite{Birrell:1982}) where the time dependence of eigenfrequencies is the only source of particle creation. This acceleration effect makes it, apart from exceptional cases, impossible to find analytical solutions to the field equations.  
               
A particular example for which analytical solutions are known and the particle production can be described by closed form expressions is the scenario of uniformly moving boundaries \cite{Moore:1970} (see also \cite{Fulling:1976,Castagnino:1984}). Moore \cite{Moore:1970} found that the mode functions $\phi_k(t,x)$ inside a one-dimensional dynamical cavity $0 \le x \le l(t)$ satisfying both, the wave equation $\left[\partial_t^2-\partial_x^2\right]\phi_k(t,x)=0$ with $\phi_k(t,0)=0$ and the non-stationary Dirichlet boundary condition $\phi_k[t, l(t)]=0$, can be written (up to a normalization) as $\phi_k(t,x)=\exp\{ik\pi R(t+x)\} - \exp\{ik\pi R(t-x)\}$ provided that $R(z)$ satisfies the equation $R[t+l(t)]-R[t-l(t)]=2$. This equation can be solved and $R(z)$ evaluated exactly, for instance, in the scenario in which $l(t)$ undergoes a uniform motion. 

Another particularly interesting scenario are so-called vibrating cavities \cite{Lambrecht:1996}. Thereby the field is confined between two parallel walls whose distance $l(t)$ changes periodically in time: 
\begin{equation}   
l(t)=l_0\left[1+\epsilon\,\delta(t)\right]\; , \;{\rm with}\,\, \delta(t+T)=\delta(t).
\label{l of t}
\end{equation}
In the case where the frequency of the wall oscillations $\omega_{\rm cav}$ is twice the frequency of some quantum mode inside the unperturbed cavity, resonant particle (photon) creation occurs which makes this scenario the most promising candidate for an experimental verification of this pure quantum effect \cite{Dodonov:1996}. 

\begin{figure}
\begin{center}
\begin{tabular}{cc}
\includegraphics[height=5.2cm]{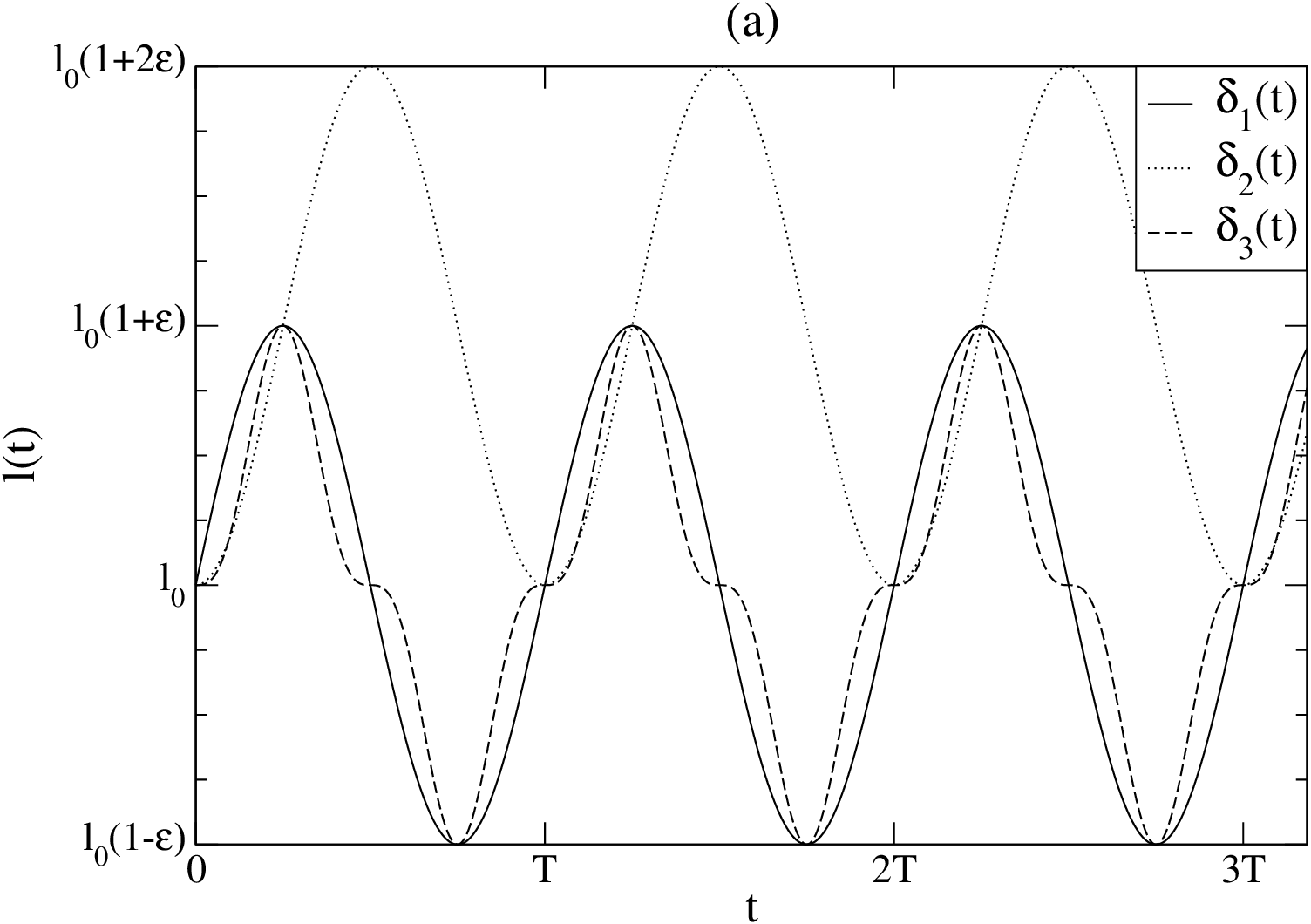}
&
\includegraphics[height=5.2cm]{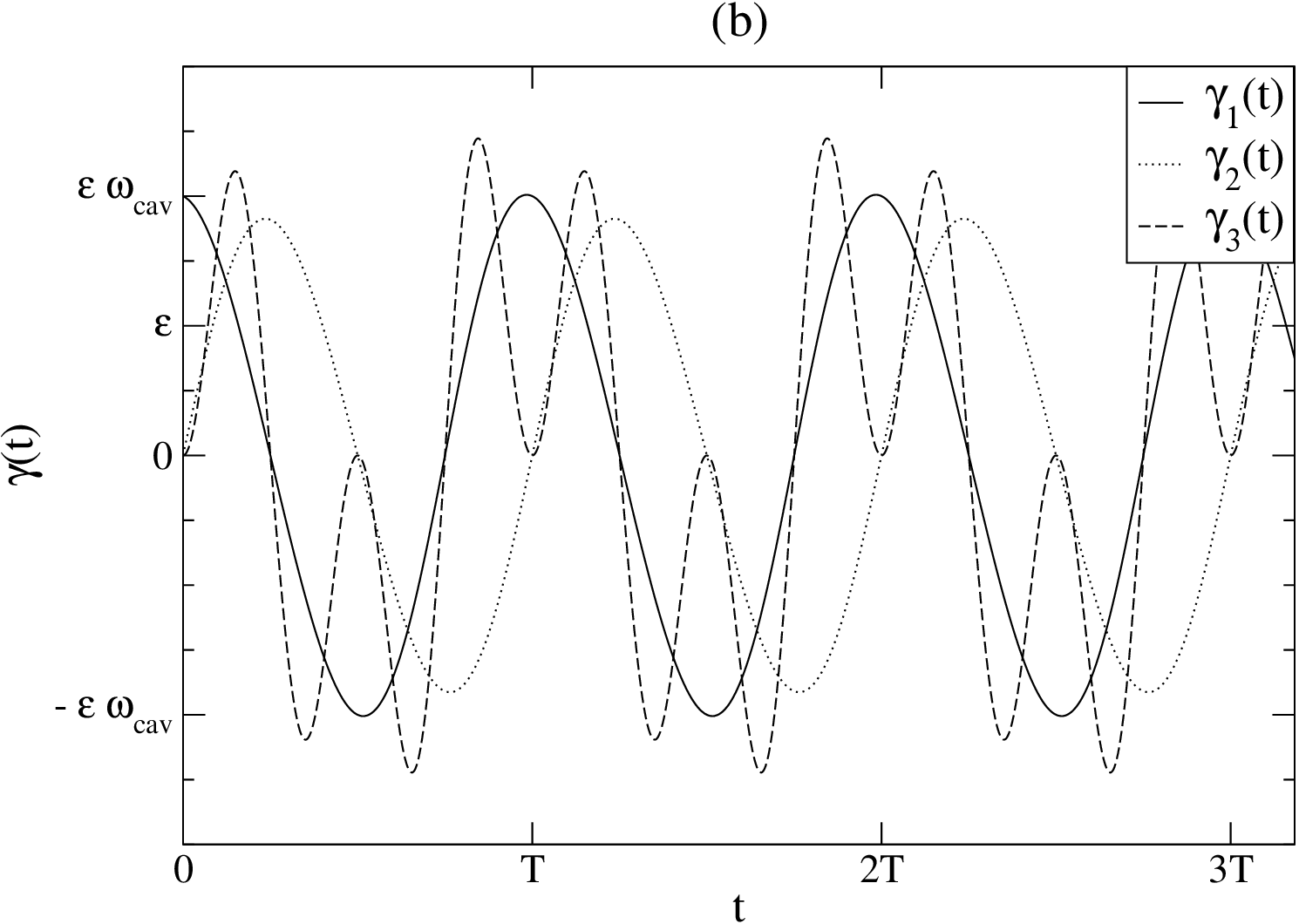}
\end{tabular}
\caption{Illustration of (a) the cavity vibrations $l(t)$ for $k=1,2$ and $3$ studied in the present paper and (b) the corresponding function $\gamma(t)=\frac{\dot{l}(t)}{l(t)}$ determining the time-dependent coupling of all field modes (acceleration effect).
                      \label{figure 1}}
\end{center}
\end{figure}

Let us parametrize the vibrations of the (one-dimensional) cavity as
\begin{equation}
\delta_k(t)=a_k\,\sin^k(\omega\,t)=a_k\,\sin^k\left(\frac{2}{a_k}\,\Omega_n^0\,t\right)
\label{background}
\end{equation}
with $a_k=1$ for $k$ odd and $a_k=2$ for $k$ even. Thereby $\Omega_n^0$ denotes the frequency of a quantum mode inside the unperturbed cavity of length $l_0$\footnote[1]{This parametrization ensures that the cavity frequency is always twice the frequency of an unperturbed mode, i.e. $\omega_{\rm cav}=2\Omega_n^0$, and that the total change of the size of the cavity $\Delta l$ during one period is always $2\epsilon\,l_0$ [see also Fig. \ref{figure 1}(a)].}. 

For a one-dimensional cavity oscillating with $\delta(t)=\delta_1(t)$ in the main resonance, i.e. the frequency of the cavity vibration is twice the frequency of the first field mode inside the unperturbed cavity $\omega_{\rm cav}=2\Omega_1^0$, analytical solutions for the total particle number as well as the number of particles created in the resonance mode $k=1$ valid for all times have been found in \cite{Dodonov:1996} (see also \cite{Dodonov:1996a}) under the assumption that the amplitude of the oscillations is very small compared to one ($\epsilon \ll 1$).
For the same scenario the rate of particle creation for higher frequency modes and in the limit $\frac{\epsilon}{l_0}t \gg1$ has been derived in \cite{Dodonov:1993} by evaluating an approximate solution to Moore's equation.  
 In \cite{Ji:1997} an analytical expression for the number of created particles has been found for more general cavity frequencies but short times only, i.e. $\frac{\epsilon}{l_0}t \ll 1$. An improved analytical solution to Moore's equation is derived in the work \cite{Dalvit:1998} where the authors study the energy density inside the cavity (see also \cite{Dalvit:1999,Cole:1995, Law:1994,Wegrzyn:2001}) and show that the total energy in the cavity increases exponentially which was also derived in \cite{Dodonov:1996}.

The interaction (backreaction) between the cavity motion and the quantum vacuum inside the cavity has been studied in, e.g., \cite{Law:1995,Golestanian:1997,Cole:2001}. For work regarding the more realistic case of a three-dimensional cavity see, e.g.,  
\cite{Dodonov:1996,Crocce:2001,Dodonov:2003,Mundarain:1998}.

The analytical results mentioned above have been derived  by means of approximations 
such as small amplitudes of the vibrations $(\epsilon \ll 1)$ and some of them are valid only 
in particular time ranges. Thus even in the extensively studied vibrating cavity scenario there are still open questions. How does the particle creation look when the assumption $\epsilon \ll 1$ is no longer valid and the approximative calculations break down \footnote[2]{Note that because of Eqs. (\ref{l of t}) and (\ref{background}) the amplitude of the velocity of the mirror is $\frac{v}{c}\propto \epsilon\,\omega_{\rm cav}$ and thus care has to be taken by increasing the amplitude of the vibrations $\epsilon$ such that the velocity of the mirror never exceeds the speed of light. }. 
How does particle production behave for higher resonance frequencies as studied in \cite{Ji:1997} but large times? In what manner does the behaviour of the particle production change when considering different kinds of cavity vibrations, for instance $\delta_1(t)$ and $\delta_2(t)$?
What happens in the case of detuning (see, e.g.,  \cite{Crocce:2001,Dodonov:1998,Dodonov:1998a})
when the frequency of the cavity vibrations does not exactly match the resonance condition? 

In order to answer these questions and to study the vacuum decay in the dynamical Casimir effect for a variety of possibly interesting scenarios where less or even nothing is known analytically like arbitrary wall motions, cavity vibrations with time-varying amplitudes $\epsilon(t)$ and massive fields, the problem has to be attacked by using numerical methods to account for the coupling of the time dependence of all field modes. 

The goal of the paper at hand is to present a particular parametrization for the time evolution of the 
field modes of a real massless scalar field in a one-dimensional empty cavity $[0, l(t)]$ allowing for efficient numerical calculation of the particle production for arbitrary motions $l(t)$ of the boundary.  
A different numerical approach has been  recently presented in \cite{Antunes} where the author studies the creation of particles in a one-dimensional cavity subject to vibrations of the form $\delta_2$.

The structure of the paper is as follows. In the next section we briefly review the procedure of canonical quantization of a real scalar field in an empty cavity. The third section is reserved for the presentation 
of the formalism yielding a coupled system of first-order differential equations from which the number of produced particles can be deduced by means of standard numerics. In particular we focus our studies of the particle creation in a vibrating cavity on the motions $\delta_1(t)$ and $\delta_2(t)$ (see Fig. \ref{figure 1}).
Note that the cavity motion $\delta_1(t)$ exhibits a discontinuity in the velocity at the beginning of the vibrations, i.e. $\dot{\delta}_1(t=0)\neq 0$ and thus $\dot{l}(t=0)\neq 0$, which can be regarded as a pathological feature of the model. However, because of the richness of analytical results obtained for this particular scenario (see, e.g., \cite{Dodonov:1996,Dodonov:1993,Ji:1997}) we study the particle production for this cavity motion in detail in subsection $4.1$ and compare the numerical results with the analytical predictions. The impact of the initial discontinuity in the velocity of the wall motion is discussed. In $4.2$ we study the more realistic cavity motion $\delta_2(t)$ with the smooth initial condition $\dot{l}(t=0)= 0$. We compare the results with the scenario discussed in $4.1$ and comment on the results and conclusions presented in \cite{Antunes}. 
In connection with the results for the cavity motion $\delta_2(t)$ shown in $4.2$ we briefly discuss the effect of detuning in subsection $4.3$. Finally we show one example for the cavity motion $\delta_3(t)$ (see Fig. \ref{figure 1}) in $4.4$ and conclude in Section 5.

A more detailed description of the formalism as well as a presentation and discussion of numerical results for a wider range of parameters, massive fields and other cavity dynamics will be found in \cite{Ruser}. 

\section{Canonical formulation and quantization}

Let us consider a non-interacting real and massless scalar field $\Phi(t,x)$ confined to the time-dependent interval $I(t)=[0,l(t)]$. We assume that the scalar field $\Phi(t,x)$ is subject to Dirichlet conditions at the boundaries of $I(t)$, i.e. $\Phi(t,0)=\Phi[t,l(t)]=0$. The time evolution of $\Phi(t,x)$ on $I(t)$ is described by the Klein-Gordon equation $\left[\partial_t^2-\partial_x^2\right]\Phi(t,x)=0$ which also determines the evolution of the vector potential of the electromagnetic field in one space dimension 
(scalar electrodynamics; see, e.g., \cite{Dodonov:1996}).

By introducing an orthonormal and complete set of instantaneous eigenfunctions $\phi_{n}(t,x)$ obeying the eigenvalue equation $-\partial_x^2\phi_n(t,x)=\Omega_n^2(t)\phi_n(t,x)$ on $I(t)$ we may decompose the field as 
\begin{equation}
\Phi(t,x)=\sum_nq_n(t)\phi_n(t,x).
\label{mode decomposition}
\end{equation}
The eigenfunctions and time-dependent eigenvalues are explicitly given by
\begin{equation}
\phi_n(t,x)=\sqrt{\frac{2}{l(t)}}\sin\left[\frac{n\pi}{l(t)}x\right]\;,\;\;\Omega_n^2(t)=\left[\frac{n\pi}{l(t)}\right]^2
\label{eigenfrequency}
\end{equation}
where $n=1,2,...\,$ \footnote[1]{We are using units where $\hbar=c=1$.}. 
Inserting the mode decomposition (\ref{mode decomposition}) into the 
Klein-Gordon equation yields the set of coupled second-order differential equations 
\cite{Schuetzhold:1998,Dodonov:1996, Ji:1997}
\begin{equation}
\ddot{q}_n(t)+\Omega_n^2(t)q_n(t)+2\sum_mM_{mn}(t)\dot{q}_m(t)+\sum_m[\dot{M}_{mn}(t)-N_{nm}(t)]q_m(t)=0
\label{second order mode equation}
\end{equation}
where we have defined $N_{nm}=\sum_kM_{nk}M_{mk}$. The coupling matrices $M_{nm}$ are determined by the integral $M_{nm}(t)=\int_0^{l(t)}dx \dot{\phi}_{n}(t,x)\phi_{m}(t,x)$ taken over the 
time-dependent interval $I(t)$ (cavity). By using the particular expression for $\phi_n(t,x)$ given in (\ref{eigenfrequency}) one finds
\begin{equation}
M_{nm}(t)=-M_{mn}(t)=\frac{\dot{l}(t)}{l(t)}(-1)^{n+m}\frac{2nm}{m^2-n^2}
\label{coupling matrix}
\end{equation}
for $n \neq m$ and $M_{nn}(t)=0$.
The time evolution of the mode functions depends in two different ways on the dynamics of the cavity corresponding to two sources of particle creation \cite{Schuetzhold:1998}: the squeezing of the vacuum due to the non-stationary eigenfrequencies (\ref{eigenfrequency}) and the acceleration effect caused by the time-dependent coupling matrix $M_{nm}$.   

Quantization is achieved by replacing the classical mode functions by operators, 
i.e. $q_n \rightarrow \hat{q}_n$ and $p_n \rightarrow \hat{p}_n$, and demanding the usual equal-time commutation relations for position and momentum operators. Adopting the Heisenberg-picture the time evolution of the operators $\hat{q}_n$ is described by the same system of coupled differential equations (\ref{second order mode equation}) as the classical mode functions.  Assuming a static cavity with size $l_0$ for times $t \le 0$ the Hamiltonian can be diagonalized by introducing time-independent creation and annihilation operators $\hat{a}_n^\dagger$ and  $\hat{a}_n$ of particles with frequency 
$\Omega_n^0\equiv\Omega_n(t=0)$. The vacuum state $|0,t\le0\rangle \equiv |\Omega_0\rangle$ 
which is annihilated by $\hat{a}_n$, i.e.
\begin{equation}
\hat{a}_n|\Omega_0\rangle=0,
\end{equation} 
is given by the ground state of the diagonalized Hamiltonian $\hat{H}(t \le 0)=\sum_n\Omega_n^0\left[\hat{n}_n+\frac{1}{2}\right]$ where the number operator 
$\hat{n}_n=\hat{a}_n^\dagger\hat{a}_n$ counts the number of particles defined with respect to 
the initial state of the cavity, i.e. for times $t \le 0$.

\section{Time evolution and particle creation}

We may now use the ansatz 
\begin{equation}
\hat{q}_n(t) \equiv \sum_m\frac{1}{\sqrt{2\Omega_m^0}}\left[
\hat{a}_m\epsilon_n^{(m)}(t)+\hat{a}_m^\dagger\epsilon_n^{(m)*}(t)
\right]
\label{time evolution of q}
\end{equation}
to parametrize the operator $\hat{q}_n$ for times $t \ge 0$ where the particle operators $\hat{a}_n$
and $\hat{a}_n^\dagger$ defined with respect to the initial vacuum state $|\Omega_0\rangle$ have been used as "expansion coefficients" \cite{Dodonov:1996, Ruser}. The time dependence is carried by the complex functions $\epsilon_n^{(m)}(t)$ exclusively which  obey the same system of coupled second order differential equations (\ref{second order mode equation}) as the mode functions $q_n(t)$. By introducing the functions \cite{Ruser}
\begin{eqnarray}
\xi_k^{(m)}(t)&=&\epsilon_k^{(m)}(t)+\frac{i}{\Omega_k^0}
\left[\dot{\epsilon}_k^{(m)}(t)+\sum_nM_{nk}(t)
\epsilon_n^{(m)}(t)\right]\;,
\label{xi as function of epsilon}\\
\eta_k^{(m)}(t)&=&\epsilon_k^{(m)}(t)-\frac{i}{\Omega_k^0}
\left[\dot{\epsilon}_k^{(m)}(t)+\sum_nM_{nk}(t)
\epsilon_n^{(m)}(t)\right],
\label{eta as function of epsilon}
\end{eqnarray}  
differentiating them with respect to time $t$ and making  use of the second-order differential equation (\ref{second order mode equation}) for $\epsilon_n^{(m)}(t)$ one obtains the system of coupled 
linear first-order differential equations
\begin{eqnarray}
\label{deq for xi}
\dot{\xi}_k^{(m)}(t)=&-&i\left[a^+_{kk}(t)\xi_k^{(m)}(t)-
a^-_{kk}(t)\eta_k^{(m)}(t)\right]  \nonumber  \\ 
&-&\sum_n\left[c^-_{kn}(t)\xi_n^{(m)}(t)+c^+_{kn}(t)\eta_n^{(m)}(t)\right],
\end{eqnarray}
\begin{eqnarray}
\label{deq for eta}
\dot{\eta}_k^{(m)}(t)=&-&i\left[a^-_{kk}(t)\xi_k^{(m)}(t)-
a^+_{kk}(t)\eta_k^{(m)}(t)\right] \nonumber \\ 
&-& \sum_n\left[c^+_{kn}(t)\xi_n^{(m)}(t)+c^-_{kn}(t)\eta_n^{(m)}(t)\right].
\end{eqnarray}
Thereby we have defined the functions
\begin{equation}
a_{kk}^\pm(t)=\frac{\Omega_k^0}{2}
\left\{1 \pm \left[\frac{\Omega_k(t)}{\Omega_k^0}\right]^2\right\}
=\frac{k\pi}{2l_0}\left\{1\pm \left[\frac{l_0}{l(t)}\right]^2\right\}
\end{equation}
and
\begin{equation}
c_{kn}^\pm(t)=\frac{1}{2}\left[M_{nk}(t) \pm \frac{\Omega_n^0}{\Omega_k^0}M_{kn}(t)\right]
=\frac{\dot{l}(t)}{l(t)}(-1)^{n+k}\frac{n}{k\pm n}
\end{equation}
for $n \neq k$ and $c^{\pm}_{nn}(t)=0$. 

Assuming that after a duration $t_1$ the motion of the boundary stops and the cavity is static again 
a second set of annihilation and creation operators $\{\hat{A}_n,\hat{A}_n^\dagger\}$ 
can be introduced to diagonalize the Hamiltonian for $t \ge t_1$, i.e. $\hat{H}(t \ge t_1)=\sum_n\Omega_n^1\left[\hat{A}_n^\dagger\hat{A}_n+\frac{1}{2}\right]$. The ground state of the Hamiltonian $\hat{H}(t\ge t_1)$, i.e. the final vacuum state
$|\Omega_1\rangle\equiv |0,t_1\rangle$, is annihilated by $\hat{A}_n$ and the 
number operator  $\hat{N}_n=\hat{A}_n^\dagger\hat{A}_n$ counts the numbers of 
particles with frequency $\Omega_n^1\equiv \Omega_n(t_1)$ defined with respect to 
$|\Omega_1\rangle$. 

The Bogolubov transformation linking the set of initial state particle operators $\{\hat{a}_n, \hat{a}_n^\dagger\}$ with the set of final state particle operators  $\{\hat{A}_n, \hat{A}_n^\dagger\}$ is found to be
\begin{equation}
\hat{A}_n=\frac{1}{2}\sum_m\sqrt{\frac{\Omega_n^1}{\Omega_m^0}}\left[
\Xi_n^{(m)}(t_1)\hat{a}_m+H_n^{(m)}(t_1)\hat{a}_m^\dagger\right],
\label{Bogolubov-transformation}
\end{equation}
where the functions $\Xi_n^{(m)}$ and $H_n^{(m)}$ are linear combinations of $\xi_n^{(m)}(t)$
and $\eta_n^{(m)}(t)$ at $t=t_1$. In particular 
\begin{eqnarray}
\Xi_n^{(m)}(t_1)&=&\Delta^+(t_1)\xi_n^{(m)}(t_1)+
\Delta^-(t_1)\eta_n^{(m)}(t_1),
\label{relation between big xi  and eta and xi}\\
H_n^{(m)}(t_1)&=&\Delta^-(t_1)\xi_n^{(m)}(t_1)+
\Delta^+(t_1)\eta_n^{(m)}(t_1)
\label{relation between big eta and eta and xi}
\end{eqnarray}
where the function 
\begin{equation}
\Delta^\pm(t)=\frac{1}{2}\left[1\pm\frac{l(t)}{l_0}\right]
\label{delta of t}
\end{equation}
is somewhat like a measure for the deviation of the final state of the cavity, characterized by the cavity length $l(t_1) \equiv l_1$, from the initial state with $l(0)=l_0$.

Starting from a vacuum state $|\Omega_0\rangle$ the Bogolubov transformation 
(\ref{Bogolubov-transformation}) has to become trivial for times $t_1=t_0=0$, i.e., $\Xi_n^{(m)}(0)=2\,\delta_{nm}$ and
 $H_n^{(m)}(0)=0$, such that $\hat{A}_n|\Omega_0\rangle=\hat{a}_n|\Omega_0\rangle=0$, which yields the set of initial conditions  
\begin{equation}
\xi_n^{(m)}(0)=2\delta_{nm}\;,\;\; \eta_n^{(m)}(0)=0
\label{initial conditions for xi and eta}
\end{equation}
for the system (\ref{deq for xi}) - (\ref{deq for eta}) of coupled differential equations
\footnote[1]{Equations (\ref{xi as function of epsilon}), (\ref{eta as function of epsilon}) and the initial conditions (\ref{initial conditions for xi and eta}) imply the initial conditions 
$\epsilon_n^{(m)}(0)=\delta_{nm }$ and $\dot{\epsilon}_n^{(m)}(0)=-i\,\Omega_n^0\,\delta_{nm} - M_{mn}(0)$ for the complex functions $\epsilon_n^{(m)}(t)$ which satisfy (\ref{second order mode equation}). Therefore, if the cavity motion does not start smoothly, i.e. with a non-zero velocity $\dot{l}(0)\neq 0$, such that $M_{mn}(0)\neq 0$ the initial conditions for $\epsilon_n^{(m)}(t)$ are not simply those of plane waves. This can be seen as well by matching $\hat{q}_n(t<0)=\frac{1}{\sqrt{2\Omega_n^0}}\left[\hat{a}_n\,e^{-i\Omega_n^0t} + \hat{a}_n^{\dagger}\,e^{i\Omega_n^0t}\right] $ to $\hat{q}_n(t>0)$ and $\hat{p}_n(t<0)=i\,\sqrt{\frac{\Omega_n^0}{2}}\left[\hat{a}^{\dagger}_n\,e^{i\Omega_n^0t} - \hat{a}_n\,e^{-i\Omega_n^0t}\right]$  to $\hat{p}_n(t>0)$ at $t=0$, where $\hat{q}_n(t>0)$ is given by (\ref{time evolution of q}) and $\hat{p}_n(t)=\dot{\hat{q}}_n(t)+\sum_mM_{nm}(t)\hat{q}_m(t)$ (see, e.g., \cite{Schuetzhold:1998}).}.
 
The number of particles created during the motion of the boundary is given by the number of final state particles, counted by $\hat{N}_n=\hat{A}_n^\dagger\hat{A}_n$ \footnote[2]{Only this particle number operator is physically meaningful for times $t\ge t_1$.}, which are contained in the initial vacuum state $|\Omega_0\rangle$, i.e.  
\begin{equation}
N_n=\langle\Omega_0|\hat{N}_n|\Omega_0\rangle=\langle\Omega_0|\hat{A}_n^\dagger\hat{A}_n|\Omega_0\rangle=\frac{1}{4}\sum_m
\frac{\Omega_n^1}{\Omega_m^0}|H_n^{(m)}|^2.
\label{particle number}
\end{equation} 
Knowing the solutions to the coupled system of linear differential equations of first order formed by Eqs. (\ref{deq for xi}) and (\ref{deq for eta}) this expectation value can be calculated by using the linear transformation  (\ref{relation between big eta and eta and xi}).

Accordingly, the total energy $E$ of the created quantum radiation is given by
\begin{equation}
E=\sum_{n=1}^{\infty}E_n=\sum_{n=1}^{\infty}\Omega_n^1\,N_n=\frac{1}{4}\sum_{n=1}^{\infty}
\left( \Omega_n^1 \right)^2\,\sum_{m=1}^{\infty}\frac{|H_n^{(m)}|^2}{\Omega_m^0}.
\label{energy}
\end{equation}

Finding solutions to a coupled system of linear first order differential equations is a standard problem in numerical mathematics. By introducing a cut-off quantum number $k_{\rm max}$ the infinite system of coupled differential equations has to be truncated to make it suitable for a numerical treatment.  
Numerical solutions to the remaining finite system can be obtained with high accuracy by using  
standard routines. Finally, the dependence of the solutions on the cut-off $k_{\rm max}$ has to be checked in order to guarantee stability of the numerical results. In addition, the quality of the numerical solutions can be assessed by testing relations like the unitarity of the Bogolubov transformation. 
Because it is beyond the scope of the present paper to go into technical details we refer 
the reader to \cite{Ruser} for a detailed discussion about the numerics which has been used to obtain the results which we show in the next section.  

\section{Numerical results}

\subsection{$\delta(t)=\delta_1(t)$} 

In this section we show some of the numerical results obtained for the scenario of the vibrating 
one-dimensional cavity with $l(t)$ given by 
\begin{equation}
l(t)=l_0\left[1+\epsilon\,\sin(2\,\Omega_n^0\,t)\right]
\label{background 1}
\end{equation}
which has been studied analytically in, for instance, \cite{Schuetzhold:1998,Dodonov:1996, Dodonov:1993,Ji:1997}.  More numerical results are collected in \cite{Ruser}. We integrate the coupled system 
formed by the linear differential equations (\ref{deq for xi}) and (\ref{deq for eta}) for a given interval $[0,t_{\rm max}]$ and calculate the expectation value (\ref{particle number}) for all time steps. 

\begin{figure}
\begin{center}
\begin{tabular}{cc}
\includegraphics[height=5.5cm]{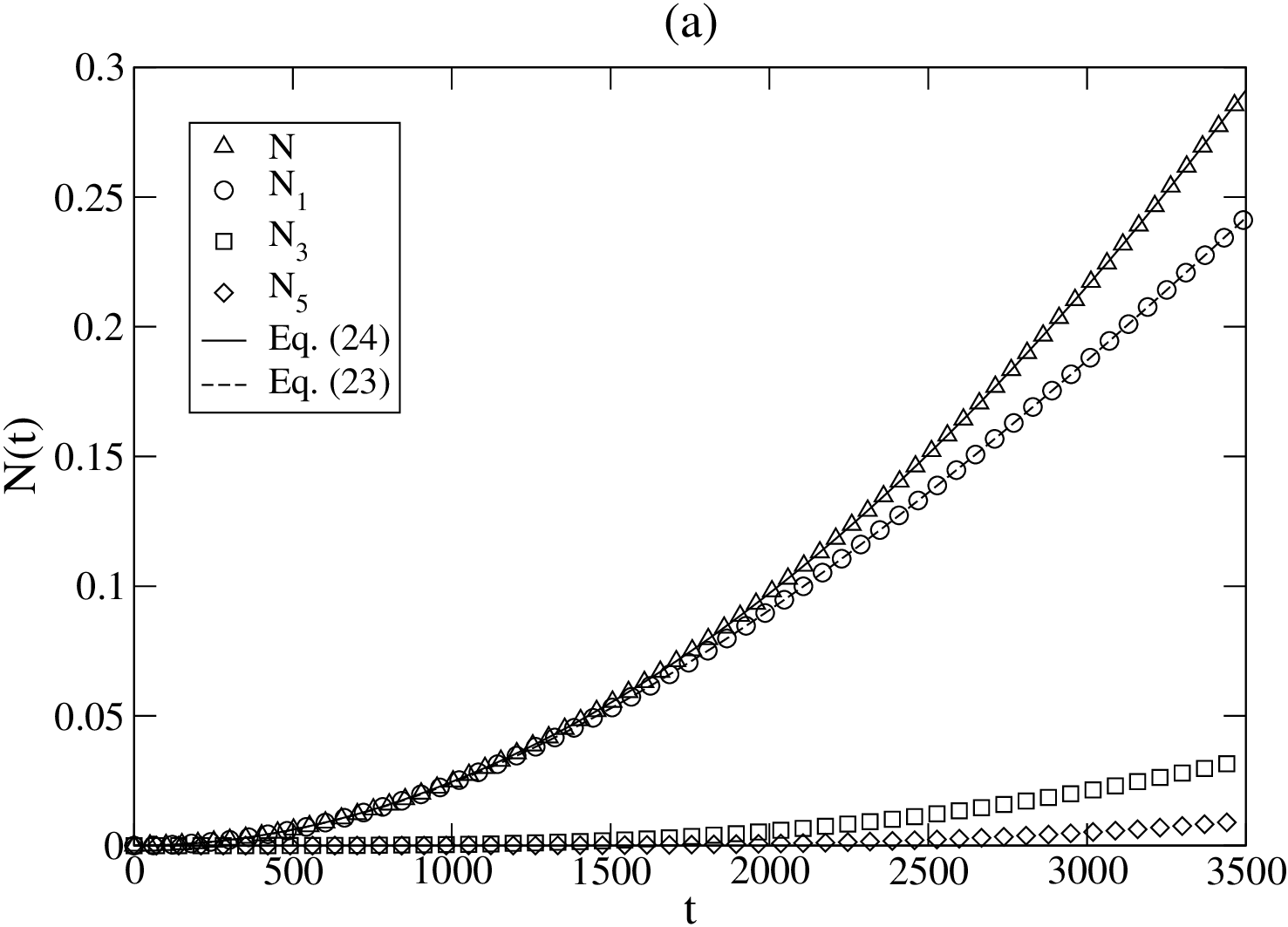}
&
\includegraphics[height=5.5cm]{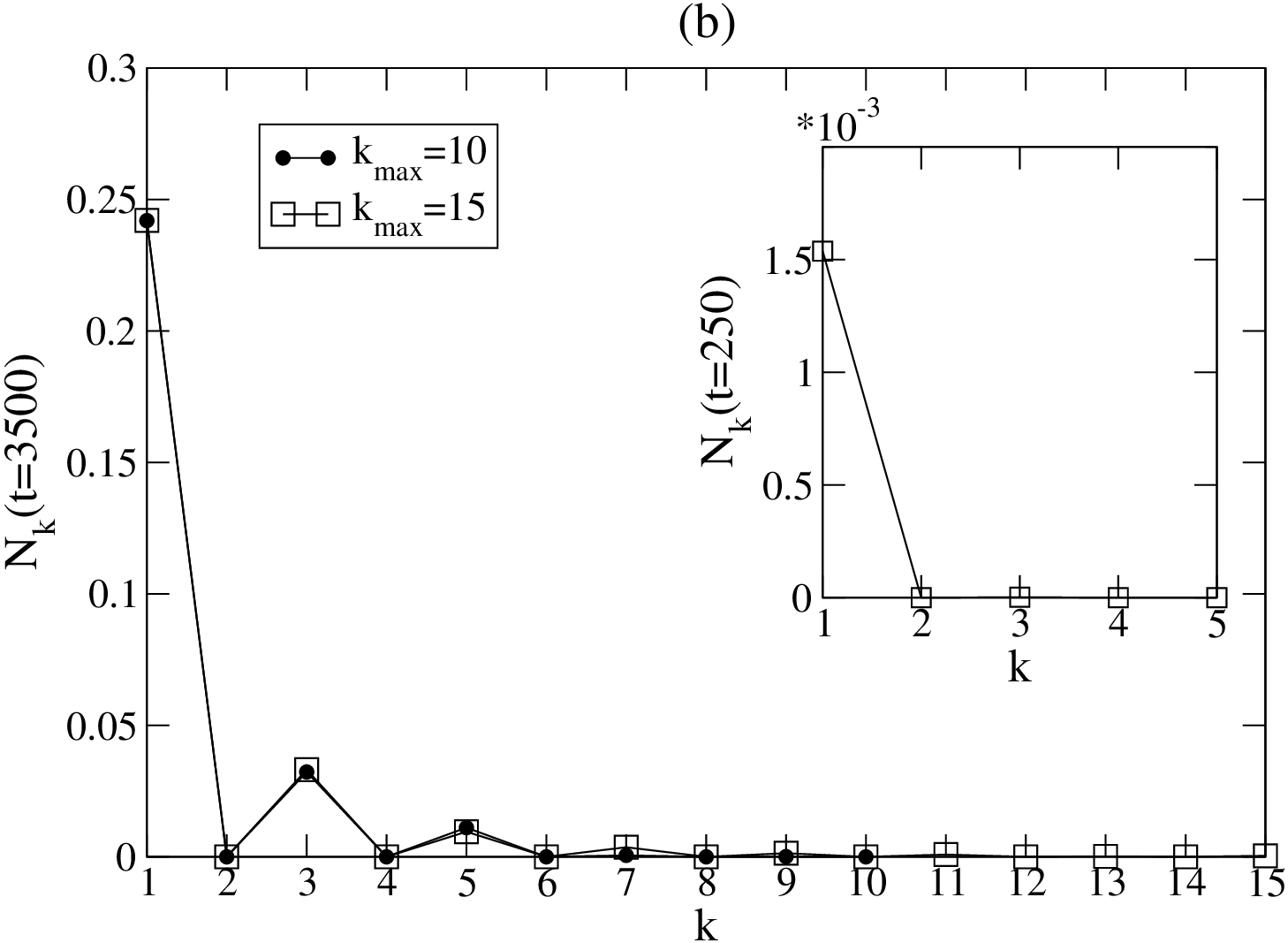}
\end{tabular}
\caption{ (a)  Number of particles produced in a cavity vibrating with (\ref{background 1}) and $n=1$, i.e. 
                        the main resonance, for parameters $l_0=0.1$, $\epsilon=0.00001$ and 
                        $k_{\rm max}=15$, together with the analytical predictions (\ref{dodonov1}) and 
                        (\ref{dodonov2}).
                 (b)  Particle spectrum corresponding to (a) for the two cut-off parameters $k_{\rm                  
                       max}=10$ and $k_{\rm max}=15$.
                      \label{figure 2}}
\end{center}
\end{figure}

In \cite{Dodonov:1996} the authors study the case in which the one-dimensional cavity performs oscillations of the form (\ref{background 1}) with $n=1$, i.e. the frequency of the cavity vibrations is twice 
the frequency of the first unperturbed field mode $\Omega_1^0=\frac{\pi}{l_0}$ inside the cavity.
Making the assumption of small amplitudes of the oscillations $\epsilon \ll 1$ (trembling cavity) they find analytical expressions for the number of created particles in the first mode $N_1(t)$ as well as for the total particle number $N(t)$ in terms of complete elliptic integrals \cite{Gradshteyn:1994}. 
In particular 
\begin{eqnarray}
N_1(t)=\frac{2}{\pi^2}E(\kappa)K(\kappa)-\frac{1}{2}, 
\label{dodonov1}\\
N(t)=\frac{1}{\pi^2}\left[\left(1-\frac{1}{2}\kappa^2\right)K^2(\kappa)-E(\kappa)K(\kappa)\right],
\label{dodonov2}
\end{eqnarray}
where $\kappa=\sqrt{1-\exp\{-8\tau\}}$ and $\tau=\frac{1}{2}\epsilon\,\Omega_1^0\,t = \frac{\pi}{2}\frac{\epsilon}{l_0}\,t$ is the so-called "slow time"
(see Eq. (6.5) and (6.10) in \cite{Dodonov:1996}). 
These expressions yield $N (\tau) = N_1 (\tau) = \tau^2$ for $\tau \ll 1$ as well as
 $N(\tau) = \tau^2$ and $N_1 (\tau) = \tau$ for $\tau \gg 1$.
In Figure \ref{figure 2} (a) we show the numerical results for a cavity with initial length $l_0=0.1$ and amplitude $\epsilon=10^{-5}$ obtained for an integration time $t_{\rm max}=3500$
and compare them to the analytical expressions (\ref{dodonov1}) and (\ref{dodonov2}).
The numerical results perfectly agree with the analytical expressions of \cite{Dodonov:1996} for all times predicting that the initial quadratic increase of both, the total particle number and the number of
particles created in the resonance mode $k=1$, devolves in a quadratic increase of the total particle number and a linear behaviour of the number of resonance mode particles. The particle spectrum at the end of the integration $t_{\rm max}$ shown in Figure \ref{figure 2} (b) for the two cut-off parameters $k_{\rm max}=10$ and $15$ indicates the stability of the numerical results. As one can infer from the spectrum only odd modes are created which was also predicted in \cite{Dodonov:1996}. From the particle spectrum shown for short times $N_{k}(t=250)$ we read off the value $N_1(t=250) \sim 1.5419 \times 10^{-3}$ which agrees perfectly with the analytical prediction $\tau^2|_{t=250} \sim 1.5421 \times 10^{-3}$.

\begin{figure}
\begin{center}
\includegraphics[height=10cm]{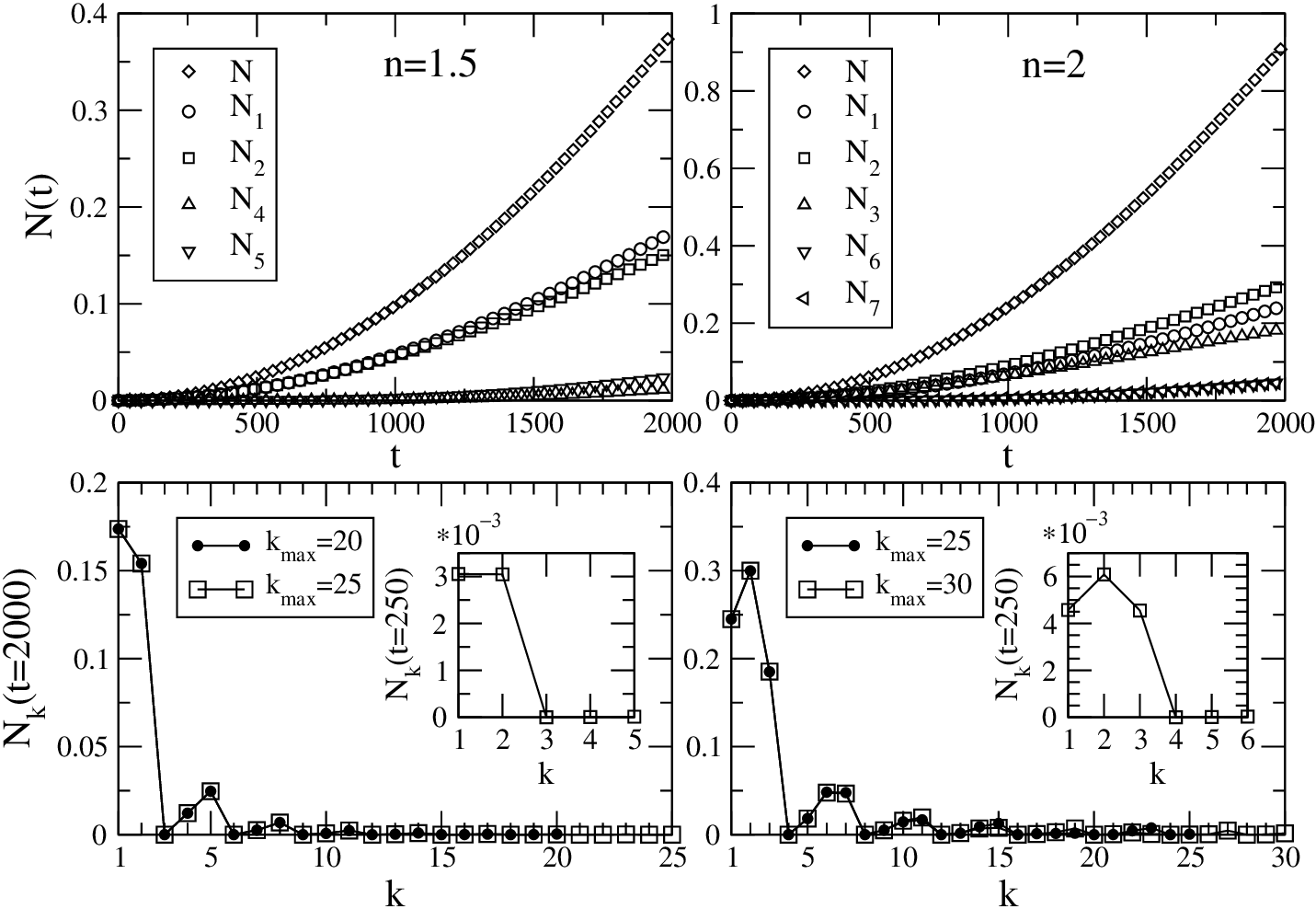}
\caption{Number and spectrum of particles created in an oscillating cavity with parameters of Figure \ref{figure 2} but for cavity frequencies given by $n=1.5$ and $n=2$. The particle spectra are shown for two integrations corresponding to two cut-off parameters $k_{\rm max}=20$ and $25$ for $n=1.5$ and  $k_{\rm max}=25$ and $30$ for $n=2$ to indicate numerical stability. 
\label{figure 3}}
\end{center}
\end{figure}

The result of \cite{Dodonov:1996} for $\tau \ll 1$ has been generalized in \cite{Ji:1997} for more general cavity frequencies to 
\begin{equation}
N_k(\tau)=(2\,n-k)\,k\,\tau^2
\label{Ji's equation}
\end{equation}
for $k<2\,n$ and $N_k(\tau)=0$ otherwise, where $n$ characterizes the frequency of the cavity vibrations. In Figure \ref{figure 3} we show the numerical results obtained for the parameters of Figure \ref{figure 2} but cavity frequencies given by $n=1.5$ and $2$ for an integration time $t_{\rm max}=2000$.

The short time particle spectra confirm the analytical predictions. 
In the first case, $n=1.5$, the only modes that become excited for times $\tau \ll 1$ are the 
$k=1$ and $k=2$ modes and particles of these frequencies are produced in the same amount. From the short time spectra shown in Figure 
\ref{figure 3} we deduce the numerical values $N_1(t=250) \sim 3.080 \times 10^{-3}$ and $N_2(t=250) \sim 3.073 \times 10^{-3}$ which are again in perfect agreement with the analytical prediction 
(\ref{Ji's equation}) yielding $N_1(t=250)=N_2(t=250) \sim 3.084 \times 10^{-3}$. 

For larger times the behaviour begins to change. The mode $k=1$ gets the upper hand and slightly more particles in this mode are produced than particles in the mode $k=2$. Higher frequency modes $k>2$ play an inferior role and, at least for this range of integration, do not significantly contribute to the total particle number. 

In the second example with $n=2$ similar statements hold. As predicted by equation (\ref{Ji's equation}) the resonance mode $k=n=2$ as well as the modes $k=1$ and $k=3$ become excited for short times $\tau\ll1$. The amount of particles created in the resonance mode $k=2$ is slightly larger than the number of particles created in the close-by modes $k=1$ and $k=3$ which are produced in the same amount. As before, this behaviour changes for large times. Then the resonance mode $k=2$ becomes excited most, followed by the  close-by modes $k=1$ and $k=3$, respectively.   

From the numerical simulations (see, e.g., the particle spectra in Fig. \ref{figure 3}) we deduce that no particles are produced in frequency modes $k=2\,n\,p$ with $p=1,2,3,...$ where $n$ characterizes the frequency of the cavity vibrations. This is a generalization of the behaviour found in \cite{Dodonov:1996} that only odd modes become excited in the main resonance scenario $n=1$  and will be discussed in more detail in \cite{Ruser}.

\begin{figure}
\begin{center}
\begin{tabular}{cc}
\includegraphics[height=5.5cm]{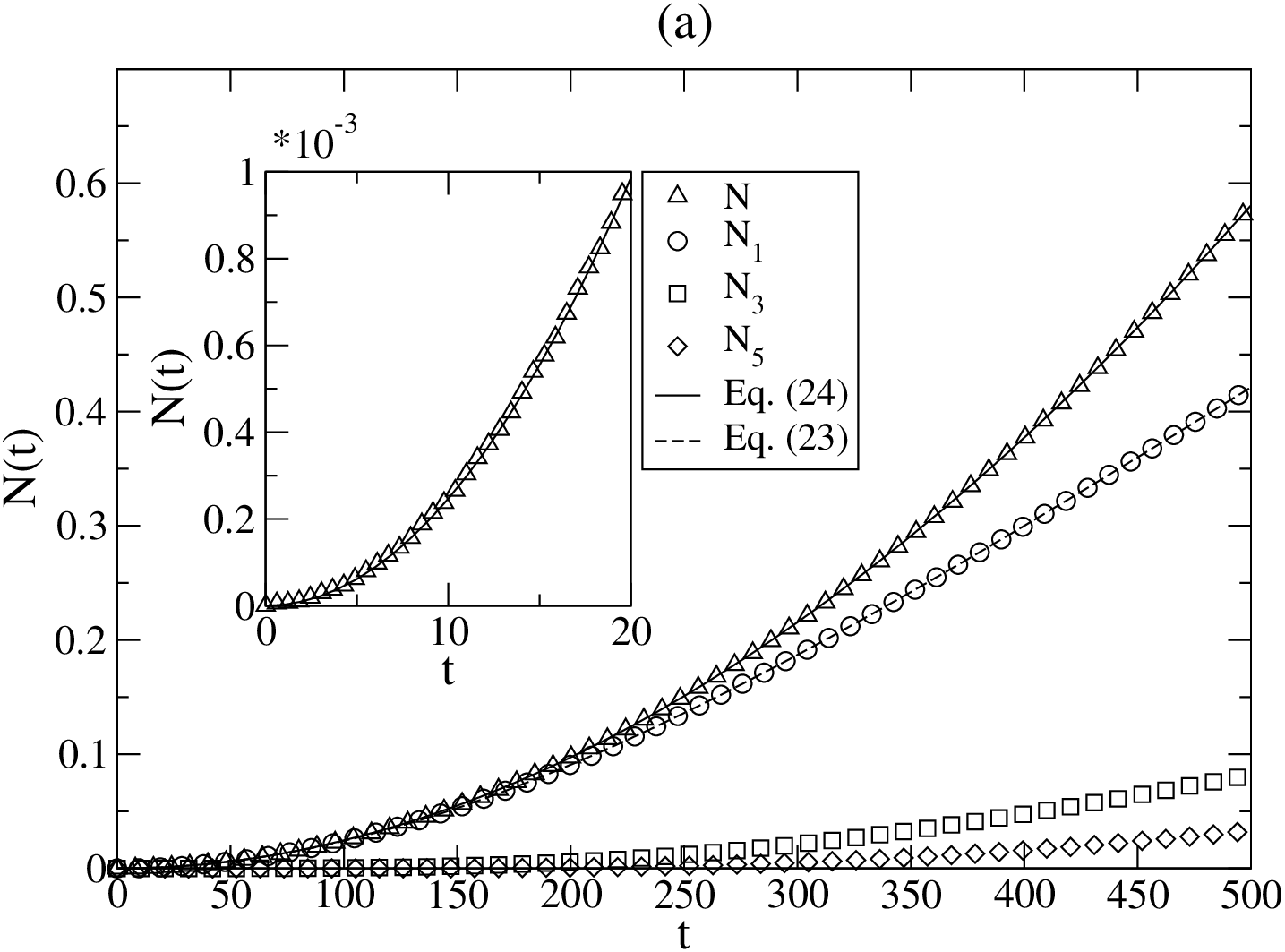}
&
\includegraphics[height=5.5cm]{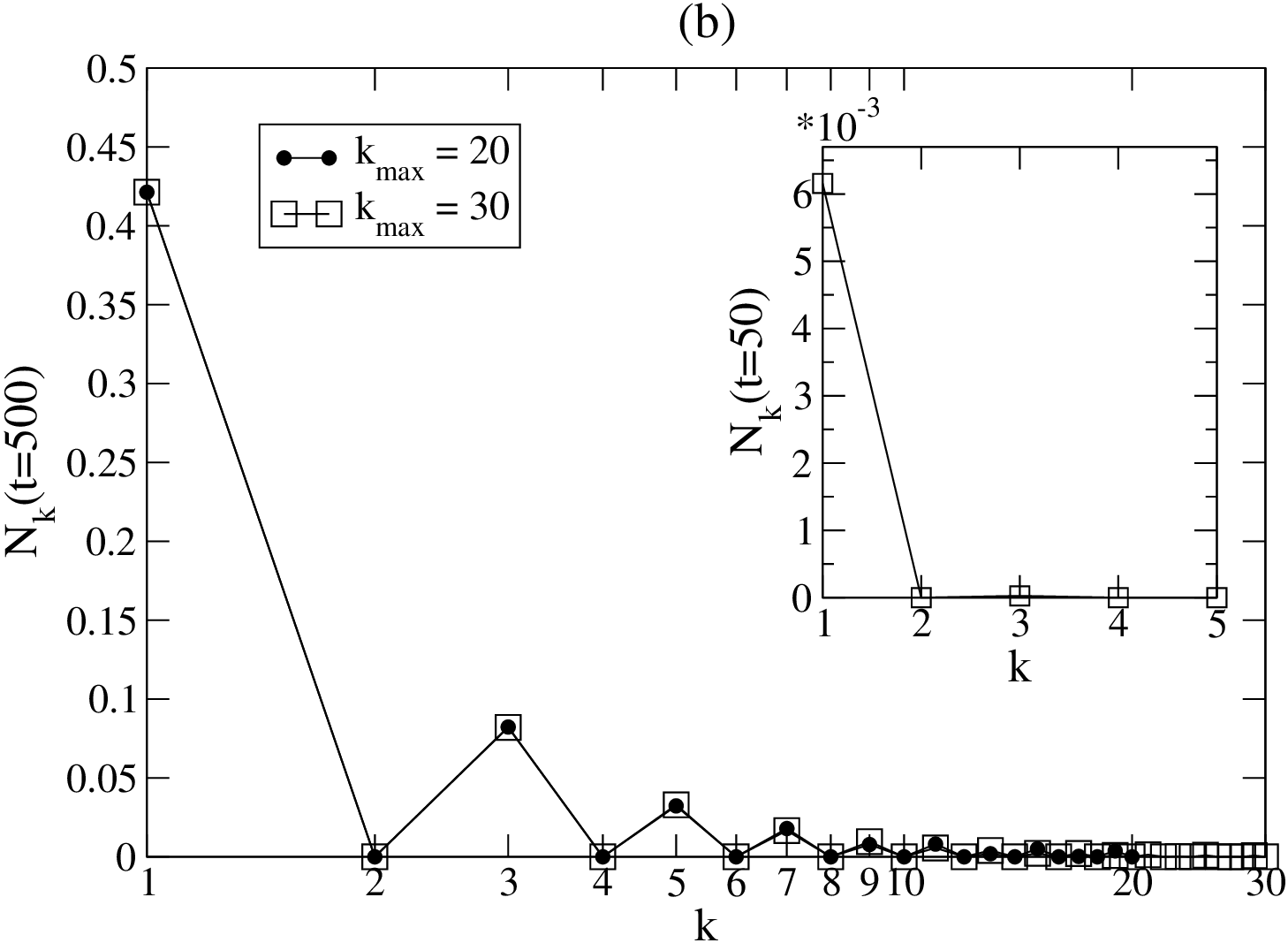}
\end{tabular}
\caption{ (a)  Number of particles produced in a cavity vibrating with (\ref{background 1}) and $n=1$, i.e. the         
                        main resonance, for parameters $l_0=1$, $\epsilon=0.001$ and $k_{\rm max}=30$,        
                        together with the analytical predictions (\ref{dodonov1}) and (\ref{dodonov2}).
                 (b)  Particle spectrum corresponding to (a) for the two cut-off parameters $k_{\rm                  
                       max}=20$ and $k_{\rm max}=30$.
                  \label{figure 4}}
\end{center}
\end{figure}

\begin{figure}
\begin{center}
\begin{tabular}{cc}
\includegraphics[height=5.5cm]{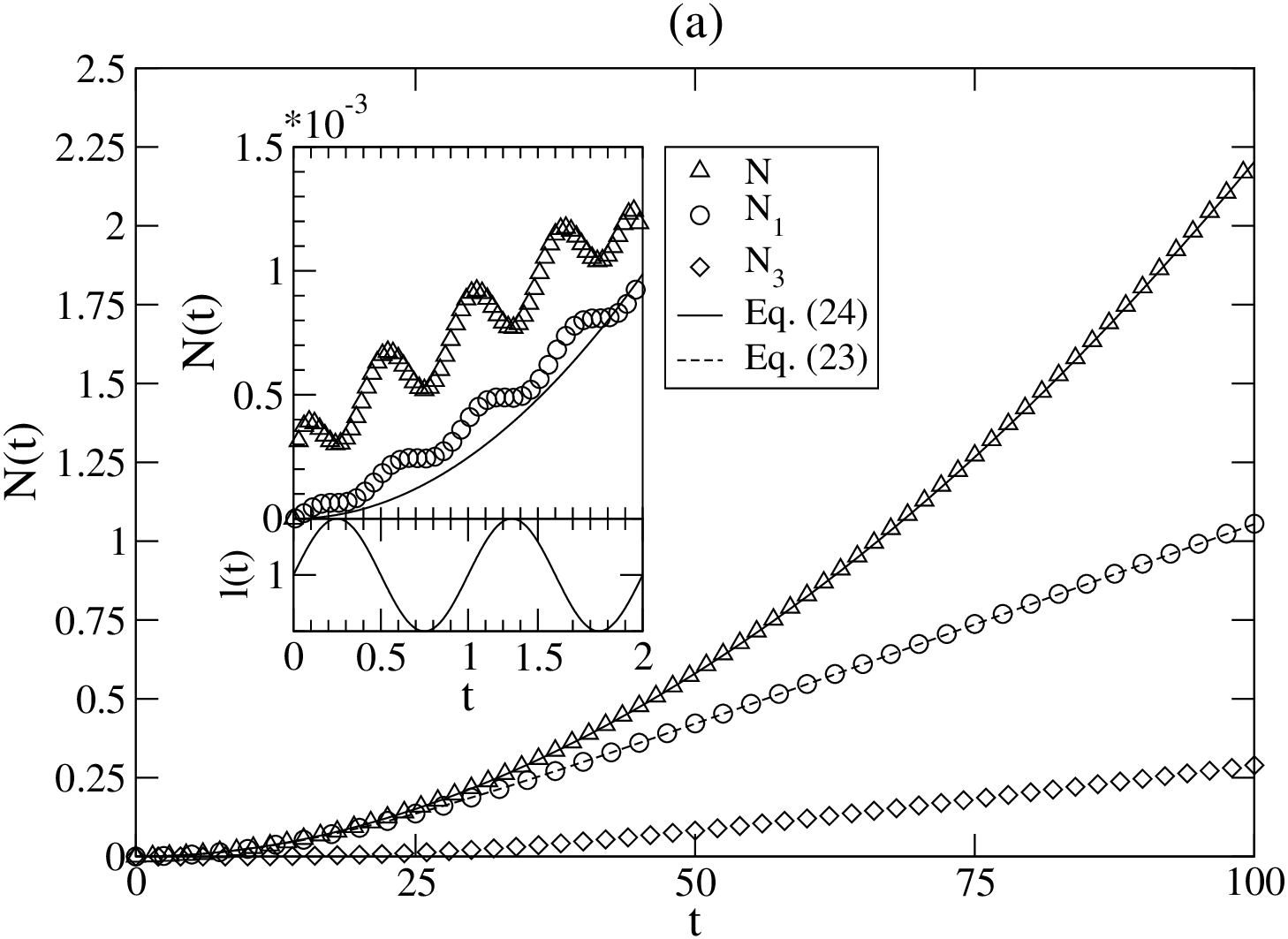}
&
\includegraphics[height=5.5cm]{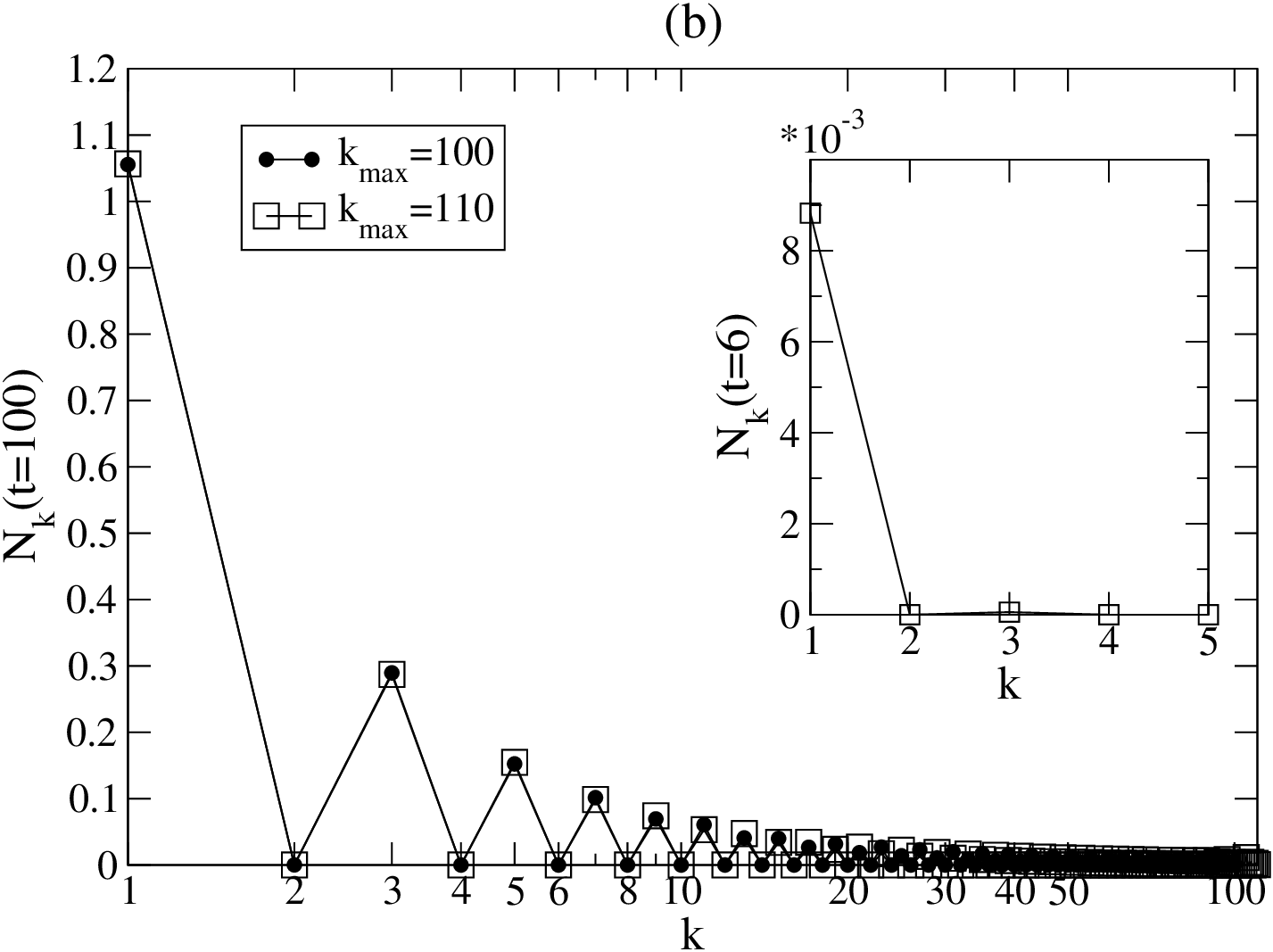}
\end{tabular}
\caption{ (a)  Number of particles produced in a cavity vibrating with (\ref{background 1}) and $n=1$, i.e. the         
                        main resonance, for parameters $l_0=1$, $\epsilon=0.01$ and $k_{\rm max}=110$,        
                        together with the analytical predictions (\ref{dodonov1}) and (\ref{dodonov2}).
                 (b)  Particle spectrum corresponding to (a) for the two cut-off parameters $k_{\rm                  
                       max}=100$ and $k_{\rm max}=110$.
                  \label{figure 5}}
\end{center}
\end{figure}

\begin{figure}
\begin{center}
\begin{tabular}{cc}
\includegraphics[height=5.5cm]{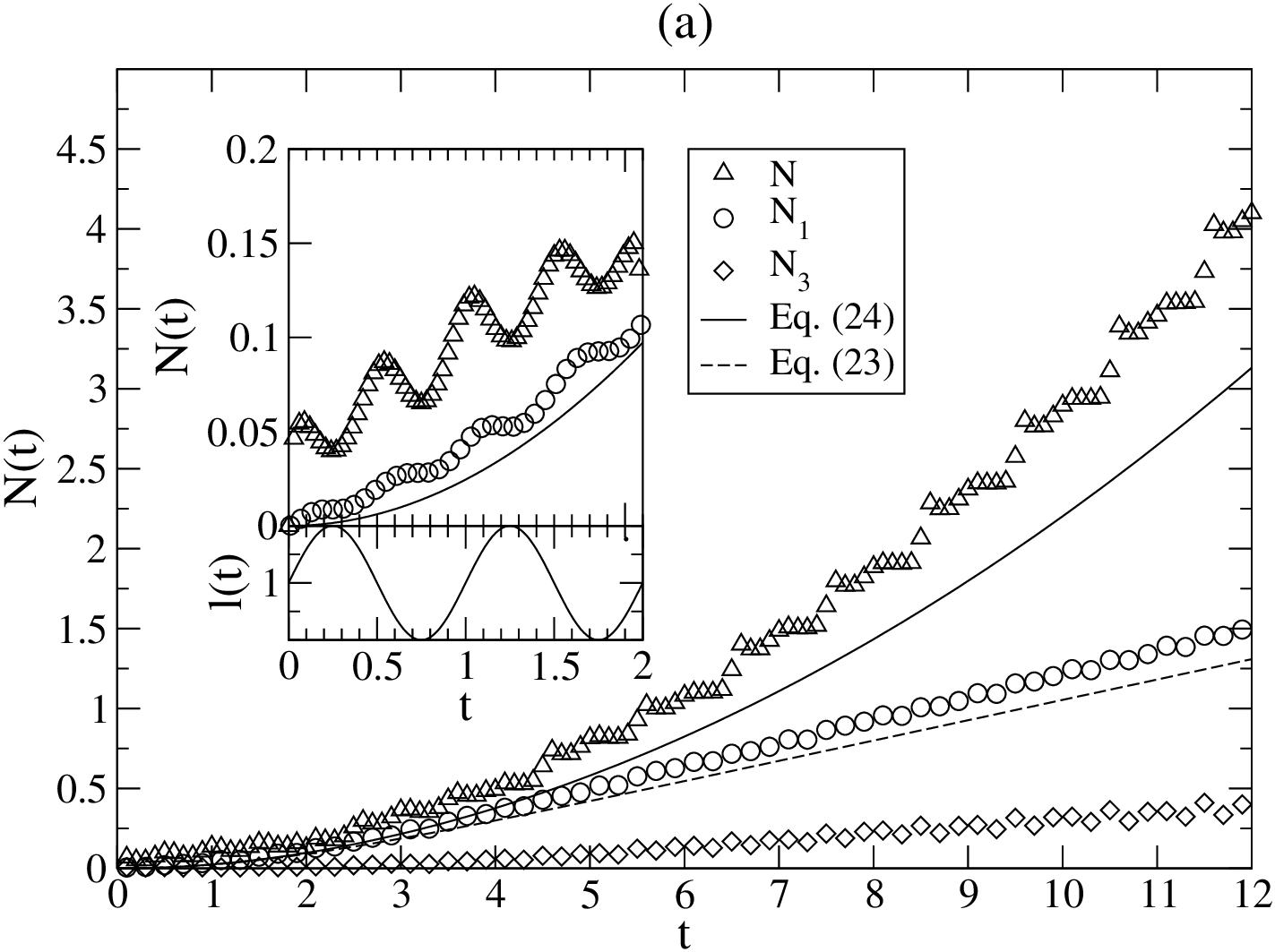}
&
\includegraphics[height=5.5cm]{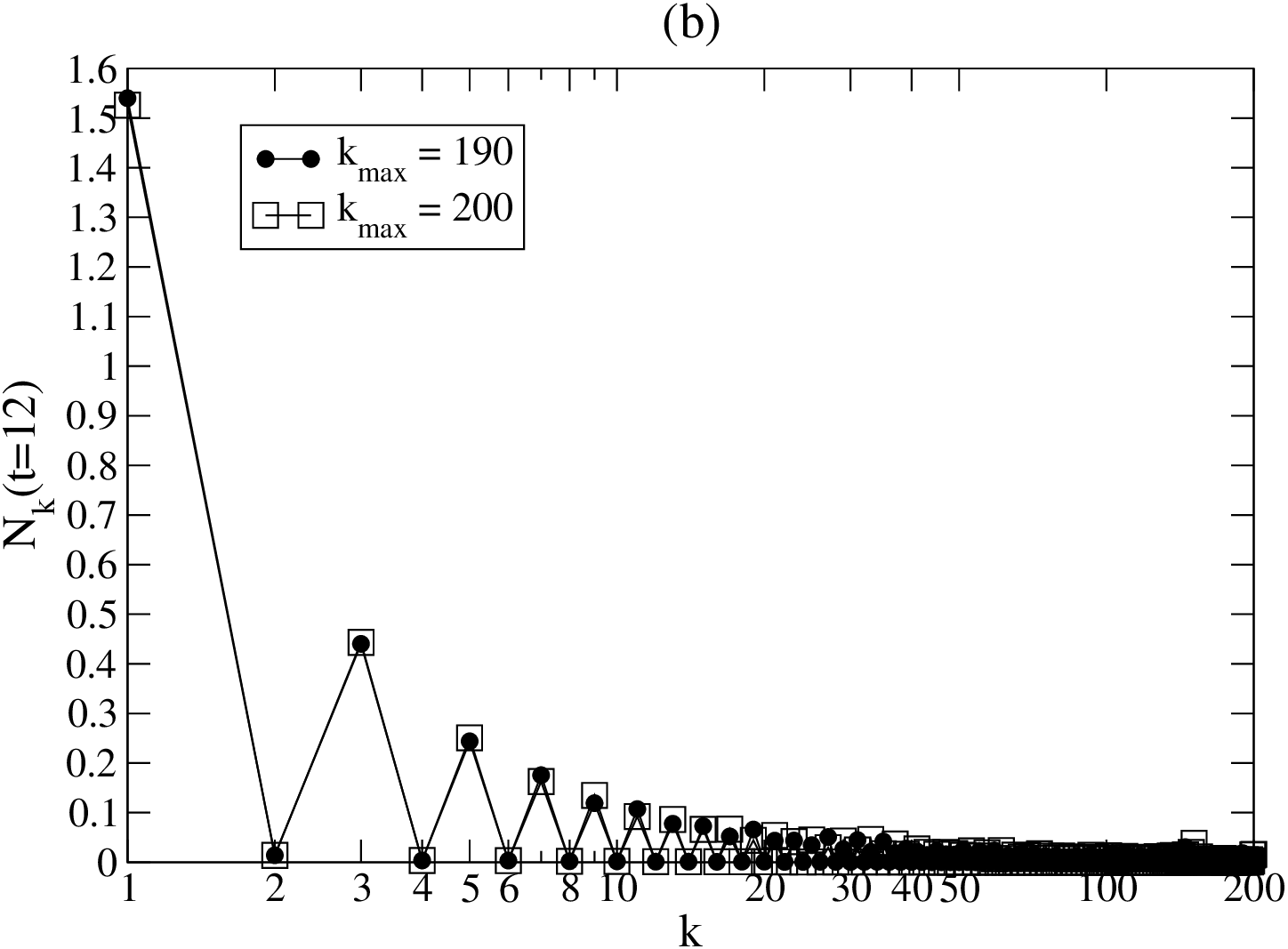}
\end{tabular}
\caption{ (a)  Number of particles produced in a cavity vibrating with (\ref{background 1}) and $n=1$, i.e. the         
                        main resonance, for parameters $l_0=1$, $\epsilon=0.1$ and $k_{\rm max}=200$,        
                        together with the analytical predictions (\ref{dodonov1}) and (\ref{dodonov2}).
                 (b)  Particle spectrum corresponding to (a) for the two cut-off parameters $k_{\rm                  
                       max}=190$ and $k_{\rm max}=200$.
                 \label{figure 6}}
\end{center}
\end{figure}

As next, let us discuss the range of validity of the analytical expressions derived in \cite{Dodonov:1996} 
with respect to the assumption $\epsilon \ll 1$. For this we consider a cavity with initial size $l_0=1$
and calculate the number of created particles for amplitudes $\epsilon$ covering three orders of magnitude. The results of the numerical calculations together with the analytical predictions (\ref{dodonov1}) and (\ref{dodonov2}) are shown in Figure \ref{figure 4} for $\epsilon=0.001$, 
in Figure \ref{figure 5} for $\epsilon=0.01$  and in Figure \ref{figure 6} for the amplitude $\epsilon=0.1$\footnote[3]{Note that in this rather extreme case with $\epsilon=0.1$, the maximum velocity of the mirror becomes relativistic with $\frac{v}{c} = 2\,\epsilon\,\pi \sim 0.63$.}.  
As one can infer from these pictures, the rate of particle creation grows rapidly on increasing the amplitude $\epsilon$ of the oscillations. While for $\epsilon=0.001$ an integration time of $t_{\rm max}=500$ is needed to obtain $N(t_{\rm max}=500)\sim 0.58$, a total particle number of the order of one is already reached for $t=6$ in the case of the large amplitude $\epsilon=0.1$. For large amplitudes like $\epsilon=0.01$ and $\epsilon=0.1$ the number of excited field modes inside the cavity increases drastically. This is reflected by the value for the cut-off parameter $k_{\rm max}$ which has to be chosen in order to obtain numerically stable solutions. Whereas for $\epsilon=0.001$ the value  $k_{\rm max}=30$ guarantees stability of the numerical solutions up to $t_{\rm max}=500$ it has to be increased to $k_{\rm max}=110$ to provide stable solutions for $\epsilon=0.01$ and $t_{\rm max}=100$.
In order to obtain stable results in the case of the large amplitude $\epsilon=0.1$ up to $t_{\rm max}=12$ already $k_{\rm max}=200$ modes have to be taken into account for this short integration range. 
Again, only odd modes become excited as predicted in \cite{Dodonov:1996}. 

The numerical results for the amplitudes $\epsilon=0.001$ and $\epsilon=0.01$ shown in Figure \ref{figure 4} and Figure \ref{figure 5}, respectively, reveal  that the expressions (\ref{dodonov1}) 
and (\ref{dodonov2}) derived in \cite{Dodonov:1996} by means of approximations for $\epsilon \ll 1$  describe the numerical solutions very well for all time scales under consideration.
For $\epsilon = 0.1$ the qualitative behaviour of both, the number of particles created in the resonance mode as well as the total particle number, seems still to be valid (at least in the shown integration range) 
but the number of created particles is larger than predicted by the analytical expressions.  

\begin{figure}
\begin{center}
\begin{tabular}{cc}
\includegraphics[height=4cm]{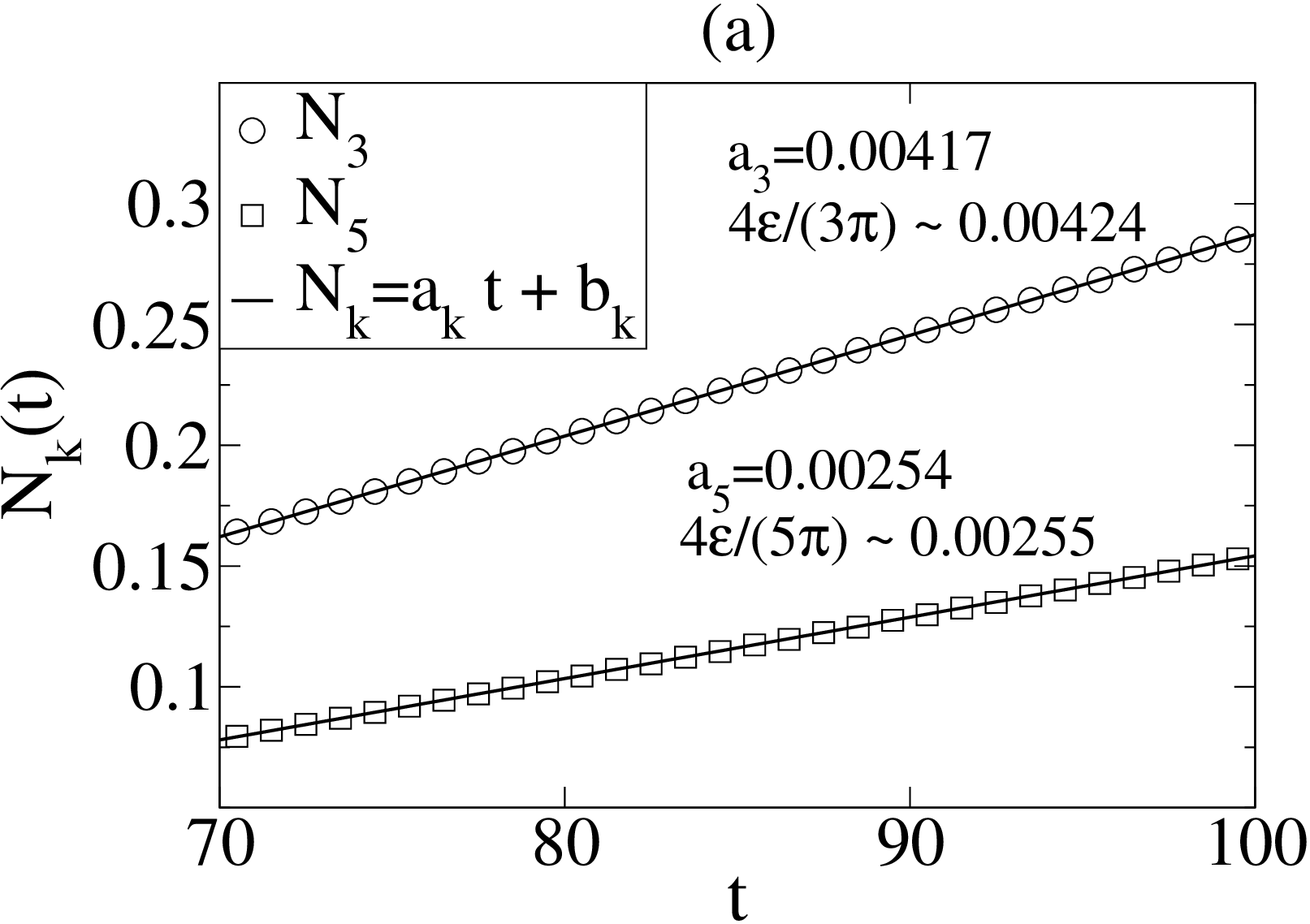}
&
\includegraphics[height=4cm]{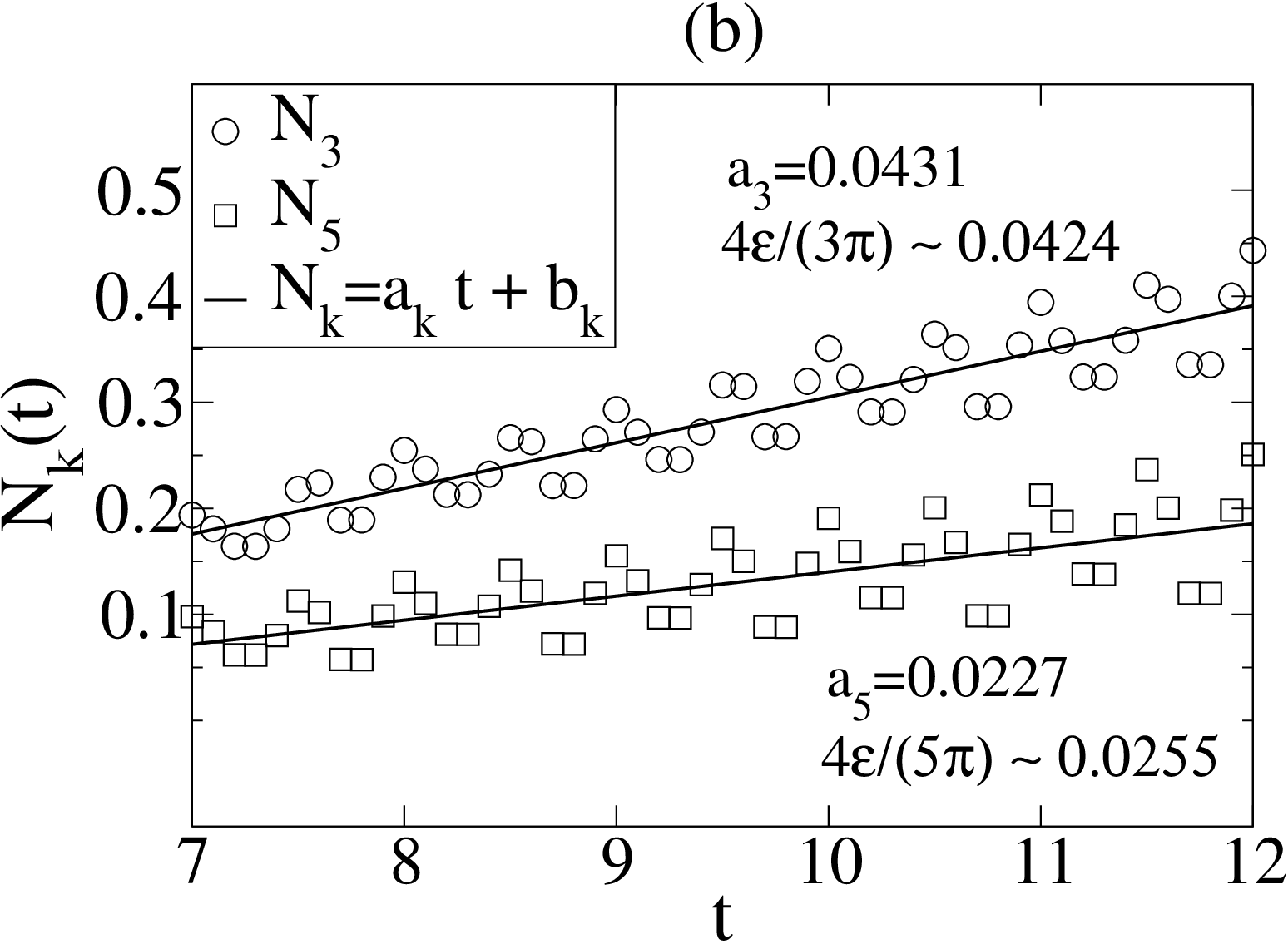}
\end{tabular}
\caption{Number of particles created in the modes $k=3$ and $k=5$ 
                for $\tau > 1$ and parameters $l_0=1$ and   
                (a) $\epsilon=0.01$ and (b) $\epsilon=0.1$, corresponding to Fig. \ref{figure 5} (a)
                and Fig. \ref{figure 6} (a), respectively, together with a fit of the data to the linear law   
                 $N_k(t)=a_k\,t +b_k$.
                 \label{figure 7}}
\end{center}
\end{figure}

For the particular case under consideration it was found that for $\epsilon \, t \gg 1$, i.e. $\tau \gg 1$, the rate of particle creation in a mode of frequency $\Omega_k^0$ (k odd) is given 
by  \cite{Dodonov:1993}
\begin{equation}
\frac{dN_k(t)}{dt}=\frac{4\,\epsilon}{\pi\,k}
\label{dodonov4}
\end{equation}
and thus, the number of particles created in the mode $k$ increases linearly for large times\footnote[1]{Note that in \cite{Dodonov:1993} a factor $2$ was missed which has been corrected in \cite{Dodonov:1996}.}. By expanding Eq. (\ref{dodonov1}) one easily recovers Eq. (\ref{dodonov4}) for the particular case $k=1$. As mentioned in \cite{Dodonov:1996} this asymptotic formula works quite well after $\tau \sim \frac{1}{2}$. Because we have already shown that the numerical results for the resonance mode agree perfectly with the analytical prediction (\ref{dodonov1}) [see Fig. \ref{figure 2} (a) for $\epsilon=0.00001$, Fig. \ref{figure 4} (a) for $\epsilon=0.001$ and \ref{figure 5} (a) for $\epsilon=0.01$] we concentrate here on the higher frequencies $\Omega_3^0$ and $\Omega_5^0$.
In Figures \ref{figure 7} (a) and (b), corresponding to $\epsilon=0.01$ and $0.1$, respectively, we show the results for the number of particles created in the modes $k=3$ and $k=5$ together with a linear fit $N_k(t)=a_k t +b_k$ to the numerical values for certain time ranges. 
For the amplitude $\epsilon=0.01$, for which the slow time value $\tau=1$ corresponds to $t\sim 62$, the rate of particle creation obtained by fitting the data for times $\tau >1$ agrees very well with the values predicted by Eq. (\ref{dodonov4}) as one infers from Fig. \ref{figure 7} (a). 
From our numerical calculations we find the values $a_{3}=0.00417$ and $a_{5}=0.00254$ 
which are in very good agreement with the values $a_3=\frac{4\,\epsilon}{3\,\pi}\sim 0.00424$ and $a_5=\frac{4\,\epsilon}{5\,\pi}\sim 0.00255$ predicted by Eq. (\ref{dodonov4}).
The integration range used in the numerical simulation for $\epsilon=0.1$ yielding the results shown in Fig. \ref{figure 6} allows us as well to compare the data to the prediction (\ref{dodonov4}) for times $\tau > 1$. In this case, where the assumption $\epsilon \ll 1$ is no longer valid, we still find a relatively good agreement of the numerically obtained rate of particle creation with the predicted one [ see Fig. \ref{figure 7} (b) \footnote[2]{The scattering of the numerical values in this case is due to the oscillations in the particle number which can be also seen in Fig. \ref{figure 6} (a).}]. Interestingly, whereas the numerical result for the number of particles created in the resonance mode $\Omega_1^0$ is not very well described by the analytical prediction (\ref{dodonov1}) for times $\tau \ge 1$ [see Fig. \ref{figure 6} (a) which shows that the rate of particle creation is larger than the (in the limit $\epsilon \ll 1$) predicted one] the numerical results for the higher frequency modes $k=3$ and $k=5$ agree comparatively well with Eq. (\ref{dodonov4}).  The range of integration used in the simulations for $\epsilon=0.00001$ and $\epsilon=0.001$ (see Figs. \ref{figure 2} and \ref{figure 4}) is not large enough to compare the numerical results with the analytical predictions for times $\tau > 1$. For the amplitude $\epsilon=0.00001$ 
[see Fig. \ref{figure 2} (a)] the number of particles created in the modes $k=3$ and $k=5$ is still in a phase of acceleration up to the maximum integration time $t_{\rm max}=3500$ which corresponds to $\tau \sim 0.55$. Similar statements hold for the numerical results obtained for $\epsilon=0.001$ [see Fig. \ref{figure 4} (a)] where the maximum integration time corresponds to $\tau \sim 0.79$. However, as discussed before, we have found a perfect agreement of the numerical results obtained for $\epsilon=0.01$ with the analytical expression (\ref{dodonov4}) for times $\tau \ge 1$ [see Fig. \ref{figure 7} (a)]. Even for the large amplitude $\epsilon=0.1$ the numerical results coincide with the predictions quite well [see Fig. \ref{figure 7} (b)]. Therefore, our numerical simulations show that the rate of particle creation for higher frequency modes is very well described by the analytical expression (\ref{dodonov4}) beginning at $\tau \sim 1$ \footnote[1]{It was noted in \cite{Dodonov:1993} that Eq. (\ref{dodonov4}) is valid only for not very large numbers $k$ due to limitations of the used approximations. For $\epsilon=0.01$ we have found that Eq. (\ref{dodonov4}) perfectly describes the rate of particle creation for $k=7$ and $k=9$ as well.}. 

Now, let us have a more detailed look at the process of particle creation.
In Figures \ref{figure 4}(a) - \ref{figure 6}(a) we have included additional pictures
showing the total particle number and the number of created resonance mode particles
for short times in high time resolution. For $\epsilon=0.01$ and $0.1$ we have illustrated the background dynamics $l(t)$ as well. For $\epsilon=0.001$ [see Fig. \ref{figure 4}(a)], again, the agreement of the numerical results with the analytical prediction $N(\tau)=\tau^2$ is convincingly illustrated. In these high resolution pictures oscillations in the particle number correlated with the motion of the boundary become visible for $\epsilon=0.01$ and $0.1$ \footnote[2]{For $\epsilon=0.001$ oscillations in $N(t)$ appear as well but with a rather tiny amplitude, not visible in Figure \ref{figure 4}(a).}. 
This observation relies on the fact that in the numerical calculation we 
evaluate the expectation value (\ref{particle number}) at every time step of the integration and not only for times at which the instantaneous state of the cavity equals the initial one \footnote[3]{If we ask for the number of created particles at times $t=n\,T$ only, where $n$ is the number of cavity oscillations of period $T$, we can use the function $\eta_k^{(m)}$ directly to evaluate the expectation value 
(\ref{particle number}), because $\Delta^+(n\,T)=1$ and $\Delta^-(n\,T)=0$ [see Eqs. (\ref{relation between big eta and eta and xi}) and (\ref{delta of t})]. }. Clearly, the larger the amplitude of the cavity vibrations, the larger the amplitude of the oscillations in the particle number.
For $\epsilon=0.01$ the numerical result for the number of particles created in the resonance mode  agrees well with the analytical expression (\ref{dodonov1}) but the numerical values for the total particle number exceed the analytical prediction. Similar statements hold in the case of the large amplitude $\epsilon=0.1$ [see Fig. \ref{figure 6}(a)].
In both cases one observes a jump in the total particle number from zero to a much larger value at the first time step of the integration. Consequently, field modes of higher frequencies $k=3,5,..$ become excited even at the first step of integration and contribute to the total particle number from the very beginning (remember that for $\epsilon=0.1$, $k_{\rm max}=200$ modes had to be taken into account to provide numerical stability for $t_{\rm max}=12$). 

The excitation of high frequency modes from the very beginning may be due to the fact that the cavity motion (\ref{background 1}) is not smooth at $t=0$ but starts with a non-zero velocity which is proportional to the amplitude $\epsilon$ \footnote[4]{In \cite{Dalvit:1998,Cole:1995} it was found, that due to such an initial discontinuity in the velocity of the wall  the energy density inside a vibrating cavity develops $\delta$ function singularities. Furthermore, the discontinuous change in the velocity of the uniformly moving  mirror discussed in \cite{Moore:1970} leads to a logarithmically divergent particle number (see also \cite{Razavy:1985}).}. The discontinuity in the velocity of the mirror at $t=0$ then induces the excitation of modes of higher frequencies and therefore acts as a source for spurious particle creation which manifests itself in a kind of particle background which is present right from the beginning. This effect is nicely illustrated in Figures \ref{figure 5} (a) and \ref{figure 6} (a) which show that the total particle number oscillates on top of this background. Being proportional to $\epsilon$, this effect should play a secondary role for tiny amplitudes $\epsilon \ll 1$ but become more and more important as $\epsilon$ increases.  
This is confirmed by our numerical simulations showing that for tiny amplitudes like $\epsilon=0.00001$, the total particle number for short times is practically identical to the number of particles created in the resonance mode and therefore the excitation of higher frequency modes does not appear. In contrast, instantaneous particle creation takes place at the first integration step for the amplitudes $\epsilon=0.01$ and $\epsilon=0.1$ [see Figs. \ref{figure 5} (a) and \ref{figure 6} (a)] due to contributions from modes of higher frequencies to the total particle number, induced by the discontinuity in the velocity of the mirror at $t=0$. 
 
\begin{figure}
\begin{center}
\begin{tabular}{cc}
\includegraphics[height=5.3cm]{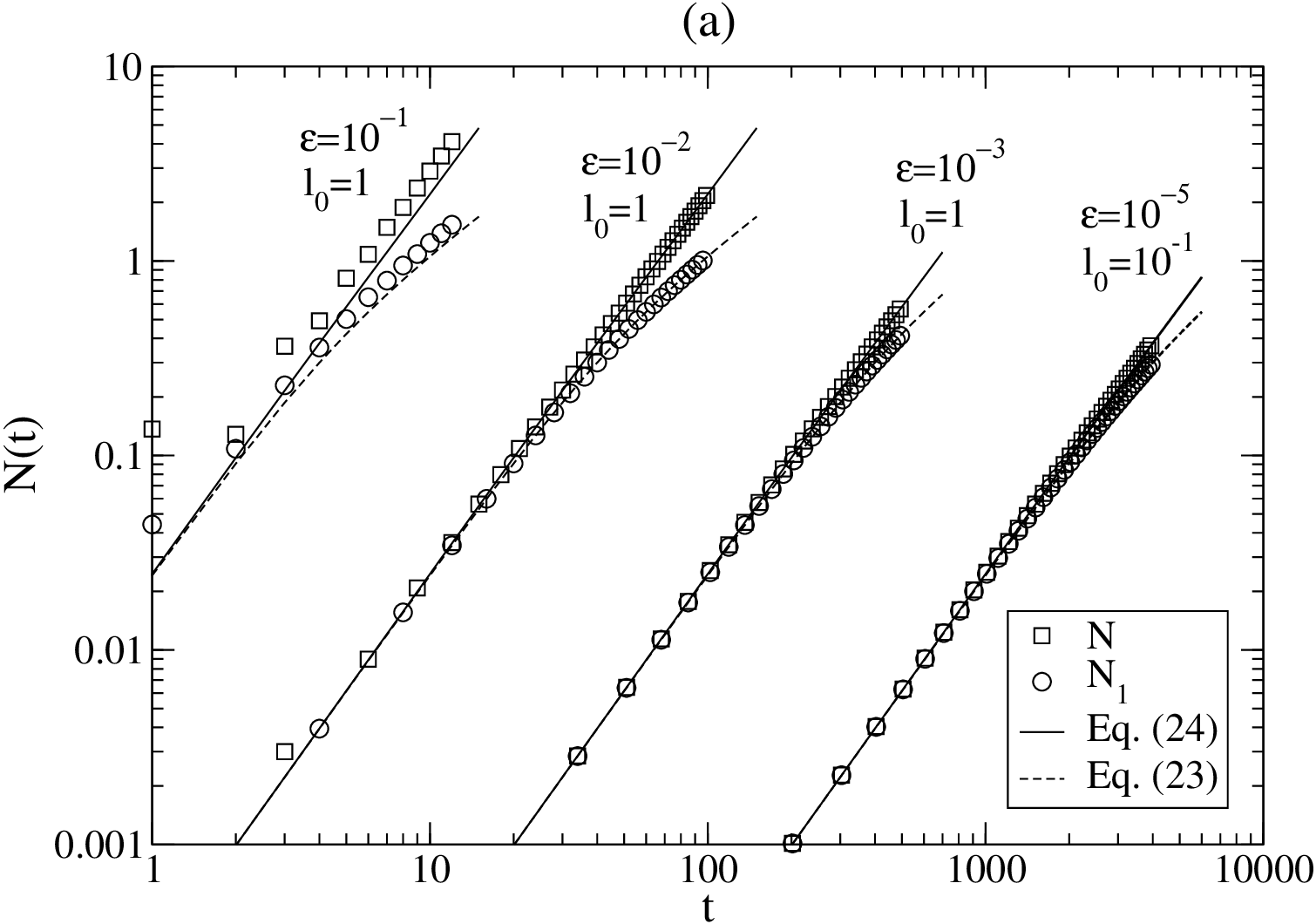}
&
\includegraphics[height=5.3cm]{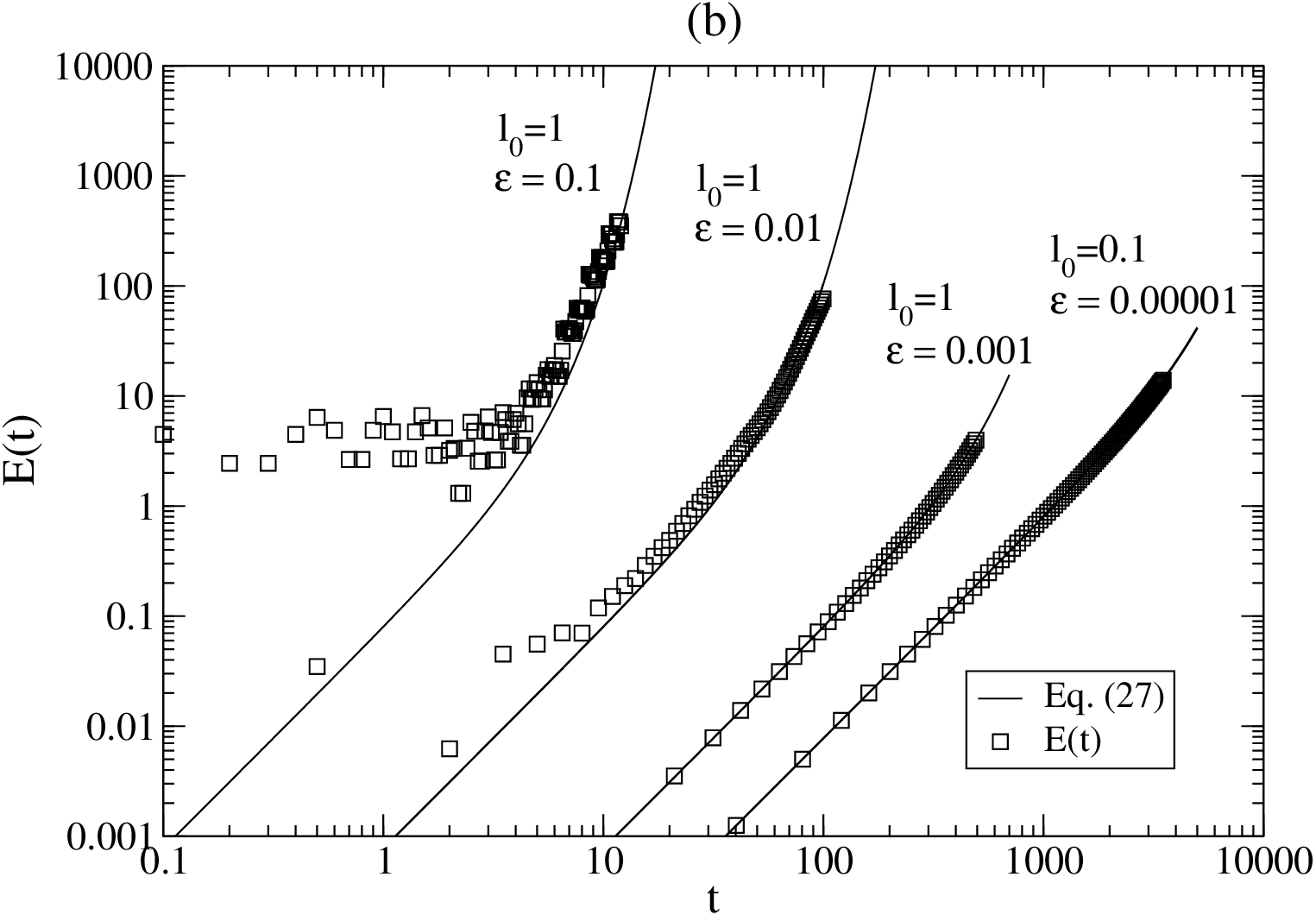}
\end{tabular}
\caption{ (a)  Summary of numerical results obtained for the particle production in a cavity oscillating with (\ref{background 1}) and $n=1$, i.e. the main resonance. We show the total particle number as well as the number of particles created in the resonance mode for amplitudes $\epsilon=0.00001$ for an initial cavity size $l_0=0.1$ corresponding to Fig. \ref{figure 2}, and $\epsilon=0.001$, $0.01$ and $0.1$ for $l_0=1$, corresponding to Figs. \ref{figure 4} -  \ref{figure 6} and compare them with the analytical expressions (\ref{dodonov1}) and (\ref{dodonov2}).
                 (b) Comparison of the numerical results obtained for the energy associated with the created   
                       quantum radiation for the parameters shown in part (a), with the analytical  
                       expression (\ref{dodonov3}).
                  \label{figure 8}}
\end{center}
\end{figure}

Let us summarize the results obtained for the cavity motion (\ref{background 1}). For this purpose, the numerical results for the main resonance scenario, i.e. $\omega_{\rm cav}=2\,\Omega_1^0$, have been arranged in Figure \ref{figure 8} (a) together with the analytical predictions (\ref{dodonov1}) and (\ref{dodonov2}). The numerical calculations reveal that the analytical expressions (\ref{dodonov1}) and (\ref{dodonov2}) describe the numerical results perfectly for the amplitudes $\epsilon=0.00001$ and $\epsilon=0.001$ for all time scales under consideration. Furthermore, for $\epsilon=0.01$ where in high time resolution the effect of the instantaneous particle creation at the first integration step becomes visible [see the high resolution picture in Figure \ref{figure 5} (a)] the behaviour of the total particle number as well as the number of particles created in the resonance mode $\Omega_1^0$ is still in very good agreement with the analytical predictions. For the certainly somewhat artificial scenario with $\epsilon=0.1$ where the mirror starts its oscillations instantaneously with a relativistic velocity (recall that in this case $v/c \sim 0.63$) the analytical expressions (\ref{dodonov1}) and (\ref{dodonov2}) do not match the numerical results. 

In \cite{Dodonov:1996} the authors found, in addition to the analytical expressions (\ref{dodonov1})
and (\ref{dodonov2}), the closed form expression 
\begin{equation}
E(\tau)=\frac{1}{4}\Omega_1^0\sinh^2(2\,\tau)
\label{dodonov3}
\end{equation}
for the energy of the created quantum radiation which  grows much faster than the total particle number. For a calculation of the energy density inside a vibrating cavity see, e.g.,  
\cite{Dalvit:1998, Cole:1995,Wegrzyn:2001}. In Figure \ref{figure 8} (b) we compare the analytical prediction (\ref{dodonov3}) with the energy of the created quantum radiation calculated numerically by means of Eq. (\ref{energy}). For the small amplitudes $\epsilon=0.00001$ and $\epsilon=0.001$  the numerical results agree perfectly with the analytical prediction (\ref{dodonov3}). In the case of $\epsilon=0.01$ the numerically calculated energy of the created quantum radiation deviates slightly from the analytical prediction for small as well as large times where the total particle number is still in very good agreement with the analytical expression (\ref{dodonov2}) [cf. Figs. \ref{figure 5}(a) and \ref{figure 8}(a)] \footnote[5]{Note that the energy of the created quantum radiation (\ref{energy}) is much more sensitive to contributions from higher frequency modes than the total particle number $N=\sum_n\,N_n$, due to the multiplication of the number of particles created in the mode $n$ with the frequency $\Omega_n^0$.}. The energy exceeds the predicted value for small times indicating again that in this case the effect of instantaneous excitation of higher frequency modes due to the non-smooth beginning of the cavity motion (\ref{background 1}) seems to become important. For $\epsilon=0.1$ the numerically calculated energy of the created quantum radiation deviates drastically from the analytical prediction for short times showing the effect of instantaneous excitation of high frequency modes caused by the discontinuity in the velocity of the mirror at $t=0$ quite impressively. Right from the very beginning much of the energy of the cavity motion is transfered to quantum modes of higher frequencies yielding a comparatively large energy of the quantum radiation even for small times. 

Let us now proceed to the investigation of a (more realistic) scenario in which the vibrations of the cavity start with zero velocity. 

\subsection{$\delta(t)=\delta_2(t)$ }

\begin{figure}
\begin{center}
\begin{tabular}{cc}
\includegraphics[height=5.5cm]{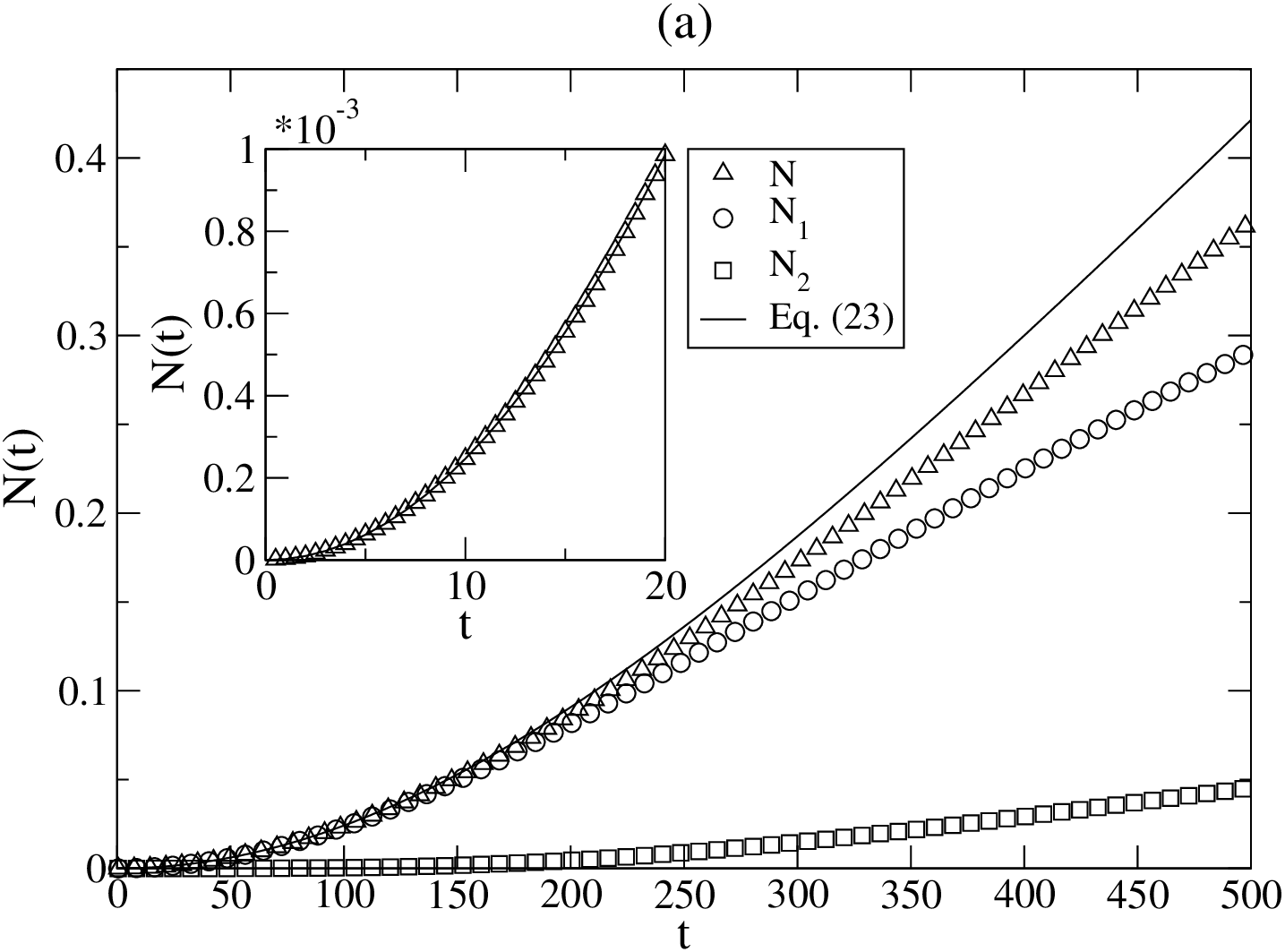}
&
\includegraphics[height=5.5cm]{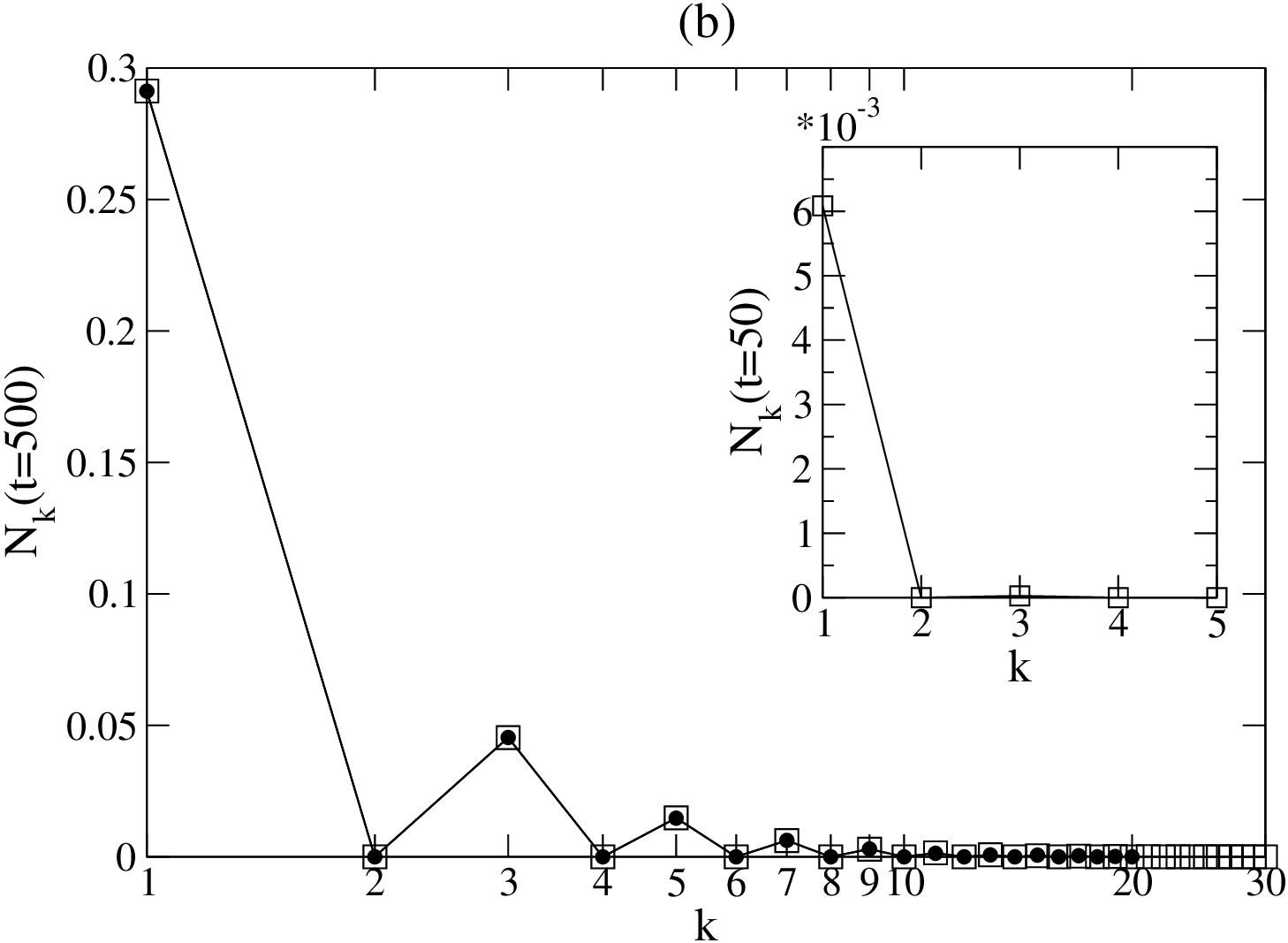}
\end{tabular}
\caption{ (a)  Number of particles produced in a cavity vibrating with (\ref{background 2}) and $n=1$ for parameters $l_0=1$, $\epsilon=0.001$ and $k_{\rm max}=30$,        
                        together with the analytical prediction (\ref{dodonov1}).
                 (b)  Particle spectrum corresponding to (a) for the two cut-off parameters $k_{\rm                  
                       max}=20$ and $k_{\rm max}=30$.
                  \label{figure 9}}
\end{center}
\end{figure}

\begin{figure}
\begin{center}
\begin{tabular}{cc}
\includegraphics[height=5.5cm]{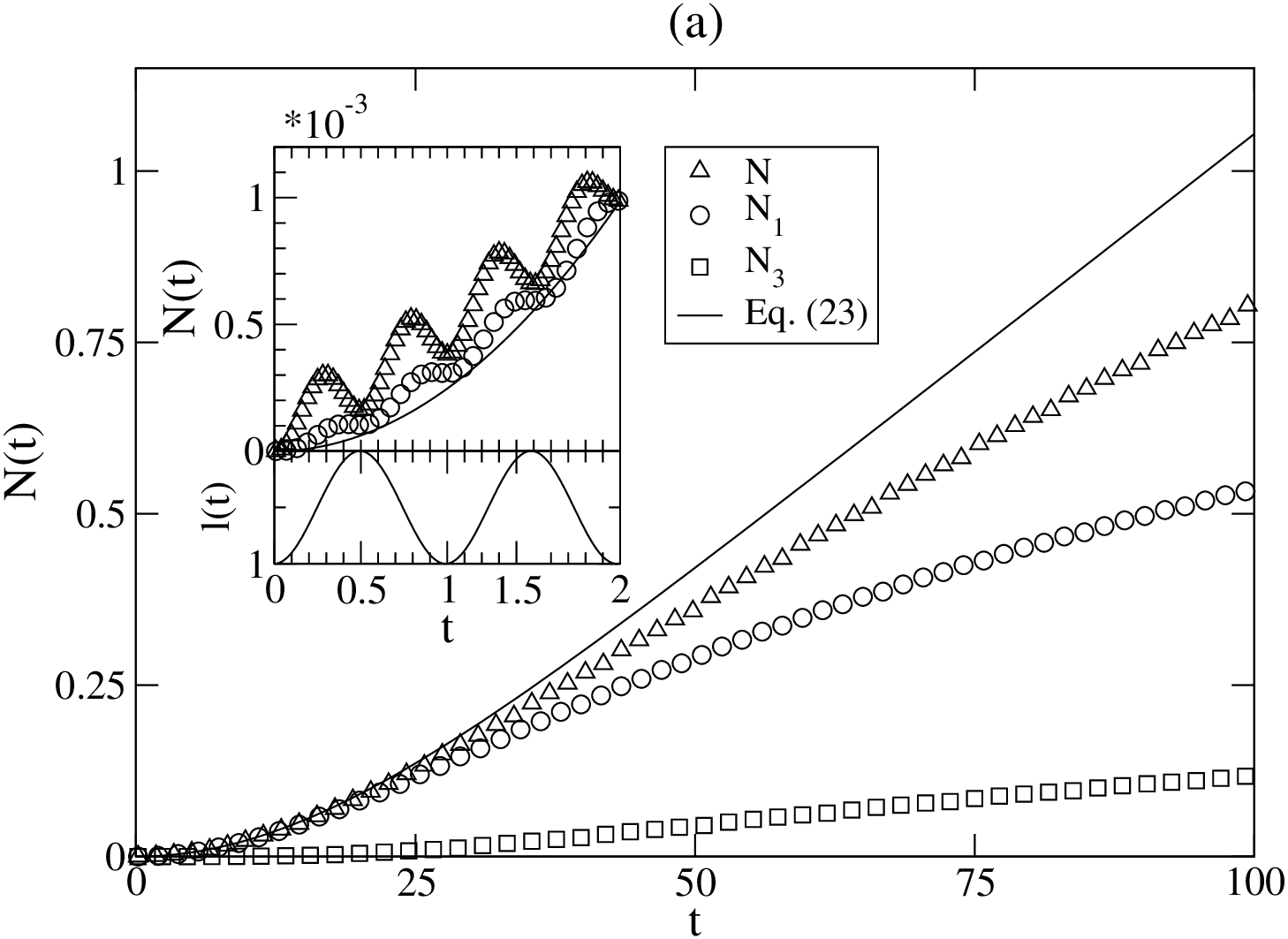}
&
\includegraphics[height=5.5cm]{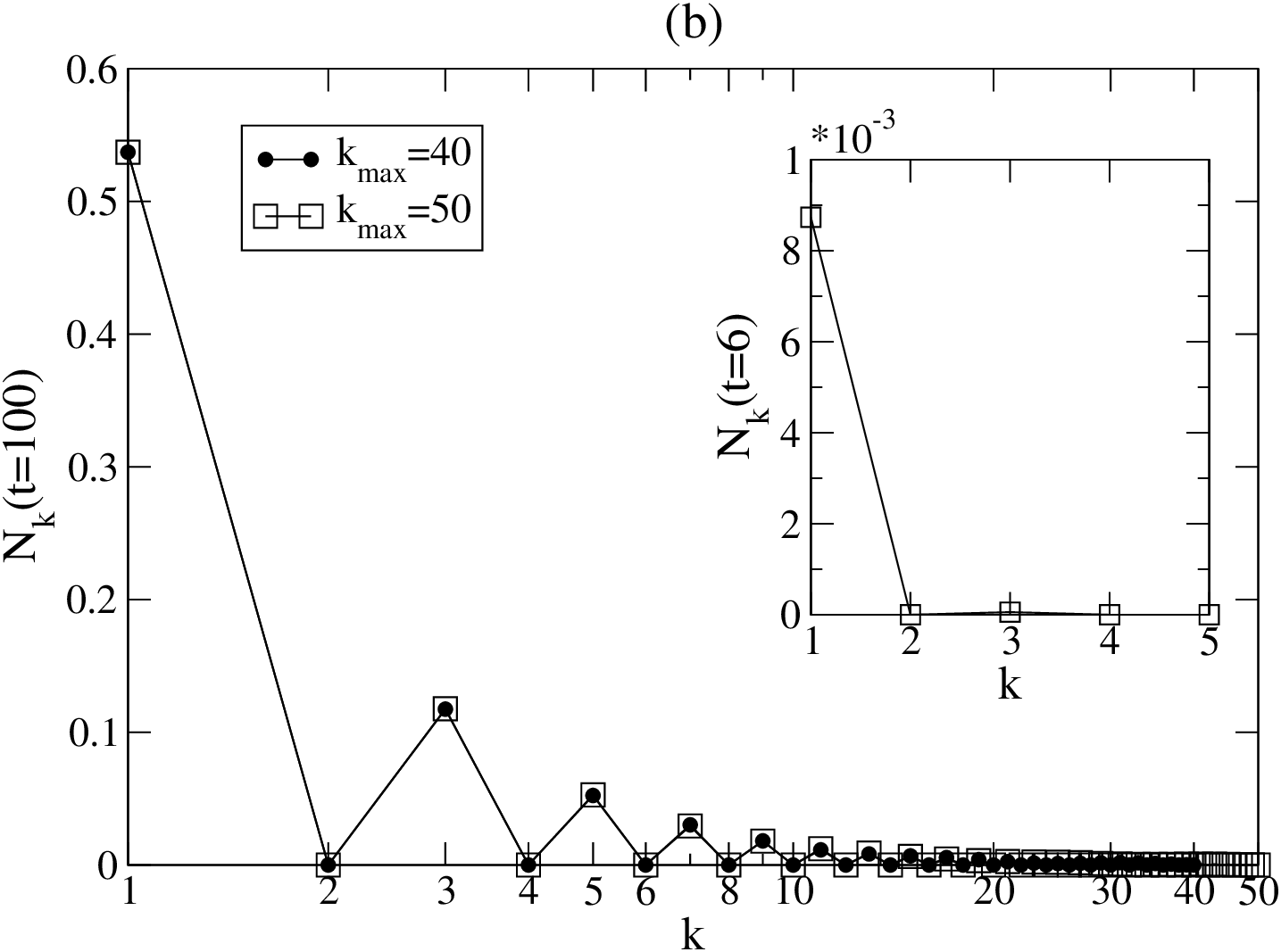}
\end{tabular}
\caption{ (a)  Number of particles produced in a cavity vibrating with (\ref{background 2}) and $n=1$ for parameters $l_0=1$, $\epsilon=0.01$ and $k_{\rm max}=50$,        
                        together with the analytical prediction (\ref{dodonov1}).
                 (b)  Particle spectrum corresponding to (a) for the two cut-off parameters $k_{\rm                  
                       max}=40$ and $k_{\rm max}=50$.
                  \label{figure 10}}
\end{center}
\end{figure}

\begin{figure}
\begin{center}
\begin{tabular}{cc}
\includegraphics[height=5.5cm]{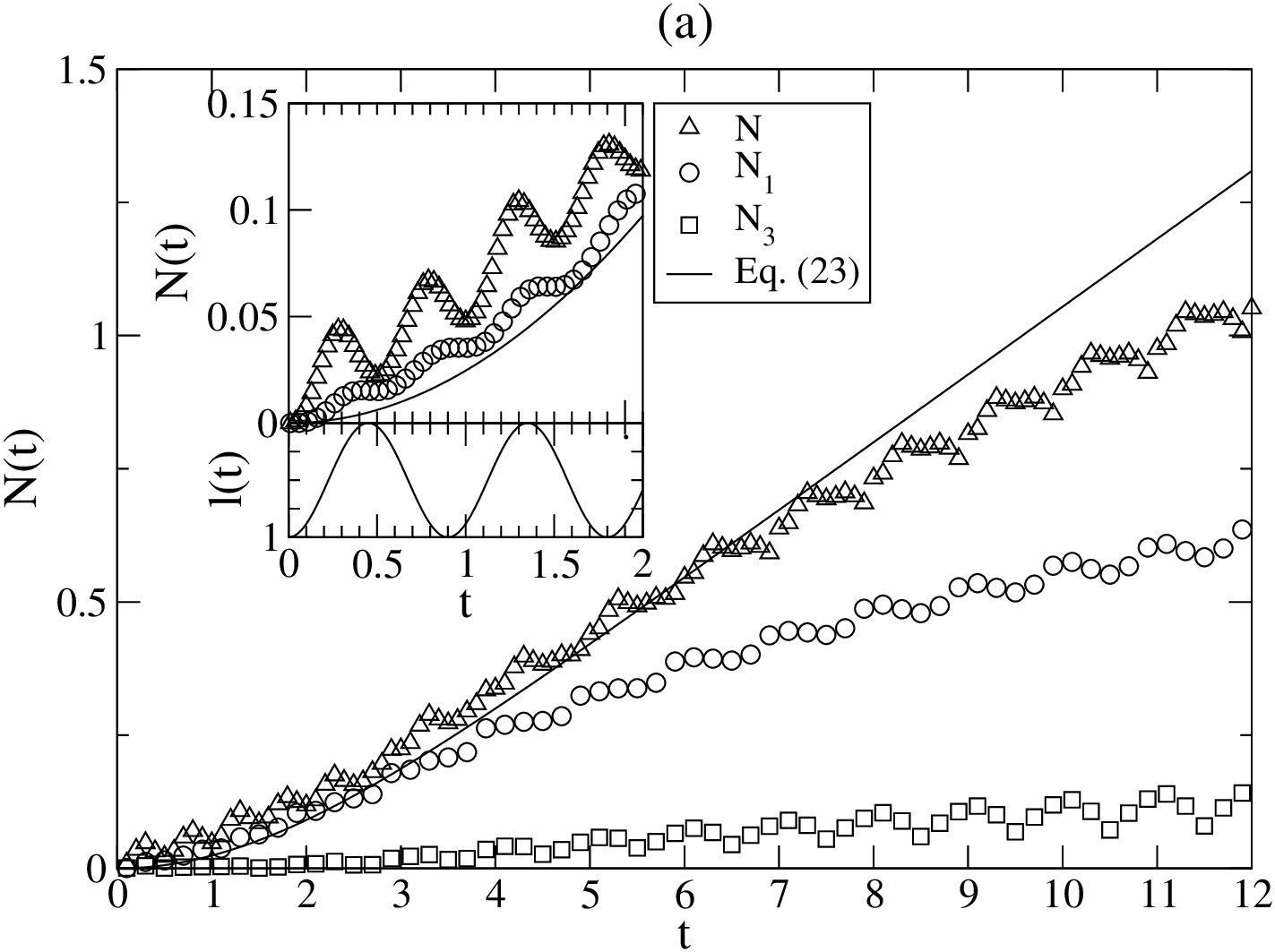}
&
\includegraphics[height=5.5cm]{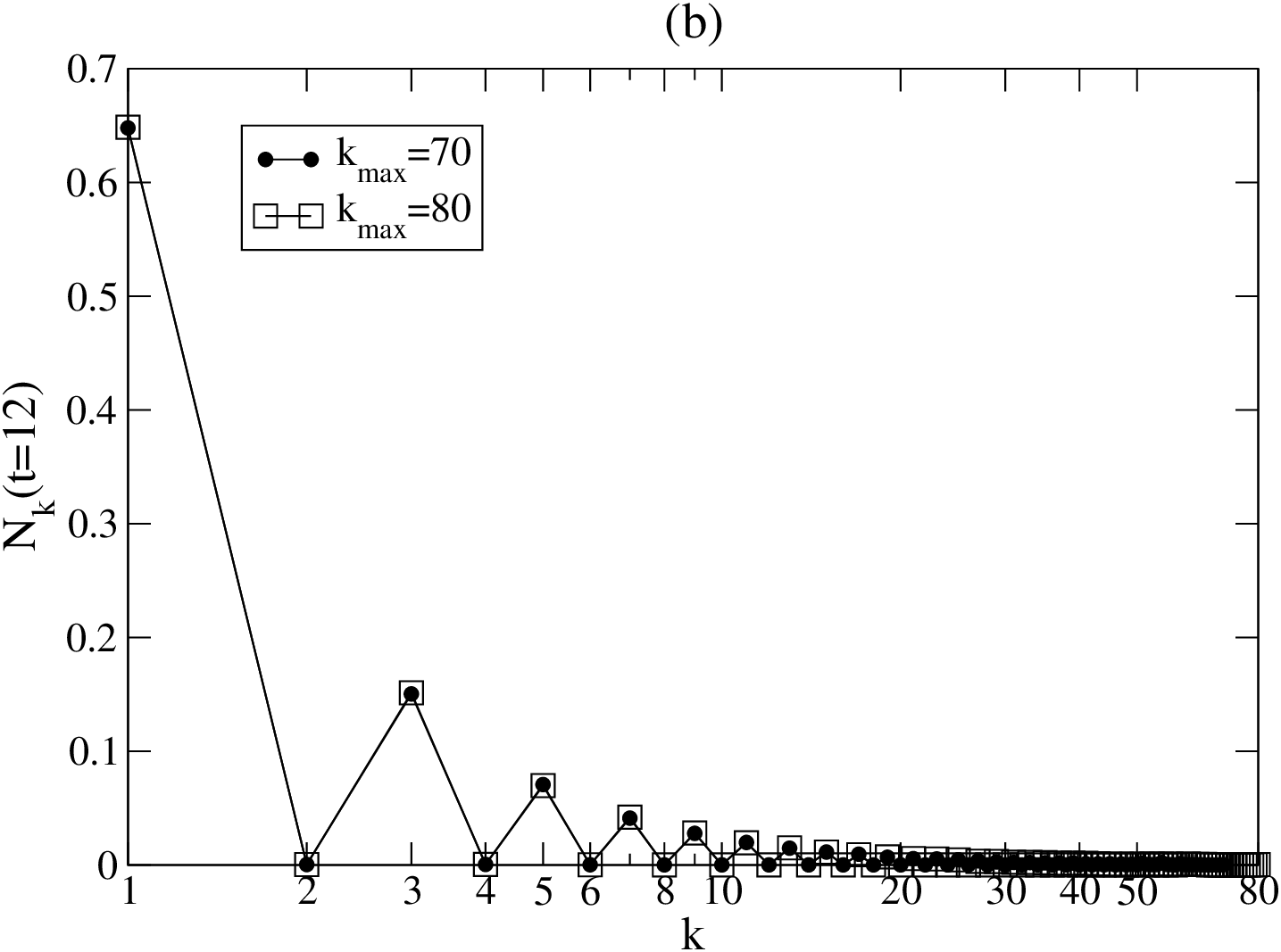}
\end{tabular}
\caption{ (a)  Number of particles produced in a cavity vibrating with (\ref{background 2}) and $n=1$ for parameters $l_0=1$ ,$\epsilon=0.1$ and $k_{\rm max}=80$,        
                        together with the analytical prediction (\ref{dodonov1}).
                 (b)  Particle spectrum corresponding to (a) for the two cut-off parameters $k_{\rm                  
                       max}=70$ and $k_{\rm max}=80$.
                  \label{figure 11}}
\end{center}
\end{figure}

In this section we study the cavity motion
\begin{equation}
l(t)=l_0\left[1 + 2\,\epsilon \sin^2(\Omega_n^0\,t)\right]=l_0\left[1+\epsilon\, (1-\cos(2\,\Omega_n^0\,t)\,)\right]
\label{background 2}
\end{equation}
with smooth initial conditions $l(0)=l_0$ and $\dot{l}(0)=0$ which has also been studied numerically in \cite{Antunes}. In the following we compare the process of particle creation in a one-dimensional cavity vibrating with (\ref{background 2}) with the results obtained for the cavity motion (\ref{background 1}). Therefore we set $l_0=1$ and restrict ourselves to amplitudes $\epsilon=0.001$, $0.01$ and $0.1$. The results of the numerical simulations are shown in Figures \ref{figure 9} - \ref{figure 11} where, for reasons of comparison,  we have included the analytical expression (\ref{dodonov1}) as well.

We observe that the process particle creation caused by the cavity motion (\ref{background 2}) is different from the one driven by (\ref{background 1}). The qualitative behaviour of the total particle number shows no differences for relatively short times compared to the background motion (\ref{background 1}), i.e. $N(t)$ grows quadratically with time. But for larger times this behaviour begins to change. In the case of $\epsilon=0.001$ [see Figure \ref{figure 9} (a)] the initial quadratic increase of the particle number is followed by a slowing down of the rate of particle creation with time and the particle number enters a region in which it is effectively described by a linear behaviour. Thereby the total particle number is always less than the number of resonance mode particles created in a cavity vibrating with (\ref{background 1}). For $\epsilon=0.01$ [see Figure \ref{figure 10} (a)] and $\epsilon=0.1$ 
[see Figure \ref{figure 11} (a)] the same qualitative behaviour is observed. But for larger times \footnote[2]{Note that the expressions "short times" and "large times" extensively used throughout the paper refer to the so called "slow time" $\tau=\frac{1}{2}\,\epsilon\,\Omega_1^0\,t$ as introduced in \cite{Dodonov:1996} rather than to the time variable $t$ used in the numerical simulations. Roughly spoken, we use the term "short time" ("large time") for $\tau < 1$ ($\tau > 1$).},  the particle number seems to have a tendency to leave the linear regime due to a further deceleration in the rate of particle production. The number of particles created in the mode $k=1$ shows this 
slowing down very clearly. 

\begin{figure}
\begin{center}
\begin{tabular}{ccc}
\includegraphics[height=3.6cm]{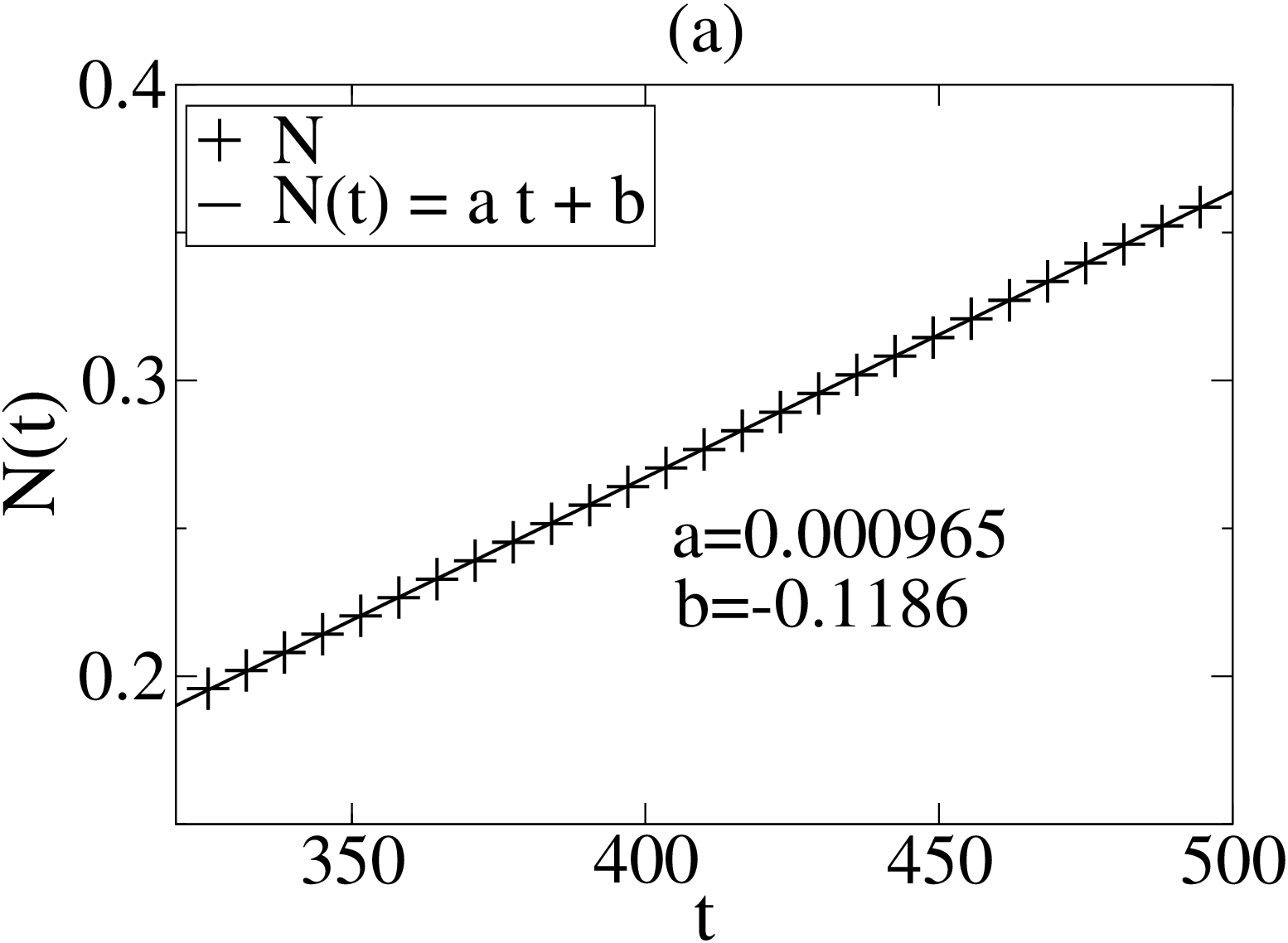}
&
\includegraphics[height=3.6cm]{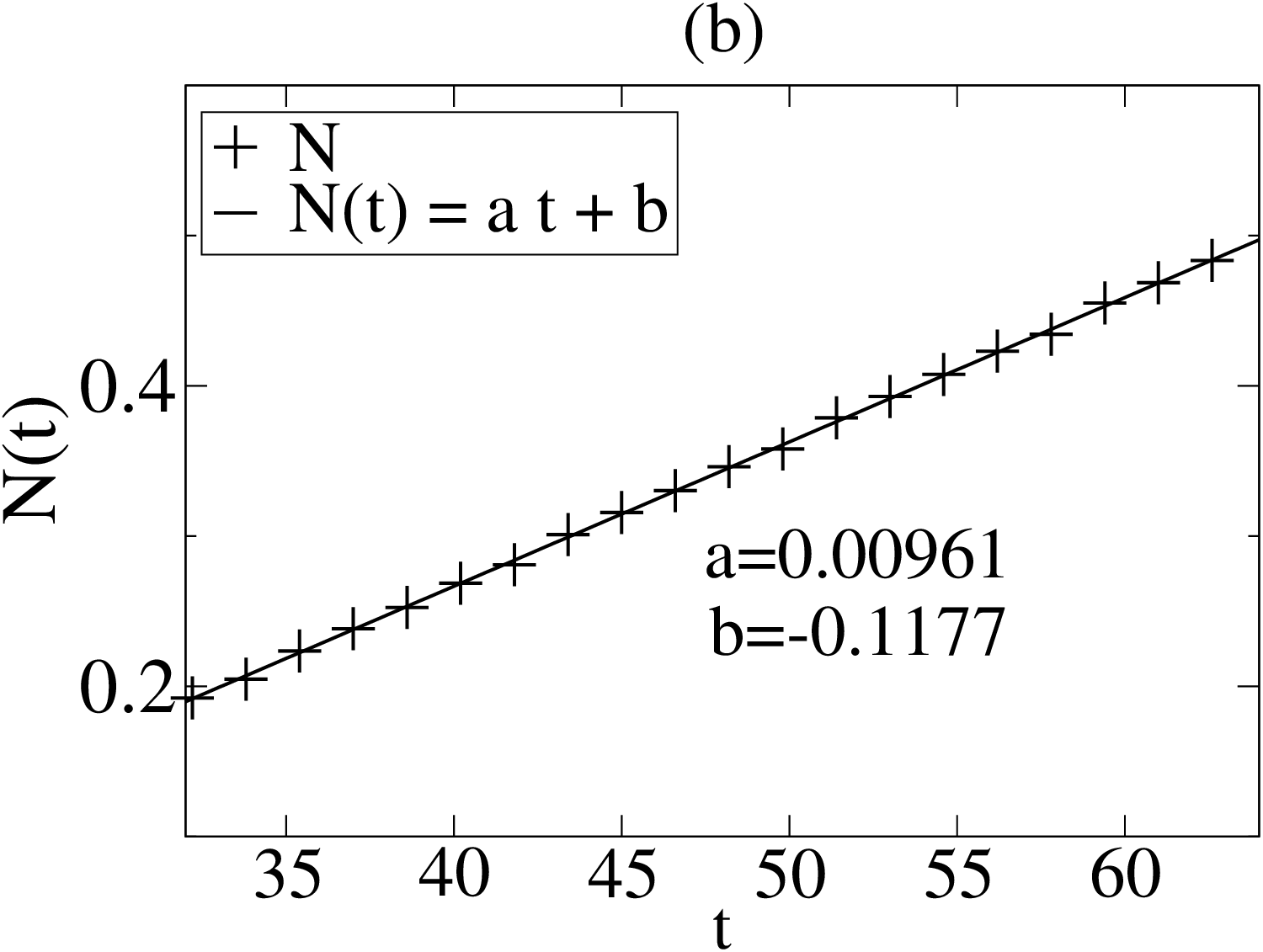}
&
\includegraphics[height=3.6cm]{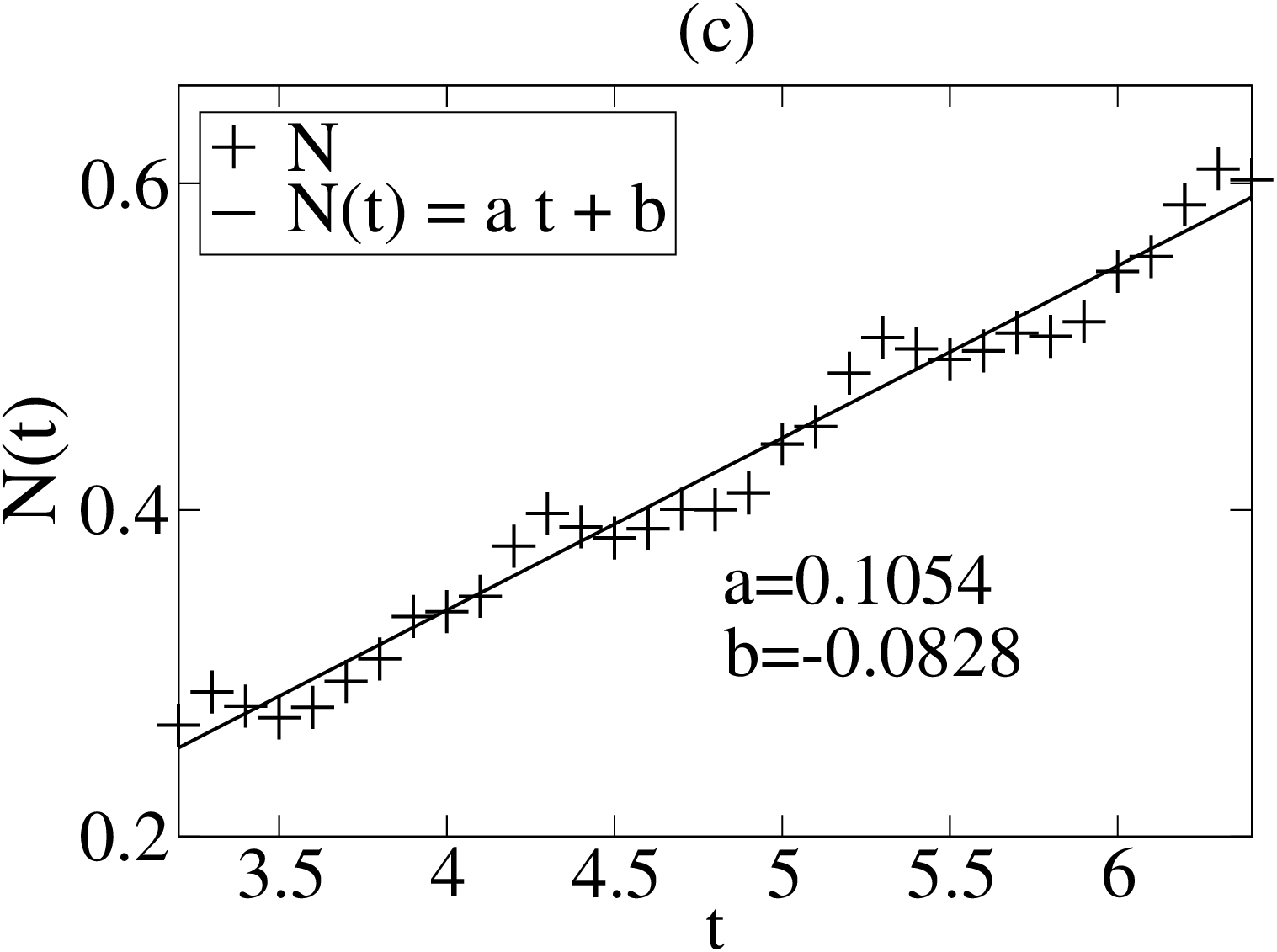}
\end{tabular}
\caption{ Total particle number in the time range  $\frac{1}{2}\le \tau \le 1$        
                 for parameters $l_0=1$ and   
                  (a) $\epsilon=0.001$, (b) $\epsilon=0.01$ and (c) $\epsilon=0.1$,  corresponding to Figs.   
                   \ref{figure 9} (a) - \ref{figure 11} (a), together with a fit of the data to the linear law   
                   $N(t)=a\,t +b$.
                  \label{figure 12}}
\end{center}
\end{figure}

We have found that the qualitative behaviour of the total particle number can be
described very well by a linear law in the time range $1/2 \le \tau \le 1$. This is illustrated in Fig. \ref{figure 12} where we show fits of the numerical results to $N(t)=a\,t + b$ in the corresponding time ranges. From these plots the linear behaviour of the total particle number in the time range $\frac{1}{2} \le \tau \le 1$ becomes evident \footnote[3]{From the parameters $a$ obtained by fitting the numerical results in the time range $1/2 \le \tau \le 1$ one may deduce, at least as a good approximation, the dependence $a\sim\epsilon$. Later on we will see that, more generally, $a=\frac{\epsilon}{l_0}$ seems to hold such that, in the linear regime, $\frac{dN(\tau)}{d\tau}=\frac{2}{\pi}$ (for $n=1$, i.e. $\omega_{\rm cav}=2\Omega_1^0$).}.

\begin{figure}
\begin{center}
\begin{tabular}{cc}
\includegraphics[height=5.5cm]{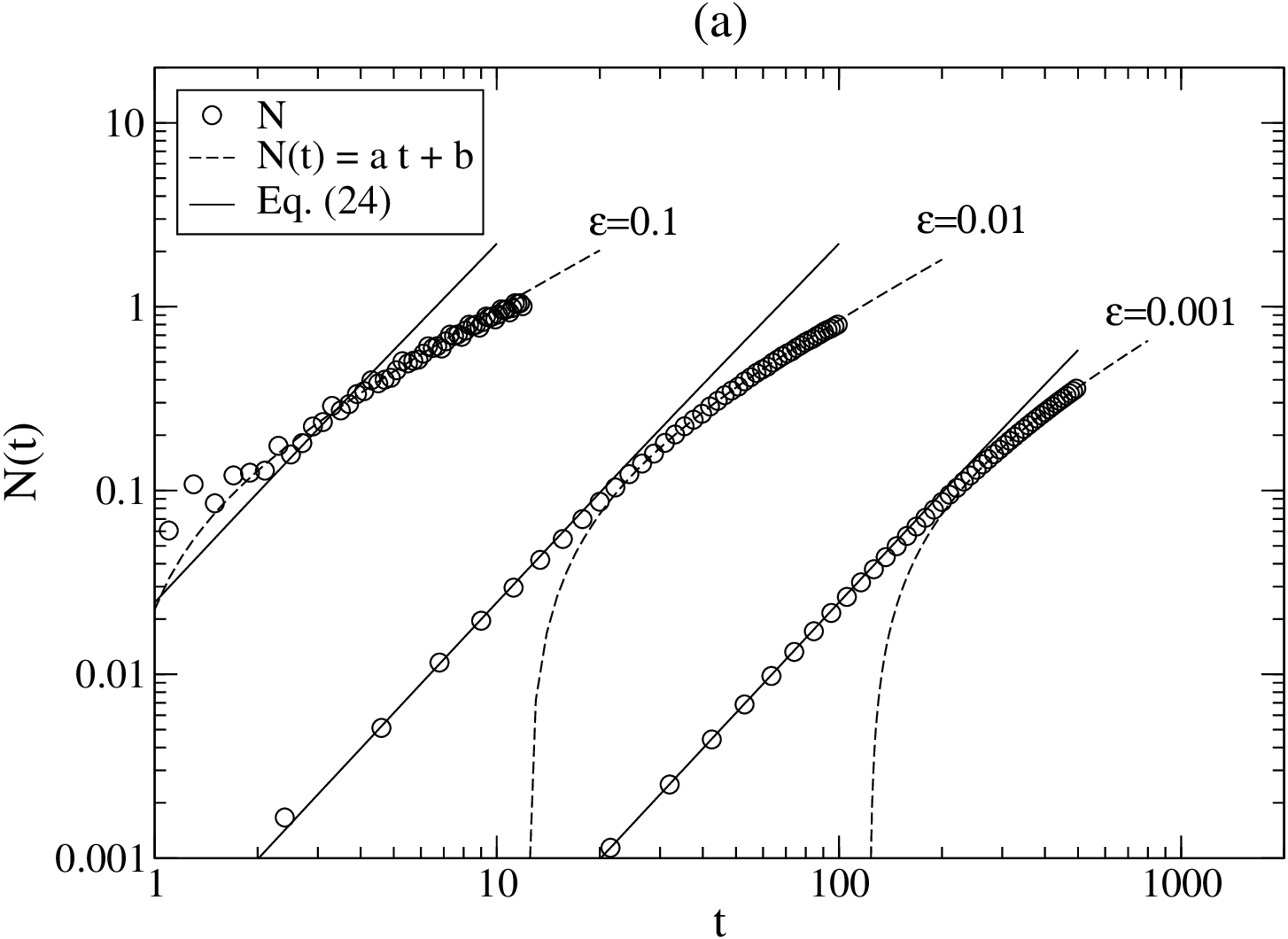}
&
\includegraphics[height=5.5cm]{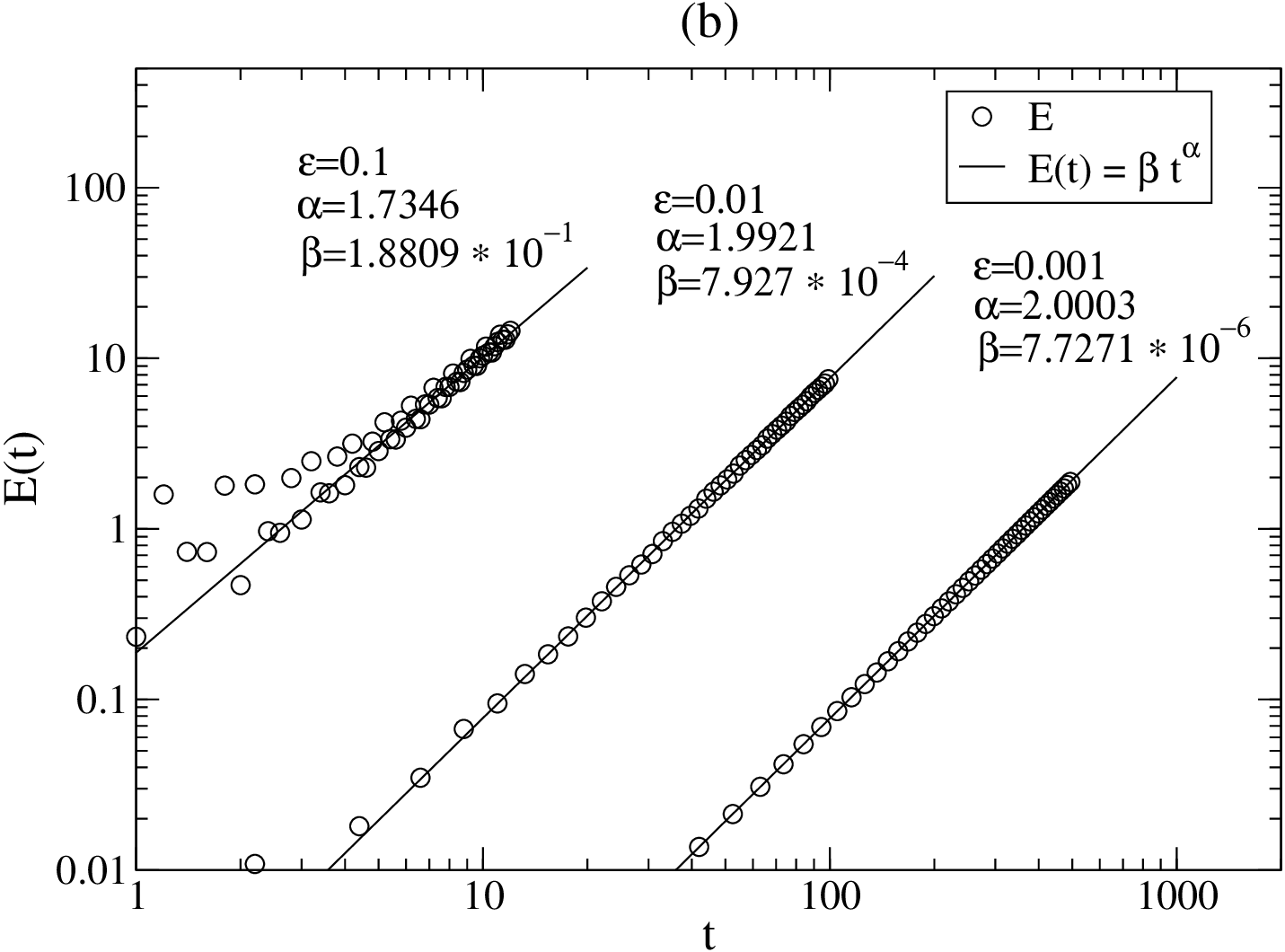}
\end{tabular}
\caption{ (a)  Summary of the numerical results obtained for the particle production in a cavity oscillating with (\ref{background 2}) and $n=1$. We show the total particle number for $l_0=1$ and amplitudes $\epsilon=0.001$, $0.01$ and $0.1$, corresponding to Figures. \ref{figure 9} (a) -  \ref{figure 11} (a) and compare them for short times with the analytical expression (\ref{dodonov2}) and for larger times with the linear law $N(t)=a\,t + b$ with parameters $a$ and $b$ given in Figures \ref{figure 12} (a) - (c).  (b) Time evolution of the energy associated with the created quantum radiation corresponding to the parameters shown in part (a) together with a fit of the numerical data to the power law $E(t)=\beta \,t^\alpha$.                  
\label{figure 13}}
\end{center}
\end{figure}

To demonstrate the transition from the quadratic to the linear behaviour of the total particle number more carefully we show a summary of the numerical results for the total particle number in Fig. \ref{figure 13} (a)  together with the analytical prediction (\ref{dodonov2}) [derived for the cavity motion (\ref{background 1})] as well as linear functions $N(t)=a\,t+b$ with parameters $a$ and $b$ corresponding to Fig. \ref{figure 12}. As on infers from Fig. \ref{figure 13} (a), the total particle number is very well described by Eq. (\ref{dodonov2}) up to $t \sim 100$ for $\epsilon=0.001$ and up to $t \sim 10$ for $\epsilon=0.01$. Then the deceleration in the rate of particle creation sets in and the time evolution of the total particle number enters the linear regime $1/2 \le \tau \le 1$. While for $\epsilon=0.001$ the particle number stays in the linear regime by reaching the end of the shown integration range the further deceleration of the rate of particle creation becomes visible for $\epsilon=0.01$ and $0.1$, and the particle number deviates more and more from the linear behaviour by approaching the maximum integration time. To investigate this deceleration in the rate of  particle creation in more detail, one clearly has to consider larger integration times. We will address this question later on.

The particle spectra shown in Figures \ref{figure 9} (b) - \ref{figure 11} (b) reveal that the characteristic features remain unchanged compared to the cavity motion (\ref{background 1}), e.g., only odd modes are created. But the number of particles produced in each mode is smaller compared to Figures \ref{figure 4} (b) - \ref{figure 6} (b) yielding a smaller total particle number as described above. 
While for $\epsilon=0.001$ the same cut off parameter $k_{\rm max}$ has to be used for both cavity dynamics, the number of modes which have to be taken into account to ensure numerical stability for amplitudes $\epsilon=0.01$ and $0.1$ is smaller for the cavity motion (\ref{background 2}) compared to 
(\ref{background 1}), indicating that fewer modes of higher frequencies are excited in a cavity
oscillating with (\ref{background 2}) compared to cavity vibrations of the form (\ref{background 1})
\footnote[2]{Note that the cavity motion (\ref{background}) [and therefore (\ref{background 2})] was chosen such that the total change $\Delta l$ of the cavity length is $2\epsilon$ for all $k$ to have comparable situations. See also Figure \ref{figure 1}. If $\Delta l$ were different, $\Delta l=2\epsilon$ for (\ref{background 1}) and $\Delta l=\epsilon$ for (\ref{background 2}), say, the total number of particles created in both cases would differ drastically (up to an order of magnitude) as we have observed in our simulations}. 

In Figures \ref{figure 9} (a) - \ref{figure 11} (a) we have included pictures showing the numerical results for short times and with a high time resolution which we now compare with the corresponding results 
obtained for the cavity motion (\ref{background 1}) shown in Figures \ref{figure 4} (a) - \ref{figure 6} (a).
By comparing the high resolution picture in Fig. \ref{figure 9} (a) with the one in Fig. \ref{figure 4} (a) we 
clearly see (in this resolution) no difference. For both cavity motions, the total number of created particles is well described for short times by $N(\tau)=\tau^2$. For $\epsilon=0.01$ and $\epsilon=0.1$ the oscillations in the particle number, caused by the fact that we evaluate the particle number at every integration step as explained in the former subsection, now mimic the background motion (\ref{background 2}). They start smoothly and differ from the oscillations in a cavity driven by (\ref{background 1}) only by a phase. Apart from that, the number of particles created in the mode of frequency $\Omega_1^0$ behaves exactly the same in both cases.
For the total particle number the situation is now different. Without the discontinuity in the velocity of the mirror at the beginning of the integration the total particle number does not exhibit the jump at the first step of integration, which is characteristic in the case of the background motion (\ref{background 1}). 
By comparing the short time pictures in Fig. \ref{figure 5} (a) and Fig. \ref{figure 10} (a) for $\epsilon=0.01$ as well as in Fig. \ref{figure 6} (a) and Fig. \ref{figure 11} (a) for $\epsilon=0.1$ we clearly recognize the effect of spurious particle creation caused by the discontinuity in the velocity of the mirror motion (\ref{background 1}) at $t=0$. The amplitudes of the oscillations in the total particle number itself are of the same height but for the cavity motion (\ref{background 1}) the oscillations sit on top of a kind of particle background which is present from the first step of integration as described in the last subsection. For the background motion (\ref{background 2}) with the smooth initial condition $\dot{l}(0)=0$ no instantaneous particle creation takes place, the particle background does not appear and thus, the total particle number starts to increase smoothly during the first integration steps.

In Figure \ref{figure 13} (b) we show the energy associated with the produced quantum radiation corresponding to Figures \ref{figure 9} - \ref{figure 11} calculated by means of Eq. (\ref{energy})
together with fits of the numerical data to a power law $E(t)=\beta \, t^{\alpha}$. 
We find that in the time ranges under consideration the energy $E(t)$ produced in a cavity 
subject to vibrations of the form (\ref{background 2}) and amplitudes $\epsilon=0.001$ and $0.01$ is very well fitted by $\alpha=2$ \footnote[1]{For $\epsilon=0.001$ and $0.01$ the parameters $\alpha$ and $\beta$ were obtained by fitting the power law to the numerical data in the range $t_{\rm max}/10 \le t \le t_{\rm max}$.}.  Thus the energy increases quadratically for $\epsilon=0.001$ and $0.01$ in contrast to the exponential behaviour of the energy for the cavity motion (\ref{background 1}) [see also Eq. (\ref{dodonov3}) and Fig. \ref{figure 8} (b)]. For the amplitude $\epsilon=0.1$  
the rate of energy production is even less (in the discussed time range) and described by the exponent $\alpha=1.7346$ \footnote[2]{In this case the fit to the numerical data was done in the range $5 \le t \le t_{\rm max}=12$.}. Therefore, the rate of energy production is much smaller in a cavity vibrating with (\ref{background 2}) compared to (\ref{background 1}). Partly, this is due to the fact that, as mentioned above, fewer  modes are excited in the case of the background motion (\ref{background 2}). Furthermore, as one can infer from the presented particle spectra $N_k(t_{\rm max})$, the number of particles $N_k$ created in each excited mode with frequency $\Omega_k^0$ at a given time in a cavity vibrating with (\ref{background 2}) is less compared to the corresponding number of particles created in a cavity subject to the background motion (\ref{background 1}) \footnote[3]{This statement does not hold, of course, for short times for which the particle number behaves qualitatively similar for both cavity motions.}. 

\begin{figure}
\begin{center}
\begin{tabular}{cc}
\includegraphics[height=5.5cm]{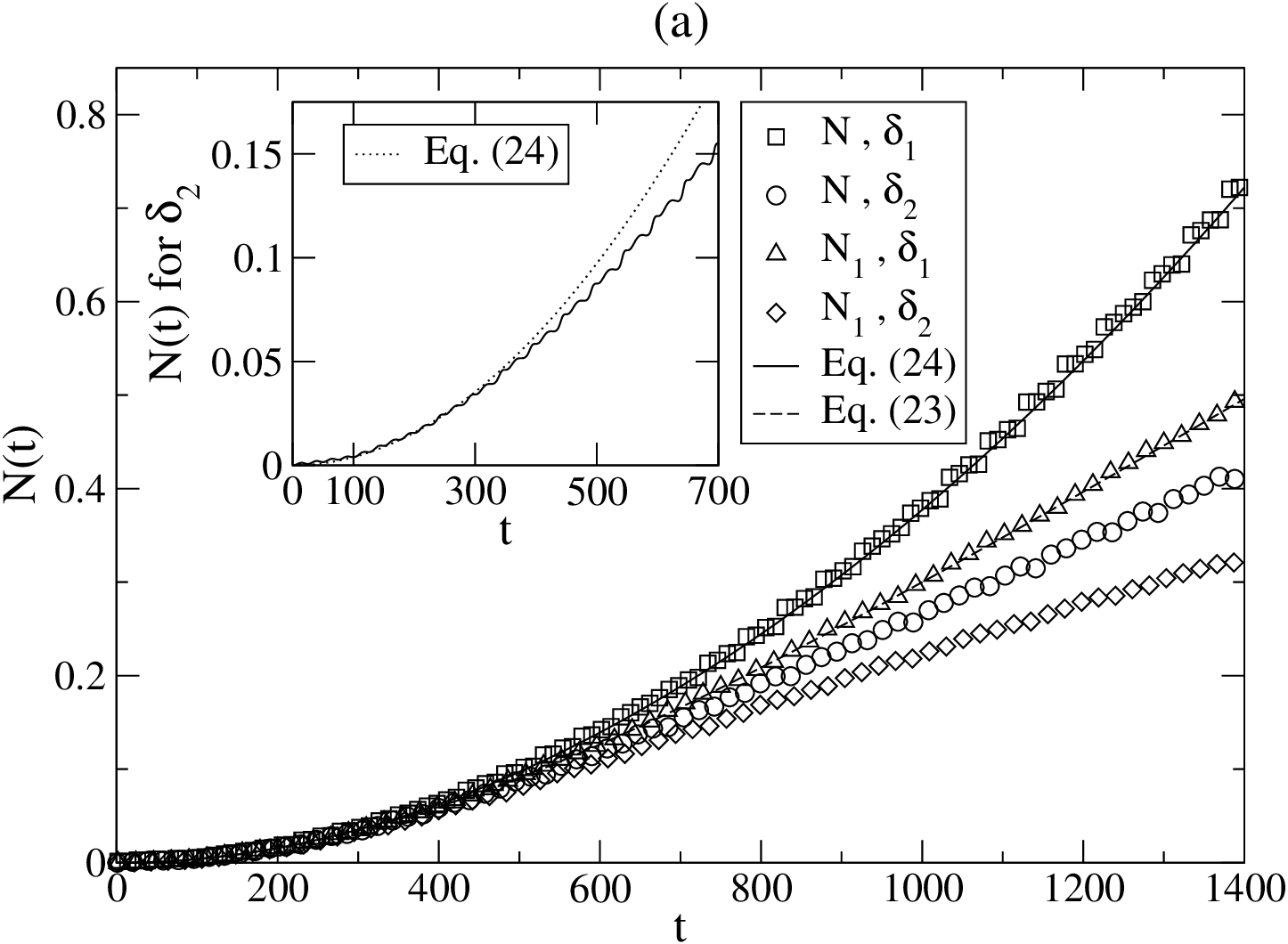}
&
\includegraphics[height=5.5cm]{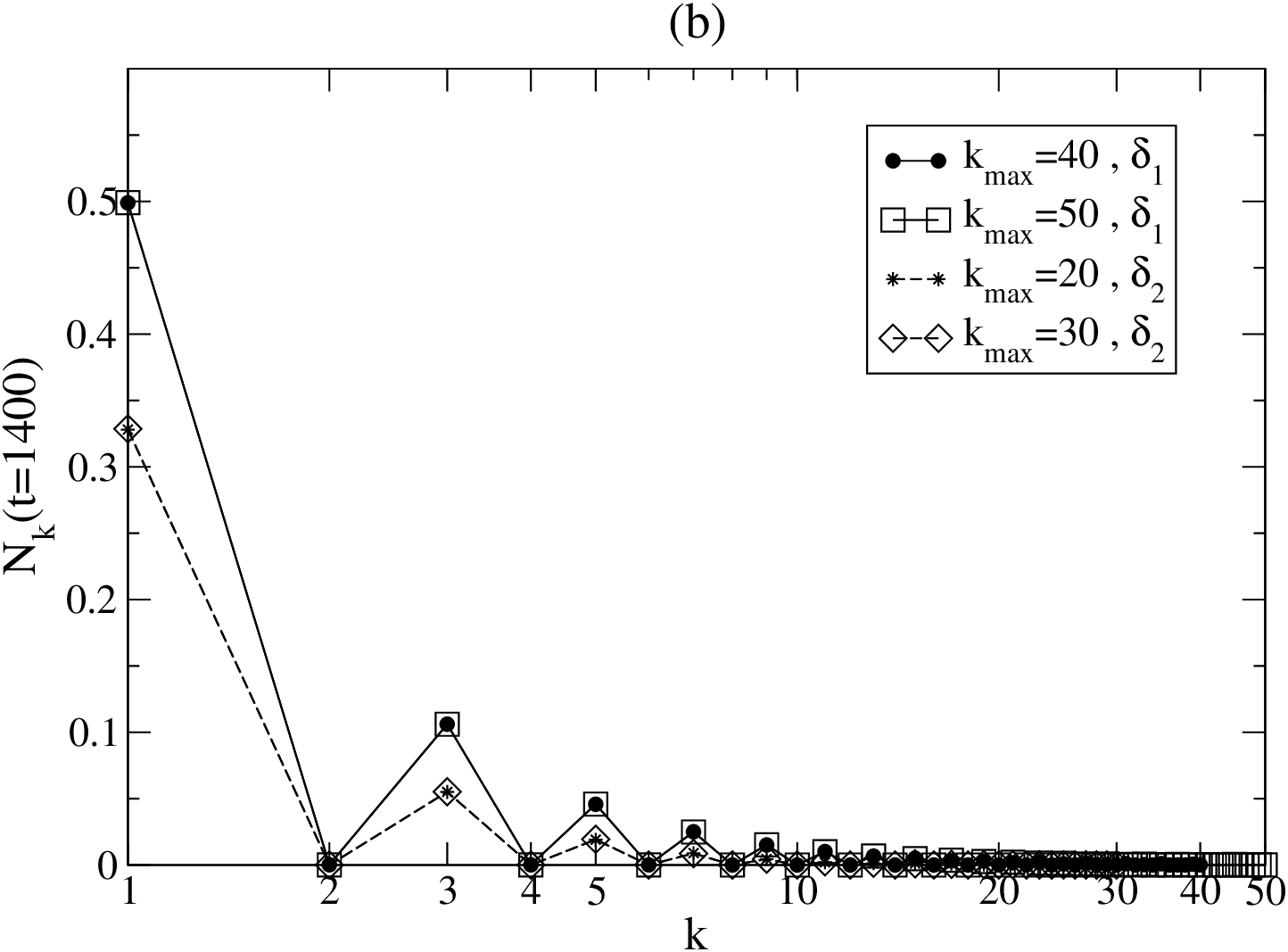}
\end{tabular}
\caption{ (a)  Comparison of the particle production for the two cavity motions (\ref{background 1}) and (\ref{background 2}) with $n=1$, i.e. the main resonance, and parameters $l_0=50$ and $\epsilon=0.02$
corresponding to the parameters studied in \cite{Antunes}. The analytical predictions (\ref{dodonov1}) and (\ref{dodonov2}), valid for the background motion (\ref{background 1}), are shown as well. 
                 (b)  Particle spectra corresponding to (a). 
                       \label{figure 14}}
\end{center}
\end{figure}

Now let us compare our method with a completely different numerical approach. 
In \cite{Antunes} the author studies the creation of massless scalar particles in a one-dimensional cavity 
subject to the motion (\ref{background 2}) by solving the Klein - Gordon equation using 
an improved Leap Frog algorithm (consult the paper for details).  
Figure \ref{figure 14} summarizes our numerical results obtained for $n=1$, i.e. $\omega_{\rm cav}=2\frac{\pi}{l_0}$, and the parameters $l_0=50$ and $\epsilon=0.02$ which correspond to the parameters used in the simulations of \cite{Antunes}. We performed our calculation for an integration time $t_{\rm max}=1400$ which is twice the one considered in \cite{Antunes}. For a better comparison, we depict our numerical results for the shorter integration range [$0 \le t \le 700$] in an additional picture in Fig. \ref{figure 14} (a) which should be compared with Fig. 5 in \cite{Antunes} and demonstrates that our results perfectly agree with the results presented in \cite{Antunes}. 
Furthermore, the outcome of the numerical simulations obtained for the same parameters but for the cavity motion (\ref{background 1}) is shown in Figure \ref{figure 14} (a) as well, which perfectly agrees with the analytical expressions (\ref{dodonov1}) and (\ref{dodonov2}) also indicated in this plot. 

The result for total number of created particles shown in Fig. 5 of \cite{Antunes} for an integration time $t_{\rm max}=700$ is fitted by a power law $\propto t^{\alpha}$ with $\alpha=1.85$ (in the range $100 \le t \le 700$) which is interpreted by the author as in reasonable agreement with the prediction $\alpha=2$ derived in \cite{Dodonov:1996} for the cavity dynamics (\ref{background 1}), but not (\ref{background 2})!  As we have shown with our numerical simulations, the two cavity motions (\ref{background 1}) and (\ref{background 2}), apart from comparatively short times, do not yield the same behaviour for the number of created particles  \footnote[3]{Recall that the analytical expression (\ref{dodonov2}) for the total number of created particles derived for the cavity motion (\ref{background 1}) behaves like $t^2$ for small as well as large times.}. As one infers from Figure \ref{figure 14} (a) the behaviour of the total number of created particles is very well described by Eq. (\ref{dodonov2}), i.e. a quadratic increase, up to $t\sim 300$. After that, the rate of particle creation slows down and the particle number shows the transition to the linear regime as discussed before in detail. 
Thus, the exponent $\alpha=1.85$ found in \cite{Antunes} by fitting the total particle number to a power law for $100 \le t \le 700$ is explained by the fact, that this time range contains the transition from the initial quadratic increase of the particle number to a linear behaviour which becomes visible only for larger integration times as confirmed by our simulations [see Figure \ref{figure 14}(a)]. 

\begin{figure}
\begin{center}
\begin{tabular}{cc}
\includegraphics[height=5.5cm]{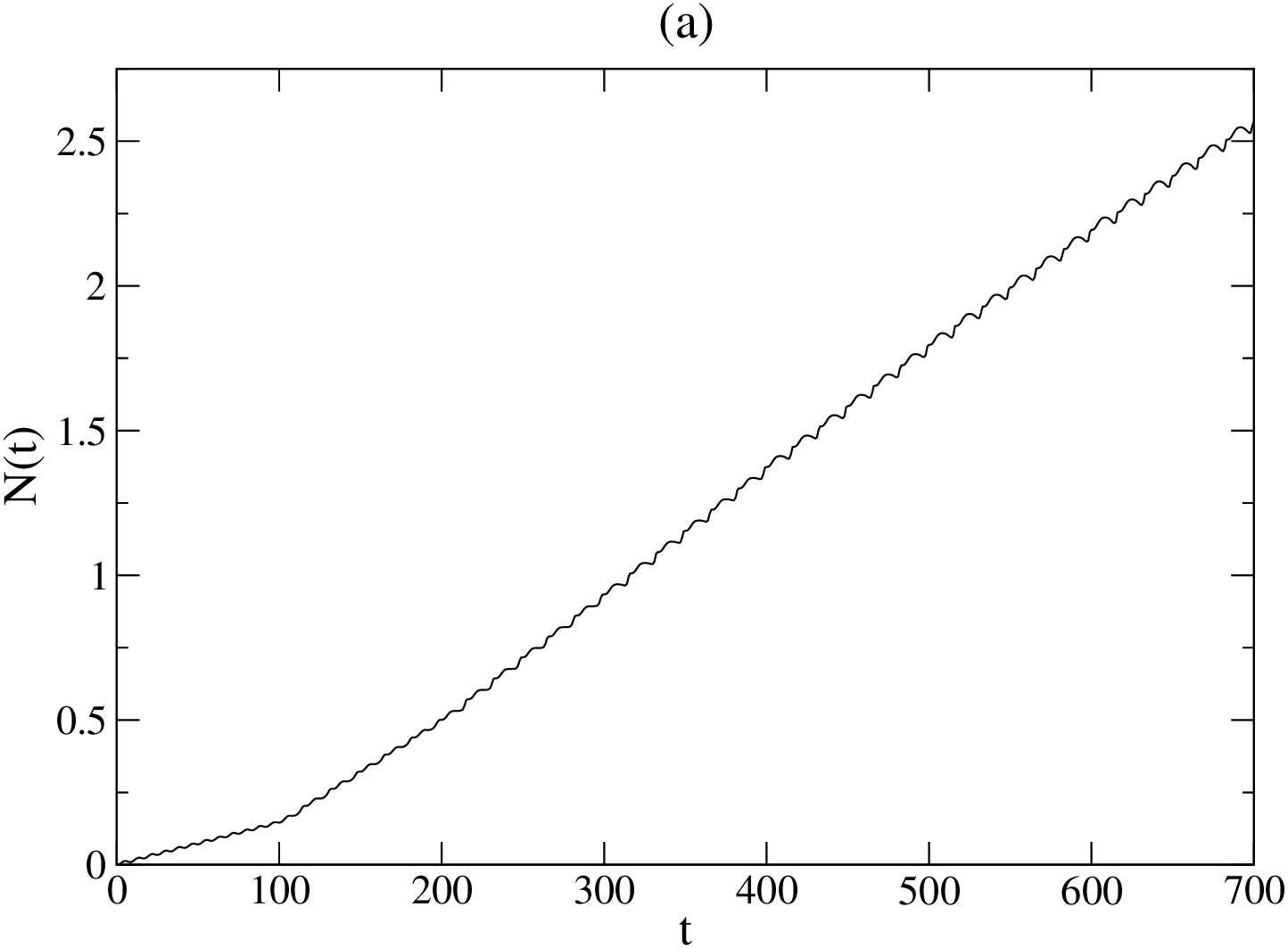}
&
\includegraphics[height=5.5cm]{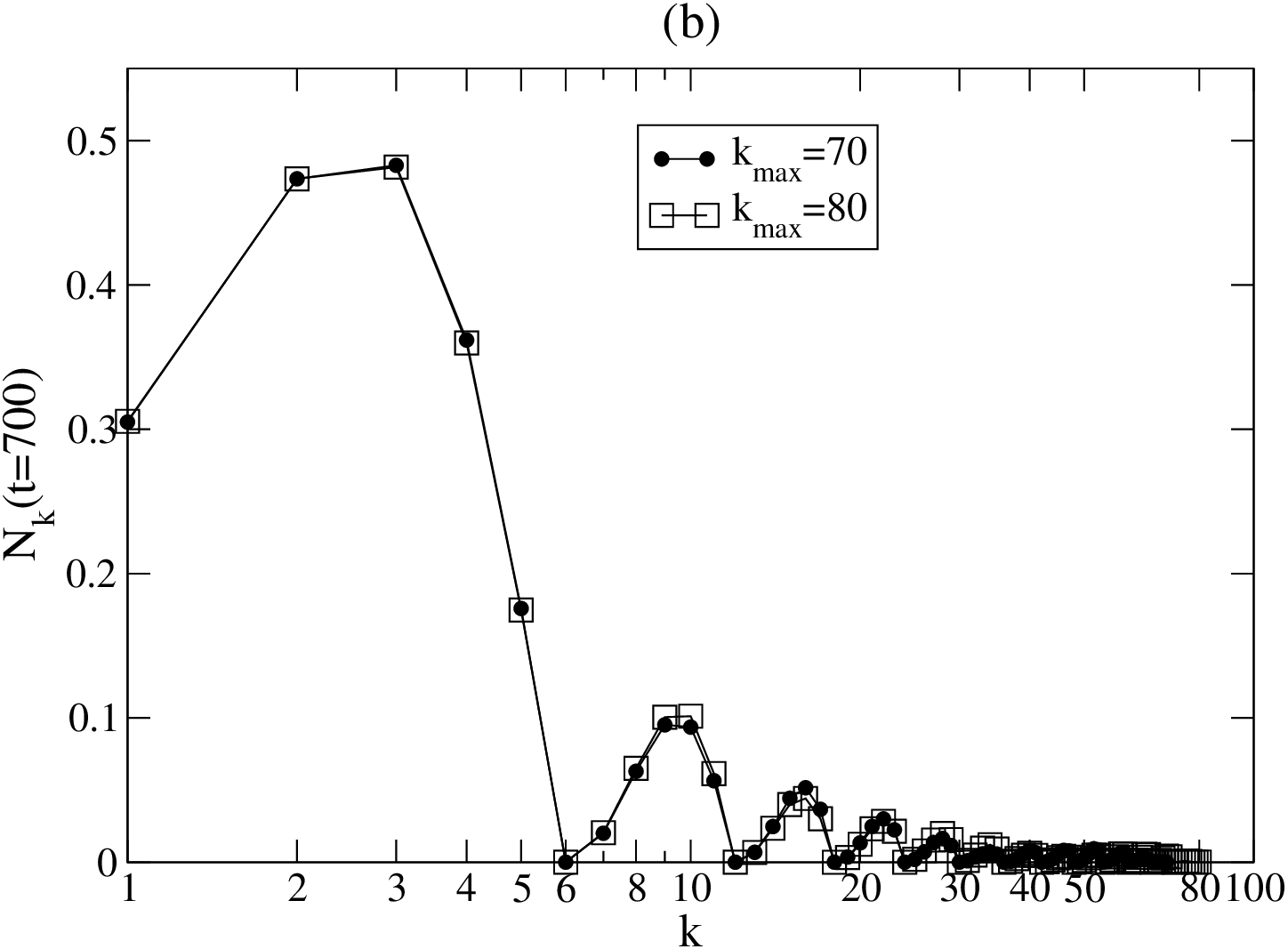}
\end{tabular}
\caption{ (a)  Total number of particles produced in a cavity vibrating with (\ref{background 2}) and 
                        $n=3$, i.e. 
		     $\omega_{\rm cav}=2\,\Omega_3^0$,  
	               for parameters $l_0=50$, $\epsilon=0.02$ and $k_{\rm max}=80$.       
                 (b)  Particle spectrum corresponding to (a) for the two cut-off parameters $k_{\rm                  
                       max}=70$ and $k_{\rm max}=80$.
                  \label{figure 15}}
\end{center}
\end{figure}

Nevertheless, the work \cite{Antunes} provides a good possibility for checking our method against a completely different numerical approach. To illustrate this by means of one more example we show the numerical result for $n=3$ , i.e. $\omega_{\rm cav}=2\,\Omega_3^0$ (which corresponds to the case $n=6$ in \cite{Antunes}), and the same parameters $l_0=50$ and $\epsilon=0.02$ in Figure \ref{figure 15} which perfectly agrees with the corresponding result obtained in \cite{Antunes} \footnote[5]{Compare Figure \ref{figure 15} (a) with the case $n=6$ in Fig. 6 of \cite{Antunes} and Figure \ref{figure 15} (b) with Fig. 7 of \cite{Antunes}.}. 
As mentioned by the author of \cite{Antunes} this behaviour of the particle number does not fit any simple expression. 

\begin{figure}
\begin{center}
\begin{tabular}{cc}
\includegraphics[height=5.5cm]{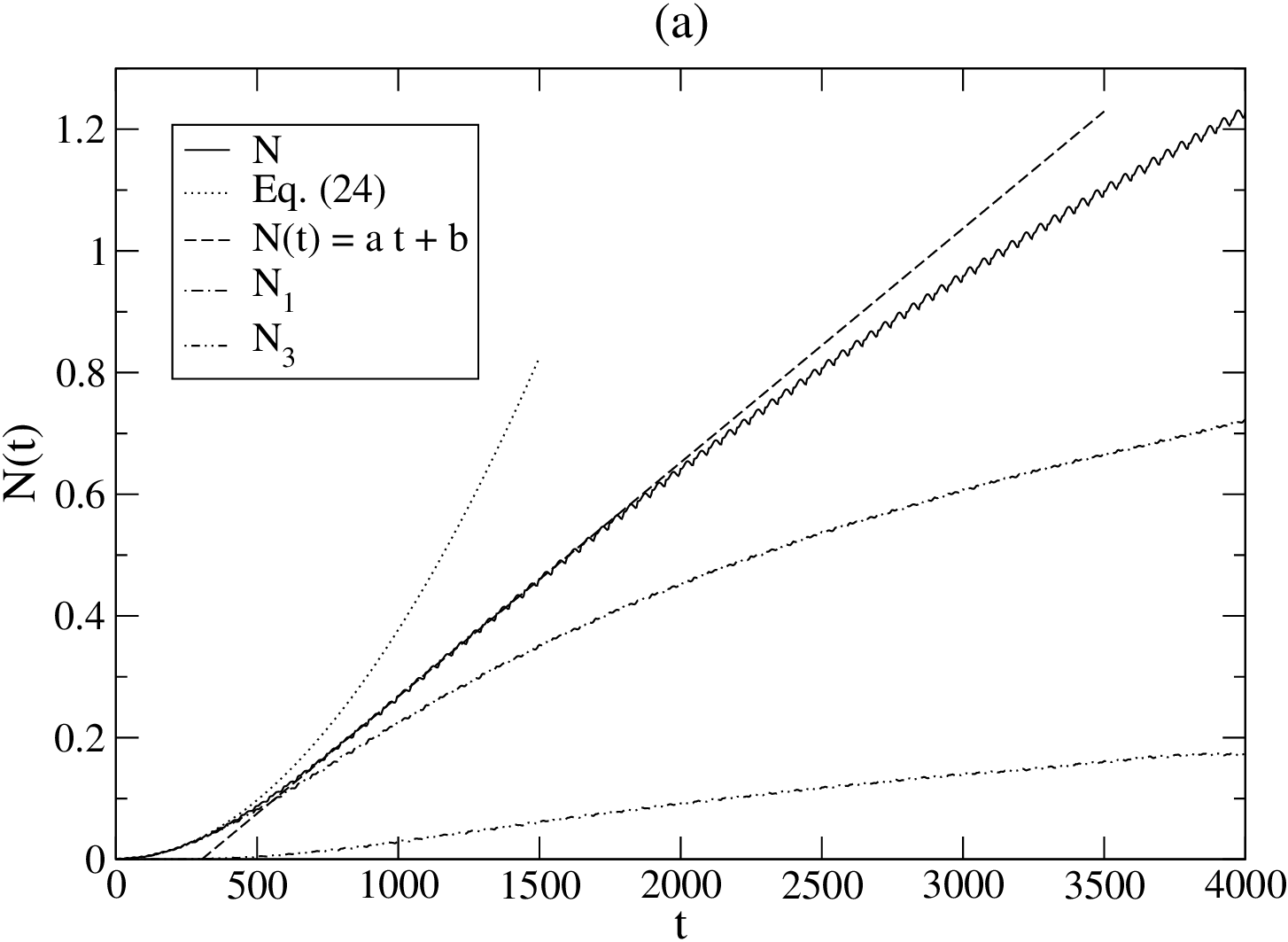}
&
\includegraphics[height=5.5cm]{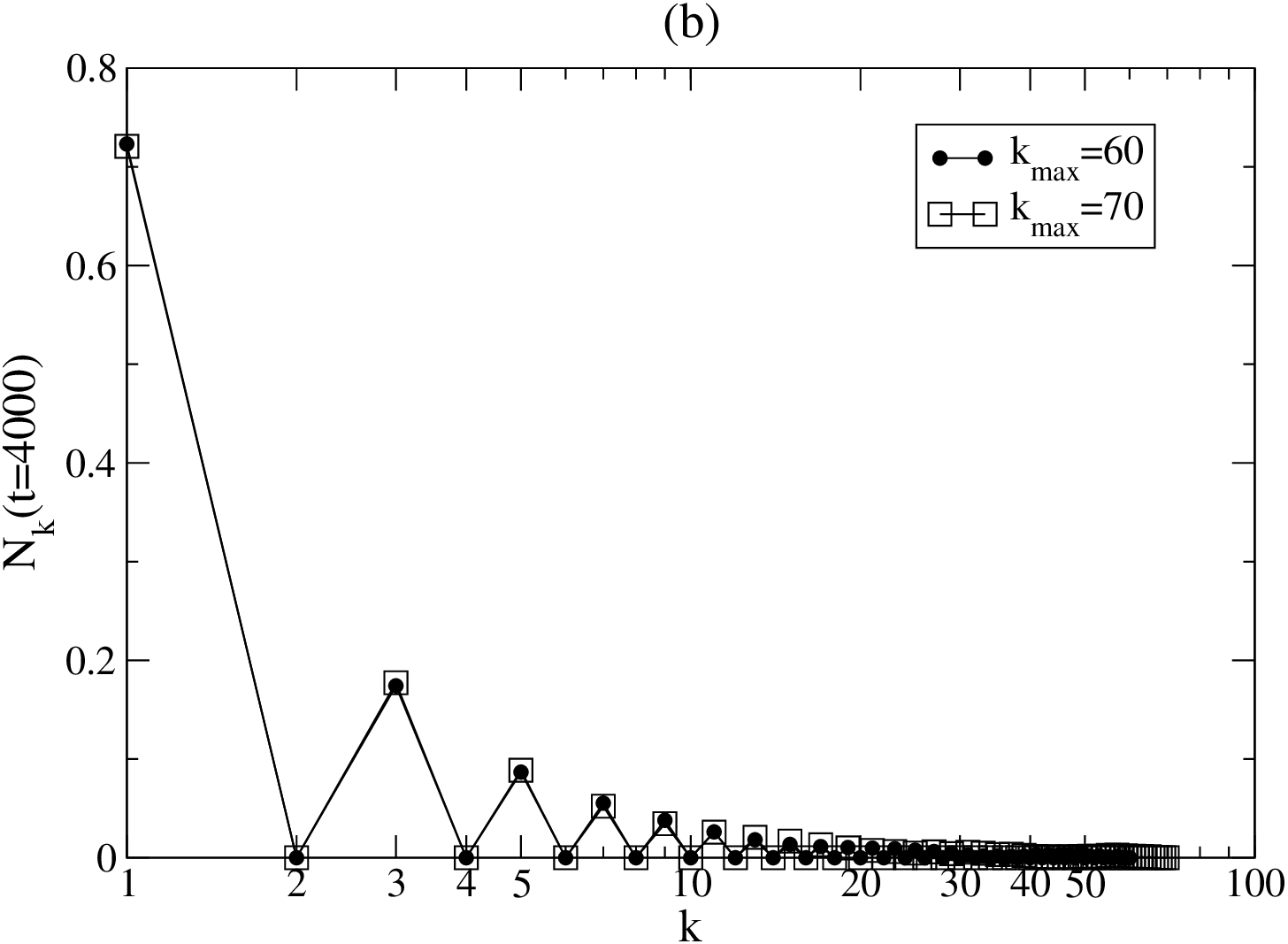}
\end{tabular}
\end{center}
\caption{ (a)  Number of particles produced in a cavity vibrating with (\ref{background 2}) and $n=1$ 
		     for the parameters of     
		     Fig. \ref{figure 14}, but now for an integration time $t_{\rm max}=4000$ and 
		     $k_{\rm  max}=70$. The analytical expression (\ref{dodonov2}) and the linear law     $N(t)=a\,t+b$ with $a=0.000385$ obtained by fitting the numerical data in the time range $\frac{1}{2}\le \tau \le 1$ are shown as well (see the discussion in the text).
	       (b)  Particle spectrum corresponding to (a) for the two cut-off parameters $k_{\rm                  
                       max}=60$ and $k_{\rm max}=70$.
\label{figure 16}}
\end{figure}

Now, let us finally address the issue of further deceleration of the rate of particle creation
which yields a deviation of the time evolution of the number of created particles from the linear behaviour for larger times which is indicated in Figs. \ref{figure 10} (a) and \ref{figure 11} (a) [see also Fig. \ref{figure 13} (a)]. For this purpose it turns out that, from a numerical point of view, it is convenient to maintain the  parameters $l_0=50$ and $\epsilon=0.02$.  For these parameters, Fig \ref{figure 14} (a) nicely demonstrates the transition from the initially quadratic to the linear behaviour of the number of particles created in a cavity vibrating with (\ref{background 2}) and $n=1$, i.e. $\omega_{\rm cav}=2\Omega_1^0$. 
In Figure \ref{figure 16} we show the numerical results obtained for an integration time $t_{\rm max}=4000$ (which corresponds to $\tau_{\rm max}=\frac{4}{5}\pi >1$)  together with Eq. (\ref{dodonov2}) and a linear fit to the total particle number \footnote[1]{The time interval in which we have performed the fitting procedure is $796 \le t \le 1592$ corresponding to the "definition" of the linear regime $1/2 \le \tau \le 1$ given above. The value $a=0.000385 \sim 0.0004$ may indicate the validity of the dependence $a=\frac{\epsilon}{l_0}$, i.e. $\frac{d N(\tau)}{d\tau}=\frac{2}{\pi}$. We have performed numerical simulations for $l_0=50$, $\epsilon=0.03$ and $l_0=25$, $\epsilon=0.02$ up to $\tau=1$ yielding $a=0.00058 \sim 0.0006$ and $a=0.00077 \sim 0.0008$, respectively. These results strongly support the indication that $\frac{dN(\tau)}{d\tau}=\frac{2}{\pi}$ holds in general for $1/2 \le \tau \le 1$.}. As already observed in Fig. \ref{figure 14} (a) the quadratic increase of the total particle number up to $t\sim 300$ is followed by a transition to the linear regime. For times $t > 1600$ ($\tau > 1$) the particle number starts to leave the region in which it is effectively described by a linear behaviour and shortly after ($t \sim 2000$) the further deceleration of the rate of particle creation yields a rising deviation of the numerical values from the linear fit. The number of particles produced in the modes $k=1$ and $k=3$, shown in Fig. \ref{figure 16} (a) as well, exhibit the same qualitative behaviour. Therefore, for a cavity vibrating with (\ref{background 2}), the overall time evolution of the total particle number as well as the number particles created in the mode $k=1$ cannot be described (and fitted) by a comparably simple expression similar to the analytical results (\ref{dodonov1}) and (\ref{dodonov2}) valid for the cavity motion (\ref{background 1}). This behaviour is due to a detuning effect as we will explain in the next subsection.

\subsection{$\delta(t)=\delta_2(t)$ - Detuning and the "true resonance"}

In order to understand the long time behaviour of the number of particles produced in a cavity oscillating with (\ref{background 2}) we have to formulate the resonance condition more carefully.
For a resonance to happen the external frequency $\omega_{\rm cav}$ has to be twice the eigenfrequency of a quantum mode defined with respect to the average position $\bar{l}$ of the cavity, i.e. $\omega_{\rm cav}=2\bar{\Omega}_n$ where $\bar{\Omega}_n=n\pi/\bar{l}$. 

Whereas for the cavity motion (\ref{background 1}) the average frequency $\bar{\Omega}_n$ corresponds to the frequency $\Omega_n^0$ defined with respect to the initial vacuum state 
$|\Omega_0 \rangle$ (see section 2) the situation is different for the background motion 
(\ref{background 2}). In this case the average position of the cavity is $\bar{l}=l_0(1+\epsilon)$ (see also Fig. \ref{figure 1}) and therefore $\bar{\Omega}_n\neq \Omega_n^0$.  Thus, Eq. (\ref{background 2}) is not a real resonant cavity motion because the cavity frequency does not match the exact 
resonance condition. Such an effect is called detuning (see, e.g., \cite{Crocce:2001,Dodonov:1998}) and can be parametrized by the detuning parameter $\Delta$. In particular, for the cavity motion (\ref{background 2}) we have $\omega_{\rm cav}=2(\bar{\Omega}_n+\Delta)$ with $\Delta=\Omega_n^0-\bar{\Omega}_n=\Omega_n^0(\bar{l}-l_0)/{\bar{l}}$, which for $n=1$ and $l_0=1$ reduces to $\Delta=\pi\left(\frac{\epsilon}{1+\epsilon}\right)\sim \pi\,\epsilon$.

We have studied how detuning affects the particle production in a one-dimensional cavity   in dependence on the detuning parameter $\Delta$ in detail. For instance for the cavity motion (\ref{background 1}), we have found that the number of created particles oscillates in time with a period 
and an amplitude depending on the detuning parameter $\Delta$, i.e. particles are created and annihilated periodically in time. Therefore, detuning causes phases of acceleration as well as deceleration in the time evolution of the particle number and is the reason for the qualitative behaviour of the particle number which we have observed in the former subsection. These results will be reported and discussed elsewhere in more detail \cite{Ruser}. 

\begin{figure}
\begin{center}
\begin{tabular}{cc}
\includegraphics[height=5.5cm]{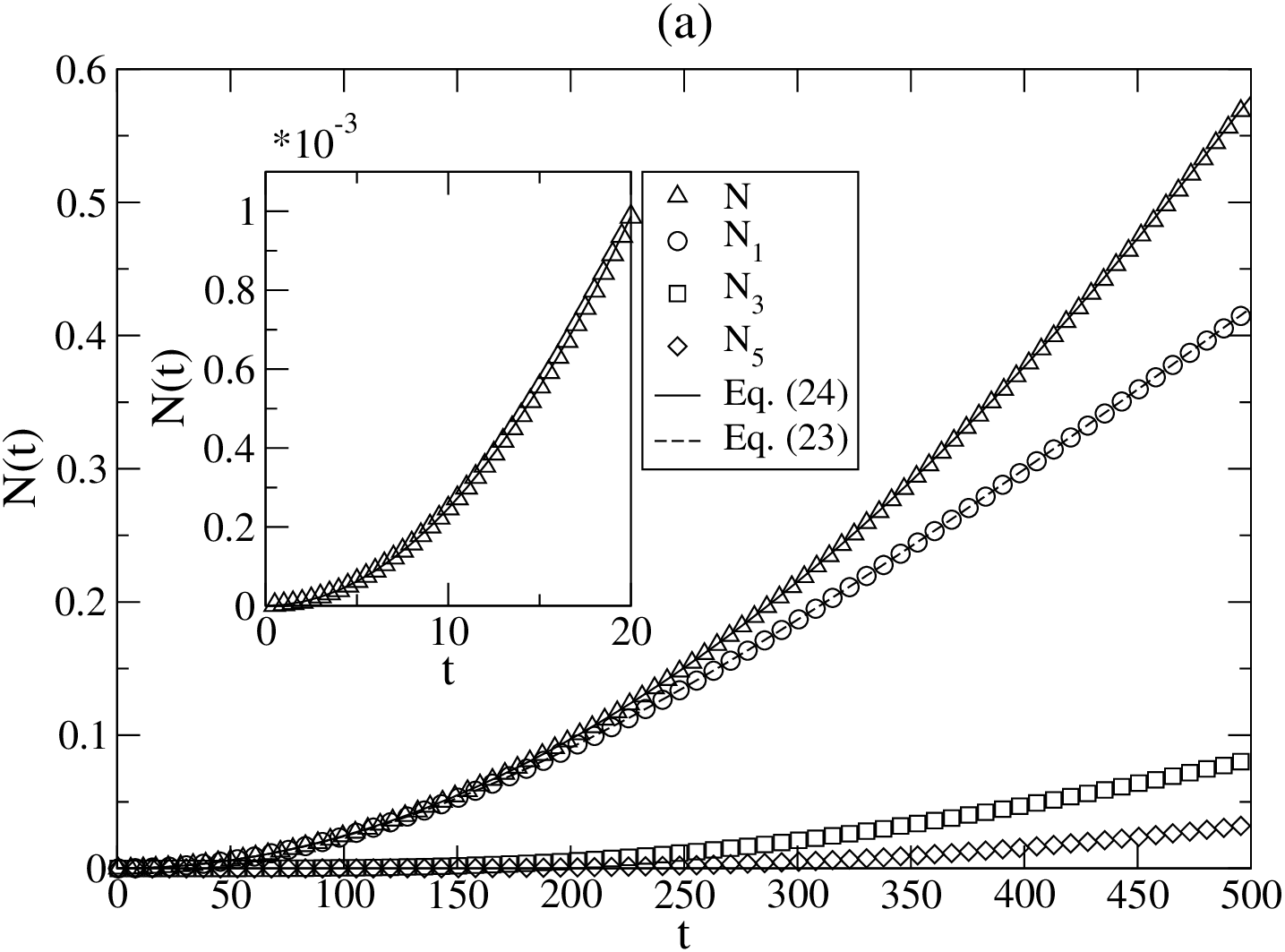}
&
\includegraphics[height=5.5cm]{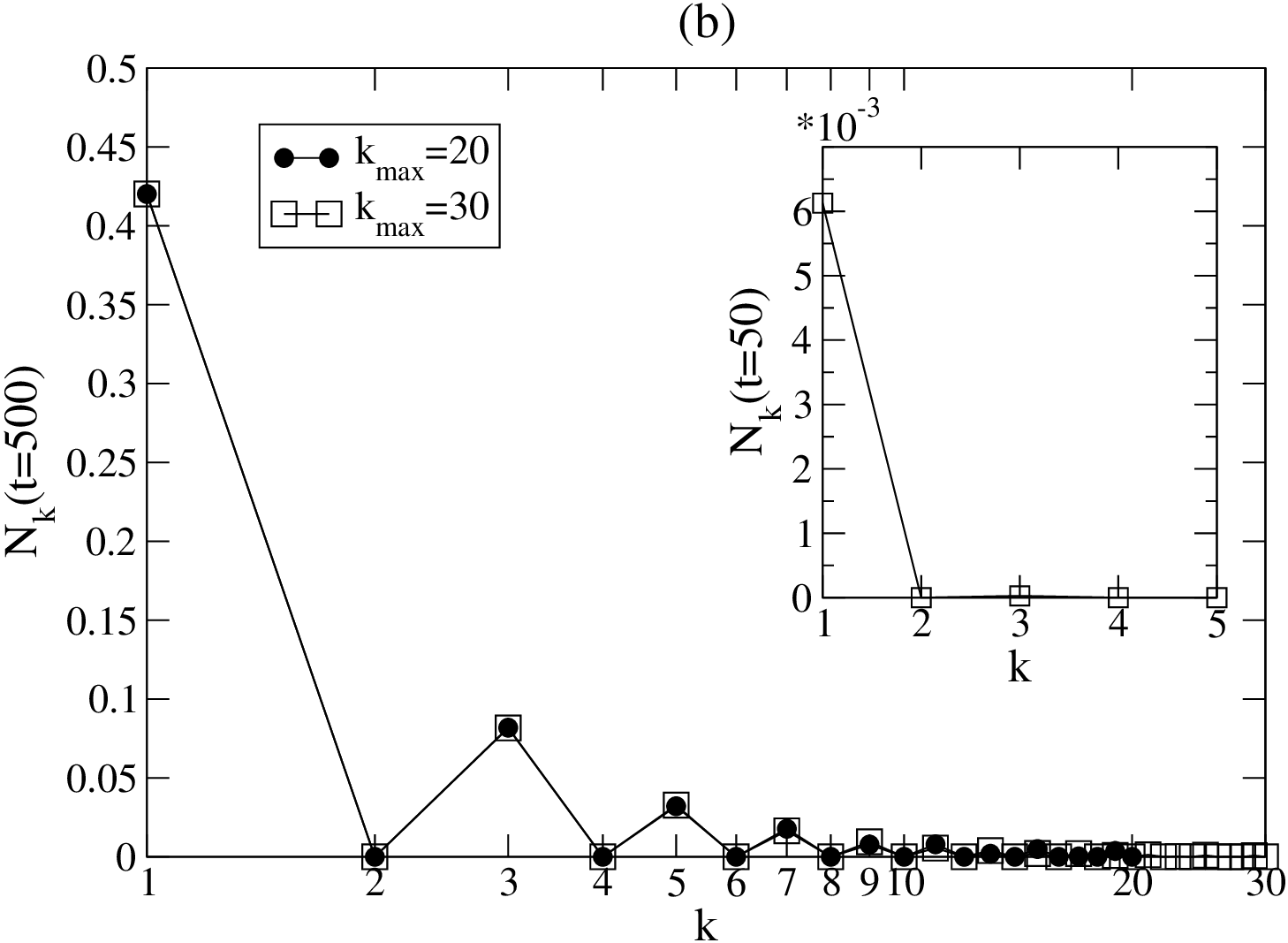}
\end{tabular}
\end{center}
\caption{ (a)  Number of particles produced in a cavity vibrating with (\ref{background 2}) but cavity frequency $\omega_{\rm cav}=2\frac{\pi}{l_0(1+\epsilon)}$, i.e. no detuning,          
                        for parameters $l_0=1$, $\epsilon=0.001$ and $k_{\rm max}=30$,        
                        together with the analytical predictions (\ref{dodonov1}) and (\ref{dodonov2}).
                 (b)  Particle spectrum corresponding to (a) for the two cut-off parameters $k_{\rm                  
                       max}=20$ and $k_{\rm max}=30$.
                       \label{figure 17}}
\end{figure}

\begin{figure}
\begin{center}
\begin{tabular}{cc}
\includegraphics[height=5.5cm]{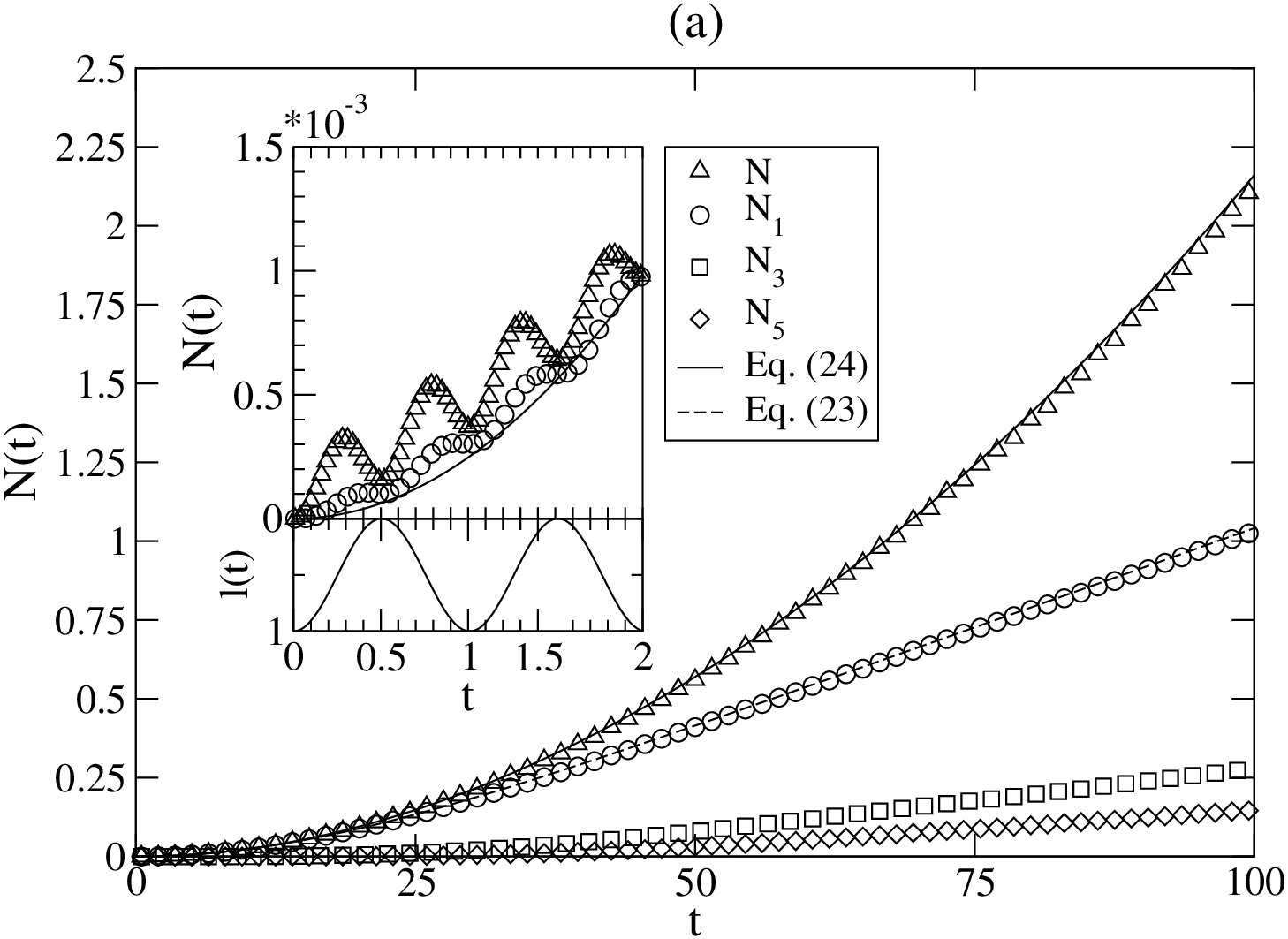}
&
\includegraphics[height=5.5cm]{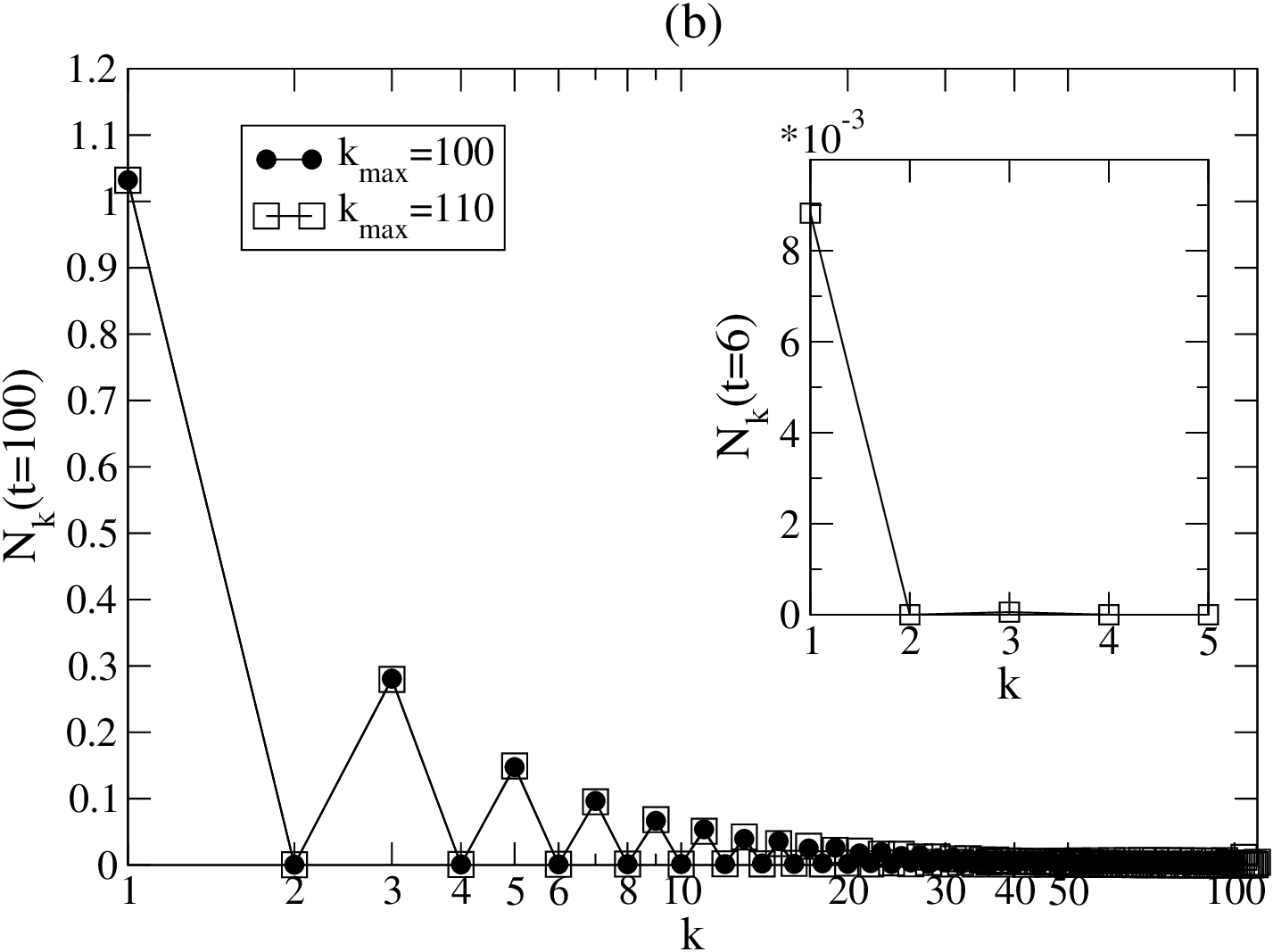}
\end{tabular}
\end{center}
\caption{ (a)  Number of particles produced in a cavity vibrating with (\ref{background 2}) but cavity   frequency $\omega_{\rm cav}=2\frac{\pi}{l_0(1+\epsilon)}$, i.e. no detuning,          
                        for parameters $l_0=1$, $\epsilon=0.01$ and $k_{\rm max}=110$,        
                        together with the analytical predictions (\ref{dodonov1}) and (\ref{dodonov2}).
                 (b)  Particle spectrum corresponding to (a) for the two cut-off parameters $k_{\rm                  
                       max}=100$ and $k_{\rm max}=110$.
                    \label{figure 18}}
\end{figure}

In this paper we restrict ourselves to the demonstration, that on replacing the cavity frequency $\omega_{\rm cav}=2\Omega_1^0=2\frac{\pi}{l_0}$ in (\ref{background 2}) by $\omega_{\rm cav}=2\bar{\Omega}_1=2\,\frac{\pi}{l_0(1+\epsilon)}$, i.e. $\Delta=0$ (no detuning), the two cavity motions (\ref{background 1}) and (\ref{background 2}) do indeed yield the same qualitative long time behaviour
for the particle number \footnote[1]{Note that we still work with the same vacuum state $|\Omega_0\rangle$ as before (the one defined with respect to the eigenfrequencies $\Omega_n^0$) and do not change the notion of particles. Instead we change only the external frequency $\omega_{\rm cav}$ which now is not twice the frequency of a quantum mode defined with respect to the vacuum state $|\Omega_0\rangle$.}. 
In Figs. \ref{figure 17} and \ref{figure 18} we show the numerical results for the parameters $l_0=1$ and $\epsilon=0.001$ and $0.01$, respectively. For a better comparison, we have depicted the analytical expressions (\ref{dodonov1}) and (\ref{dodonov2}) derived for the cavity motion (\ref{background 1}) as well.
Thereby we now use $\tau=\frac{1}{2}\epsilon \,\bar{\Omega}_1\,t$ as definition for the "slow time" to account for the modified cavity frequency.
 
As we infer from Figs. \ref{figure 17} (a) and  \ref{figure 18} (a) the time evolution of the number of created particles is now qualitatively as well as quantitatively in good agreement with the analytical predictions (\ref{dodonov1}) and (\ref{dodonov2}). 
Instead of the linear behaviour and the further deceleration in the rate of particle creation observed for the cavity motion (\ref{background 2}) with $\omega_{\rm cav}=2\Omega_1^0$ [cf.  Figs. \ref{figure 9} (a) and \ref{figure 10} (a)] the time evolution of the number of created particles shows now the typical characteristics of resonant particle creation. For this reason we call the scenario with the cavity dynamics (\ref{background 2}) and $\omega_{\rm cav}=2\bar{\Omega}_1$ the "true resonance" scenario. 

Note that the number of excited modes (the value of the cut-off parameter $k_{\rm max}$) is now equal
to the number of modes excited in a cavity vibrating with (\ref{background 1}) [compare Figs. \ref{figure 18} (b)  and \ref{figure 5} (b)]. Thus, the fact that the number of excited modes is less for the cavity motion (\ref{background 2}) with $\omega_{\rm cav}=2\Omega_1^0$ is due to the effect of detuning and not because of the smooth initial condition.  
This indicates that the discontinuity in the velocity of the cavity motion (\ref{background 1}) at the beginning of the integration does not play an important role for the long time behaviour of the particle production.

\subsection{$\delta(t)=\delta_3(t)$}

Finally, we show one example for particle creation caused by the cavity motion
\begin{equation}
l(t)=l_0\left[1+\epsilon \sin^3\left(2\,\Omega_n^0\,t\right)\right]
\label{background 3}
\end{equation}
to illustrate an interesting effect which appears in this case.
\begin{figure}
\begin{center}
\begin{tabular}{cc}
\includegraphics[height=5.5cm]{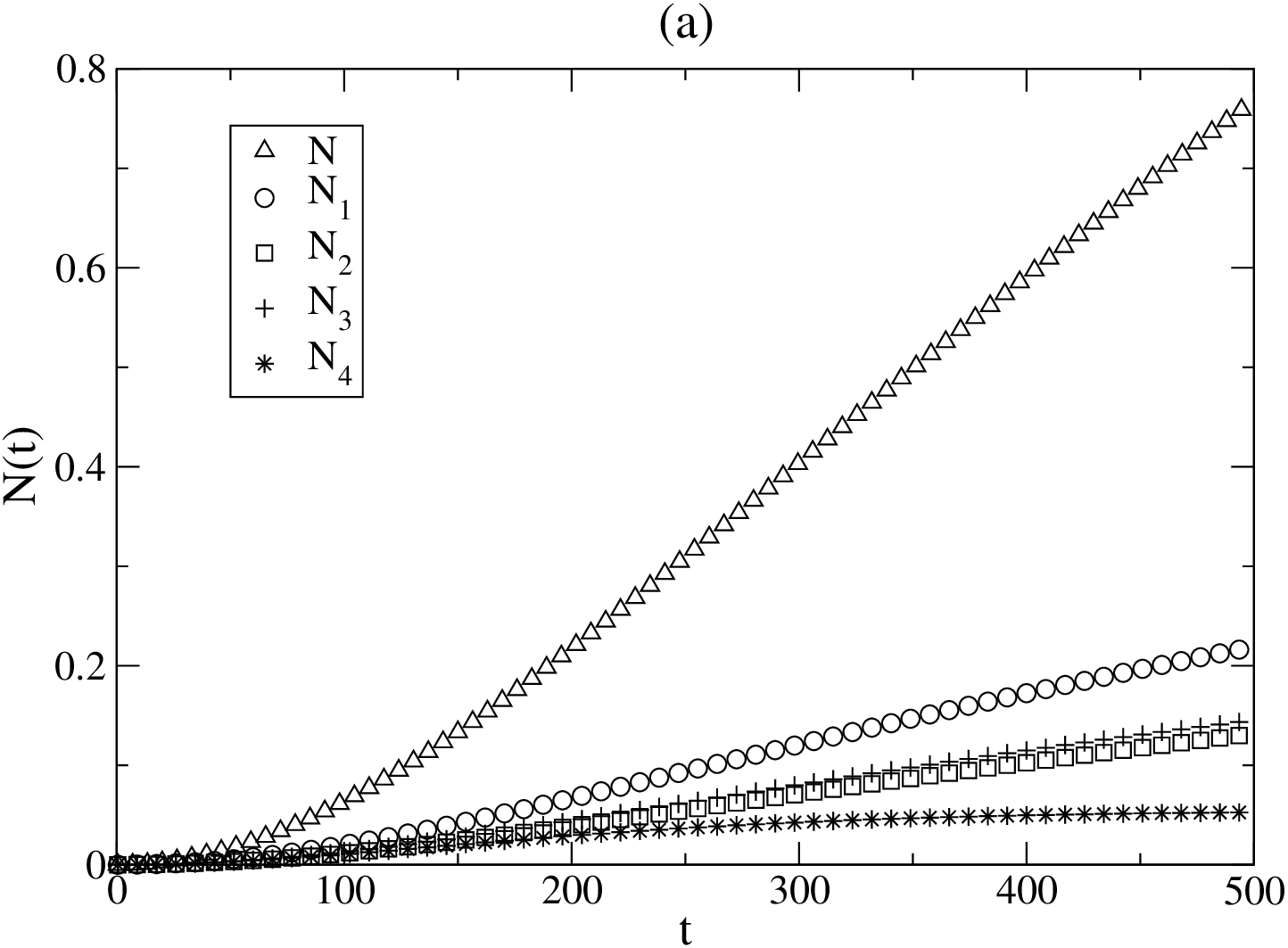}
&
\includegraphics[height=5.5cm]{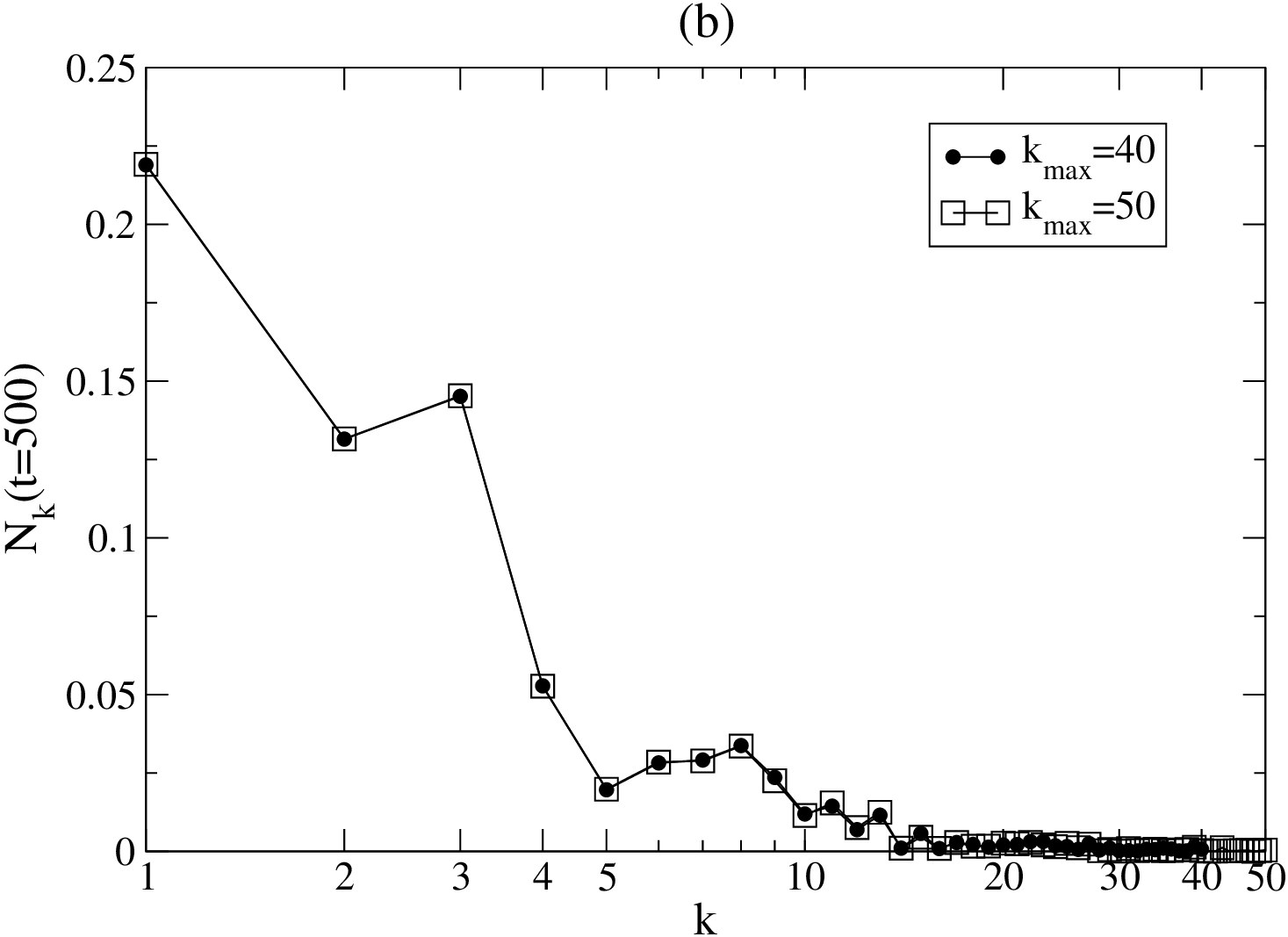}
\end{tabular}
\end{center}
\caption{ (a)  Number of particles produced in a cavity vibrating with (\ref{background 3}) and cavity   frequency $\omega_{\rm cav}=2\frac{\pi}{l_0}$           
                        for parameters $l_0=1$, $\epsilon=0.001$ and $k_{\rm max}=50$.        
                    (b)  Particle spectrum corresponding to (a) for the two cut-off parameters $k_{\rm                  
                       max}=40$ and $k_{\rm max}=50$.
                  \label{figure 19}}
\end{figure}
Figure \ref{figure 19}  shows the numerical results obtained for the parameters $l_0=1$, $\epsilon=0.001$ and $\omega_{\rm cav}=2\Omega_1^0$. The qualitative behaviour of the time evolution of the number of created particles looks similar to the one observed for the cavity motion (\ref{background 2}). But in the case of the motion (\ref{background 3}) the shape of the particle spectrum changes drastically compared to the scenarios which we have discussed before [see Fig. \ref{figure 19} (b)].
Now, not only odd but also even modes become excited. The resonance mode is of course the one which is excited most but a kind of regular structure in the spectrum seems to be absent. It is suggested that the reason for this irregular spectrum is due to the more complicated structure of the coupling function $\gamma(t)=\frac{\dot{l}(t)}{l(t)}$ for the motion (\ref{background 3}) compared to (\ref{background 1}) and (\ref{background 2}) [see Fig. \ref{figure 1} (b)].
To investigate such phenomena will be part of further work.

\section{Conclusions}

We have presented a parametrization for the time evolution of the field modes inside a dynamical cavity  allowing for efficient numerical calculation of particle production in the dynamical Casimir effect. The creation of real massless scalar particles in a one-dimensional vibrating empty cavity has been studied numerically for three particular wall motions taking the intermode coupling into account. 
The comparison of our numerical results obtained for the cavity motion (\ref{background 1}) with the analytical predictions derived in \cite{Dodonov:1996,Dodonov:1993, Ji:1997} shows that the numerical calculations are reliable and that the introduced method is appropriate to study particle creation in dynamical cavities. Furthermore, the range of validity of the analytical expressions found in \cite{Dodonov:1996} has been investigated. We have found that the Eqs. (\ref{dodonov1}) and  (\ref{dodonov2}) derived in the limit $\epsilon \ll 1$ hold to describe the numerical results up to vibration amplitudes of the order of  $\epsilon=0.01$. The impact of the discontinuity in the velocity of the mirror at the beginning of the motion yielding instantaneous particle creation has been discussed. These results have been compared with the results obtained for the cavity motion (\ref{background 2}). The numerical simulations show that the short time behaviours of the number of created particles are similar in both cases, i.e. $N(t)$ grows quadratically with time. But the long time behaviours are different. While for the cavity motion (\ref{background 1}) the total particle number increases quadratically for large times ($\tau > 1$) it passes trough a linear regime ($\frac{1}{2} < \tau < 1$) for the motion (\ref{background 2}) 
and afterwards slows down further. The different behaviours of the particle creation for the two cavity motions (\ref{background 1}) and (\ref{background 2}) yield a different behaviour for the time evolution of the energy associated with the created quantum radiation. We have found that, in accordance with \cite{Dodonov:1996},  the energy in a cavity vibrating with (\ref{background 1}) increases exponentially whereas, in the time range under consideration, it obeys a quadratic law (for $\epsilon=0.001$ and $0.01$) in the case of the cavity motion (\ref{background 2}). The explanation for the differences in the behaviour of the particle production for the cavity motions (\ref{background 1}) and (\ref{background 2}) is that, due to detuning,  Eq. (\ref{background 2}) is not a real resonant cavity motion. Without detuning, both cavity motions yield the same qualitative behaviour for the time evolution of the particle number. Taking this into account, we may conclude that the discontinuity in the velocity of the cavity motion (\ref{background 1}) at the beginning of the dynamics does not strongly affect the long time behaviour of the particle production.

An extension of the presented method to massive fields as well as to higher dimensions  is straightforward, which makes it possible to study the particle creation in the dynamical Casimir effect for scenarios where no analytical results can be deduced.

\section{Acknowledgements}

The author is grateful to Ruth Durrer, Ralf Sch{\"u}tzhold and G{\"u}nter Plunien for 
valuable and clarifying discussions and comments on the manuscript. 
Furthermore, the author would like to thank the organizers of the 
{\it International Workshop on the Dynamical Casimir Effect} in Padova/Italy 2004 
(see \cite{Padova}) for providing such a pleasant atmosphere for discussions. 
Financial support from the Swiss National Science Foundation and the Schmidheiny Foundation is gratefully acknowledged.   Finally, the author would like to thank the referee for useful suggestions.


\vspace*{1cm}


\begin{thebibliography}{widest-label}

\bibitem{Bordag:2001}
Bordag M, Mohideen U and Mostepanenko V M 2001 {\it Phys. Rept.}  {\bf 353} 1 

\bibitem{Dodonov:2001}
Dodonov V V 2001 {\it Nonstationary Casimir Effect And Analytical Solutions For Quantum Fields in Cavities With Moving Boundaries}, in Modern Nonlinear Optics, Part 1, Second Edition, Advances in Chemical Physics, Volume 119, Edited by Myron W E (John Wiley and Sons)

\bibitem{Bordag:1996}
Bordag M 1996 {\it Quantum Field Theory under the Influence of External Conditions}
(Teubner, Stuttgart)

\bibitem{Bordag:2002}
Bordag M (ed.) 2002,  {\it Quantum Field Theory under the Influence of External Conditions.} Proceedings, 5th Workshop, Leipzig, Germany, September 
10-14, 2001, Int. J. Mod. Phys. A {\bf 17} 711

\bibitem{Grib:1994}
Grib A A, Mamayev S G and Mostepanenko V M 1994, {\it Vacuum Quantum Effects in Strong Fields} (Friedmann Laboratory Publishing, St. Petersburg)

\bibitem{Fulling:1976}
Fulling S A and Davis P C W 1976 {\it Proc. R. Soc. London} {\bf A348} 393 .

\bibitem{Davis:1977}
Davis P C W and Fulling S A 1977 {\it Proc. R. Soc. London} {\bf A356} 237

\bibitem{Ford:1982}
Ford L H and Vilenkin A 1982 {\it Phys. Rev D} {\bf 25} 2569

\bibitem{Neto:1996}
Maia Neto P A and Machado L A S 1996 {\it Phys. Rev. A} {\bf 54} 3420

\bibitem{Schuetzhold:1998}
Sch{\"u}tzhold R, Plunien G and Soff G 1998 {\it Phys. Rev. A}  {\bf 57}  2311

\bibitem{Birrell:1982}
Birrell N D and Davis P C W 1982 {\it Quantum fields in curved space}
(Cambridge Univesity Press, Cambridge)

\bibitem{Moore:1970}
Moore G T 1970 {\it J. Math. Phys.}  {\bf 11}  2679

\bibitem{Castagnino:1984}
Castagnino M and Ferraro R 1984 Ann. Phys. (N.Y.) {\bf 154} 1

\bibitem{Lambrecht:1996}
Lambrecht A, Jaekel M - T and Reynaud S 1996 {\it Phys. Rev. Lett.} {\bf 77} 615   

\bibitem{Dodonov:1996}
Dodonov V V and Klimov A B 1996 {\it Phys. Rev. A } {\bf 53}  2664

\bibitem{Dodonov:1996a}
Dodonov V V 1996 {\it Phys. Lett. A} {\bf 213} 219

\bibitem{Dodonov:1993}
Dodonov V V, Klimov A B  and Nikonov D E 1993 {\it J. Math. Phys. } {\bf 34}  2742

\bibitem{Ji:1997}
Ji J Y, Jung H H, Park J W and Soh K S  1997 {\it Phys. Rev. A} {\bf 56}  4440

\bibitem{Dalvit:1998}
Dalvit D A R and Mazzitelli F D 1998 {\it Phys. Rev. A} {\bf 57} 2113

\bibitem{Dalvit:1999}
Dalvit D A R an Mazzitelli F D 1999 {\it Phys. Rev. A} {\bf 59} 3049

\bibitem{Cole:1995}
Cole C K and Schieve W C 1995 {\it Phys. Rev. A} {\bf 52} 4405

\bibitem{Law:1994}
Law C K 1994 {\it Phys. Rev. Lett.} {\bf 73} 1931 

\bibitem{Wegrzyn:2001}
Wegrzyn P and Rog T 2001 {\it Act. Phys. Pol.} {\bf 32} 129
\bibitem{Law:1995}
Law C K 1994 {\it Phys. Rev. A} {\bf 51} 2537

\bibitem{Golestanian:1997}
Golestanian R and Kardar M 1997 {\it Phys. Rev. Lett.} {\bf 78} 3421

\bibitem{Cole:2001}
Cole C K and Schieve W C 2001 {\it Phys. Rev. A} {\bf 64} 023813

\bibitem{Crocce:2001}
Crocce M, Dalvit D A R and Mazzitelli F D 2001 {\it Phys. Rev. A} {\bf 64} 013808

\bibitem{Dodonov:2003}
Dodonov A V, Dodonov E V and Dodonov V V 2003, quant-ph/0308144

\bibitem{Mundarain:1998}
Mundarain D F and Maia Neto P A 1998 {\it Phys. Rev. A} {\bf 57} 1379

\bibitem{Dodonov:1998}
Dodonov V V 1998 {\it Phys. Rev. A} {\bf 58} 4147

\bibitem{Dodonov:1998a}
Dodonov V V 1998 {\it Phys. Lett. A} {\bf 244} 517

\bibitem{Antunes}
Antunes N D 2003 hep-ph/0310131

\bibitem{Ruser}
Ruser M in preparation

\bibitem{Gradshteyn:1994}
Gradshteyn I S and Ryzhik I M 1994 {\it Tables of Integrals, Series and Products}
(Academic, New York)

\bibitem{Razavy:1985}
Razavy M and Terning J 1985 {\it Phys. Rev. D} {\bf 31} 307

\bibitem{Padova}
{\tt http://www.pd.infn.it/casimir}
\end{thebibliography}
\end{document}